\RequirePackage{fix-cm} 
\documentclass[a4paper, twoside, reqno, dvips, 12pt]{amsart}
\usepackage{fixltx2e}   

\usepackage{etex}

\usepackage[latin1]{inputenc}
\usepackage[T1]{fontenc}

\usepackage{algorithm}
\usepackage[noend]{algpseudocode}

\usepackage[titletoc,title]{appendix}

\usepackage{esint}
\usepackage{dsfont}
\usepackage{xspace}
\usepackage{amsgen}
\usepackage{amsthm}
\usepackage{amssymb}
\usepackage{amsmath}
\usepackage{wasysym}
\usepackage{upgreek}
\usepackage{amsfonts}
\usepackage{stmaryrd}
\usepackage{mathtools}

\usepackage{units}
\usepackage{nicefrac}

\usepackage{relsize}
\usepackage{textcomp}
\usepackage{textgreek}
\usepackage[mathcal, mathscr]{euscript}

\usepackage{mathrsfs}
\DeclareMathAlphabet{\mathscrbf}{OMS}{mdugm}{b}{n}

\usepackage{a4wide}

\usepackage[dvipsnames, table]{xcolor}
\definecolor{bckg}{RGB}{20.8, 20.8, 20.8}
\definecolor{oneblue}{rgb}{0.0, 0.0, 0.85}
\definecolor{Lightblue}{RGB}{214, 214, 214}
\definecolor{bluepigment}{rgb}{0.2, 0.2, 0.6}
\definecolor{charcoal}{rgb}{0.21, 0.27, 0.31}
\definecolor{denimblue}{rgb}{0.08, 0.38, 0.74}
\definecolor{Lightgray}{rgb}{0.89, 0.89, 0.89}
\definecolor{darkgrey}{rgb}{0.273, 0.281, 0.30}
\definecolor{darkelectricblue}{rgb}{0.33, 0.41, 0.47}

\usepackage{multirow}

\usepackage[sort&compress, comma, square, numbers]{natbib}

\usepackage{psfrag}
\usepackage{graphicx}
\usepackage{adjustbox}
\usepackage[tight]{subfigure}
\usepackage{morefloats}
\usepackage{indentfirst}

\usepackage[usenames, dvipsnames, pdf]{pstricks}
\usepackage{epsfig}
\usepackage{pst-grad} 
\usepackage{pst-plot} 

\usepackage{rotating}
\usepackage{pdflscape}

\usepackage{acronym}
\usepackage{microtype}
\usepackage[labelsep=period,%
            labelfont={bf,sf,color=bluepigment},%
            justification=raggedright]{caption}

\usepackage[perpage, symbol]{footmisc}

\usepackage[colorlinks,
           urlcolor=oneblue,
           linkcolor=denimblue,
           citecolor=NavyBlue,
           bookmarksopen=false,
           pdfpagemode=UseNone,
           pagebackref]{hyperref}

\usepackage[explicit]{titlesec}

\titleformat{\section}[block]
  {\color{NavyBlue}\Large\sffamily\bfseries}
  {}
  {0.0em}
  {\colorbox{bckg!5}{\strut\parbox{\dimexpr\linewidth-2\fboxsep\relax}{\thesection. #1}}}
  [\vspace*{0.33em}]

\titleformat{name=\section,numberless}[block]
  {\color{NavyBlue}\Large\sffamily\bfseries}
  {}
  {0.0em}
  {\colorbox{bckg!5}{\strut\parbox{\dimexpr\linewidth-2\fboxsep\relax}{#1}}}
  [\vspace*{0.33em}]

\titleformat{\subsection}
  {\color{NavyBlue}\large\sffamily\bfseries}
  {}
  {0.0em}
  {\colorbox{bckg!5}{\parbox{\dimexpr\linewidth-2\fboxsep\relax}{\thesubsection. #1}}}
  [\vspace*{0.33em}]

\titleformat{name=\subsection,numberless}
  {\color{NavyBlue}\Large\sffamily\bfseries}
  {}
  {0em}
  {\colorbox{bckg!5}{\parbox{\dimexpr\linewidth-2\fboxsep\relax}{#1}}}
  [\vspace*{0.33em}]

\titleformat{\subsubsection}
  {\color{bluepigment}\sffamily\normalsize\bfseries}
  {\thesubsubsection}
  {0.5em}
  {#1}
  [\vspace*{0.33em}]

\titleformat{\paragraph}[runin]
  {\color{bluepigment}\sffamily\small\bfseries}
  {}
  {0em}
  {#1}

\titlespacing{\section}{0.0em}{1.5em plus 2pt minus 2pt}%
{1.0em plus 2pt minus 2pt}[0em]
\titlespacing{\subsection}{0.5em}{1.5em plus 2pt minus 2pt}%
{1.0em}[0em]
\titlespacing{\subsubsection}{0.5em}{1.5em plus 2pt minus 2pt}%
{1.0em plus 2pt minus 2pt}[0em]

\usepackage{titletoc}

\contentsmargin{0.5em}
\newlength{\tocsep} 
\titlecontents{section}[\tocsep]
  {\addvspace{10pt}\bfseries\sffamily}
  {\contentslabel[\thecontentslabel]{\tocsep}}
  {}
  {\ \titlerule*[0.75pc]{.}\ \thecontentspage}
  []
\titlecontents{subsection}[\tocsep]
  {\addvspace{8pt}\sffamily}
  {\contentslabel[\thecontentslabel]{\tocsep}}
  {}
  {\ \titlerule*[0.5pc]{.}\ \thecontentspage}
  []
\titlecontents*{subsubsection}[\tocsep]
  {\addvspace{2pt}\footnotesize\sffamily}
  {}
  {}
  {\ \titlerule*[0.35pc]{.}\ \thecontentspage}
  [\\*]

\makeatletter
\def\@setauthors{%
  \begingroup
  \def\thanks{\protect\thanks@warning}%
  \trivlist
  \centering\footnotesize \@topsep30\p@\relax
  \advance\@topsep by -\baselineskip
  \item\relax
  \author@andify\authors
  \def\\{\protect\linebreak}%
  \textsc{\normalsize\textcolor{darkelectricblue}{\authors}}%
  \ifx\@empty\contribs
  \else
    ,\penalty-3 \space \@setcontribs
    \@closetoccontribs
  \fi
  \endtrivlist
  \endgroup
}
\def\@settitle{\begin{center}%
  \baselineskip14\p@\relax
    \bfseries
    \textsc{\Large\textcolor{charcoal}{\@title}}
  \end{center}%
}
\makeatother

\usepackage{enumitem}
\setlist[description]{%
  topsep=30pt,               
  itemsep=5pt,               
  font={\bfseries\sffamily\color{NavyBlue}}, 
}

\usepackage{fancyhdr}
\usepackage{lastpage}

\newcommand*\Title{\textcolor{bluepigment}{A spectral method for solving heat and moisture transfer}}
\newcommand*\Authors{\textcolor{bluepigment}{S.~Gasparin, D.~Dutykh \& N.~Mendes}}
\newcommand*{\plogo}{\textcolor{gray}{{\texttt{arXiv.org} / \textsc{hal}}}} 

\numberwithin{equation}{section}

\newcommand{\etal}{\emph{et al.\xspace}}

\newcommand*\egal{\ = \ }
\newcommand*\plus{\ + \ }
\newcommand*\moins{\ - \ }
\newcommand*\egalb{\, = \, }

\newcommand*{\Ox}{\Omega_{\, x}}
\newcommand*{\Ot}{\Omega_{\, t}}

\newcommand{\cM}{c_{\,M}}

\newcommand{\cT}{c_{\,T}}
\newcommand{\cw}{c_{\,w}}
\newcommand{\cz}{c_{\,0}}

\newcommand{\hM}{h_{\,M}}
\newcommand{\hML}{h_{\,M,\,L}}
\newcommand{\hMR}{h_{\,M,\,R}}
\newcommand{\hT}{h_{\,T}}
\newcommand{\hTL}{h_{\,T,\,L}}
\newcommand{\hTR}{h_{\,T,\,R}}
\newcommand{\Hl}{H_{\,l}}
\newcommand{\kl}{k_{\,l}}
\newcommand{\kM}{k_{\,M}}
\newcommand{\kMref}{k_{\,M,\,\text{ref}}}
\newcommand{\kTM}{k_{\,TM}}

\newcommand{\kT}{k_{\,T}}
\newcommand{\kTref}{k_{\,T,\,\text{ref}}}
\newcommand{\Lv}{L_{\,v}}

\newcommand{\Ps}{P_{\,s}}
\newcommand{\Pv}{P_{\,v}}
\newcommand{\Pvi}{P_{\,v, \,0}}
\newcommand{\Pvref}{P_{\,v, \,\text{ref}}}
\newcommand{\Pvinf}{P_{\,v, \, \infty}}
\newcommand{\Rv}{R_{\,v}}
\newcommand{\Ti}{T_{\,0}}
\newcommand{\Tinf}{T_{\, \infty}}
\newcommand{\Tref}{T_{\,\text{ref}}}
\newcommand{\Lref}{L_{\,\text{ref}}}
\newcommand{\tref}{t_{\,\text{ref}}}

\newcommand{\rhol}{\rho_{\,l}}

\newcommand{\rhoz}{\rho_{\,0}}

\newcommand{\BiM}{\mathrm{Bi}_{\,M}}
\newcommand{\BiML}{\mathrm{Bi}_{\,M,\,L}}
\newcommand{\BiMR}{\mathrm{Bi}_{\,M,\,R}}
\newcommand{\BiT}{\mathrm{Bi}_{\,T}}
\newcommand{\BiTL}{\mathrm{Bi}_{\,T,\,L}}
\newcommand{\BiTR}{\mathrm{Bi}_{\,T,\,R}}
\newcommand{\BiTM}{\mathrm{Bi}_{\,TM}}
\newcommand{\BiTML}{\mathrm{Bi}_{\,TM,\,L}}
\newcommand{\BiTMR}{\mathrm{Bi}_{\,TM,\,R}}
\newcommand{\cMs}{c_{\,M}^{\,\star}}
\newcommand{\cTs}{c_{\,T}^{\,\star}}
\newcommand{\kMs}{k_{\,M}^{\,\star}}
\newcommand{\kTs}{k_{\,T}^{\,\star}}
\newcommand{\kTMs}{k_{\,TM}^{\,\star}}

\newcommand{\cMsd}{c_{\,M,\,2}^{\,\star}}
\newcommand{\cTsd}{c_{\,T,\,2}^{\,\star}}
\newcommand{\kMsd}{k_{\,M,\,2}^{\,\star}}
\newcommand{\kTsd}{k_{\,T,\,2}^{\,\star}}
\newcommand{\kTMsd}{k_{\,TM,\,2}^{\,\star}}

\newcommand{\gsinf}{g^{\,\star}_{\,\infty}}

\newcommand{\Hls}{H^{\,\star}_{\,l}}

\newcommand{\uinf}{u_{\,\infty}}
\newcommand{\uinfL}{u_{\,\infty ,\,L}}
\newcommand{\uinfR}{u_{\,\infty ,\,R}}
\newcommand{\vinf}{v_{\,\infty}}
\newcommand{\vinfL}{v_{\,\infty ,\,L}}
\newcommand{\vinfR}{v_{\,\infty ,\,R}}
\newcommand{\ts}{t^{\,\star}}
\newcommand{\xs}{x^{\,\star}}

\newcommand*\pd[2]{\dfrac{\partial #1}{\partial #2}}

\newcommand{\eqdef}{\mathop{\stackrel{\,\mathrm{def}}{:=}\,}}

\renewcommand{\L}{\mathcal{L}}

\newcommand{\dt}{\Delta t}
\newcommand{\dx}{\Delta x}
\newcommand{\M}{\mathcal{M}}
\newcommand{\A}{\mathcal{A}}
\newcommand{\B}{\mathcal{B}}
\newcommand{\C}{\mathcal{C}}
\renewcommand{\b}{\mathrm{b \,}}

\newcommand{\T}{\mathsf{T}}
\renewcommand{\O}{\mathcal{O}\, }

\newcommand{\dix}[1]{ \cdot 10^{\,#1}}

\renewcommand{\unitfrac}[2]{{\mathsf{#1}}/{\mathsf{#2}}}
\newcommand{\gC}{^{\circ}\mathsf{C}}

\begin{document}

\title[\Title]{A spectral method for solving heat and moisture transfer through consolidated porous media}

\author[S.~Gasparin]{Suelen Gasparin$^*$}
\address{\textbf{S.~Gasparin:} LAMA, UMR 5127 CNRS, Universit\'e Savoie Mont Blanc, Campus Scientifique, F-73376 Le Bourget-du-Lac Cedex, France and Thermal Systems Laboratory, Mechanical Engineering Graduate Program, Pontifical Catholic University of Paran\'a, Rua Imaculada Concei\c{c}\~{a}o, 1155, CEP: 80215-901, Curitiba -- Paran\'a, Brazil}
\email{suelengasparin@hotmail.com}
\urladdr{https://www.researchgate.net/profile/Suelen\_Gasparin/}
\thanks{$^*$ Corresponding author}

\author[D.~Dutykh]{Denys Dutykh}
\address{\textbf{D.~Dutykh:} Univ. Grenoble Alpes, Univ. Savoie Mont Blanc, CNRS, LAMA, 73000 Chamb\'ery, France and LAMA, UMR 5127 CNRS, Universit\'e Savoie Mont Blanc, Campus Scientifique, F-73376 Le Bourget-du-Lac Cedex, France}
\email{Denys.Dutykh@univ-smb.fr}
\urladdr{http://www.denys-dutykh.com/}

\author[N.~Mendes]{Nathan Mendes}
\address{\textbf{N.~Mendes:} Thermal Systems Laboratory, Mechanical Engineering Graduate Program, Pontifical Catholic University of Paran\'a, Rua Imaculada Concei\c{c}\~{a}o, 1155, CEP: 80215-901, Curitiba -- Paran\'a, Brazil}
\email{Nathan.Mendes@pucpr.edu.br}
\urladdr{https://www.researchgate.net/profile/Nathan\_Mendes/}


\begin{titlepage}
\thispagestyle{empty} 
\noindent
{\Large Suelen \textsc{Gasparin}}\\
{\it\textcolor{gray}{Pontifical Catholic University of Paran\'a, Brazil}}\\
{\it\textcolor{gray}{LAMA--CNRS, Universit\'e Savoie Mont Blanc, France}}
\\[0.02\textheight]
{\Large Denys \textsc{Dutykh}}\\
{\it\textcolor{gray}{LAMA--CNRS, Universit\'e Savoie Mont Blanc, France}}
\\[0.02\textheight]
{\Large Nathan \textsc{Mendes}}\\
{\it\textcolor{gray}{Pontifical Catholic University of Paran\'a, Brazil}}
\\[0.10\textheight]

\colorbox{Lightblue}{
  \parbox[t]{1.0\textwidth}{
    \centering\huge\sc
    \vspace*{0.7cm}
    
    \textcolor{bluepigment}{A spectral method for solving heat and moisture transfer through consolidated porous media}

    \vspace*{0.7cm}
  }
}

\vfill 

\raggedleft     
{\large \plogo} 
\end{titlepage}


\newpage
\thispagestyle{empty} 
\par\vspace*{\fill}   
\begin{flushright} 
{\textcolor{denimblue}{\textsc{Last modified:}} \today}
\end{flushright}


\newpage
\maketitle
\thispagestyle{empty}


\begin{abstract}

This work presents an efficient numerical method based on spectral expansions for simulation of heat and moisture diffusive transfers through multilayered porous materials. Traditionally, by using the finite-difference approach, the problem is discretized in time and space domains (Method of lines) to obtain a large system of coupled Ordinary Differential Equations (ODEs), which is computationally expensive. To avoid such a cost, this paper proposes a reduced-order method that is faster and accurate, using a much smaller system of ODEs. To demonstrate the benefits of this approach, tree case studies are presented. The first one considers nonlinear heat and moisture transfer through one material layer. The second case -- highly nonlinear -- imposes a high moisture content gradient -- simulating a rain like condition -- over a two-layered domain, while the last one compares the numerical prediction against experimental data for validation purposes. Results show how the nonlinearities and the interface between materials are easily and naturally treated with the spectral reduced-order method. Concerning the reliability part, predictions show a good agreement with experimental results, which confirm robustness, calculation efficiency and high accuracy of the proposed approach for predicting the coupled heat and moisture transfer through porous materials.


\bigskip\bigskip
\noindent \textbf{\keywordsname:} Spectral methods; \textsc{Chebyshev} polynomials; \textsc{Tau--Galerkin} method; numerical simulation; heat and moisture transfer; reduced-order modeling \\


\end{abstract}


\newpage
\tableofcontents
\thispagestyle{empty}


\newpage
\section{Introduction}

Energy consumption of conditioned spaces is strongly related to heat transfer through porous structures, which are dependent of external (weather) and internal conditions. Moreover, moisture migration and accumulation within the porous structures considerably affect the transient hygrothermal performance of porous elements, especially in buildings. Thus, coupled hygrothermal simulations are important to estimate envelope energy performance and risks associated to the presence of moisture such as material degradation, mold growth and related health aspects of occupants.

Initially, the major hypothesis was of no coupling between governing equations of heat and moisture transfers. Although, in porous materials, moisture transfer and accumulation have a direct impact on the heat transfer, especially when there is phase change \cite{Mulay1990, Bouddour1998, Rees2001}. As presented by \textsc{Deru} \cite{Deru2003}, to precisely determine the heat losses, simultaneous calculations with moisture content are required, as they are closely interdependent. Temperature and moisture contents are highly interconnected phenomena and, hence, must be simulated together \cite{Berger2015a, Mendes1997}. For instance, moisture can affect the effective thermal conductivity by a factor of ten.

However, there are some challenges on how to characterize mathematically those physical processes, due to the complexity of the physical phenomena and to the porous structure. Studies of heat and moisture transfer have been published since $1950$'s by \textsc{Philip} and \textsc{De Vries} \cite{Philip1957} and \textsc{Luikov} \cite{Luikov1966}. They represented the coupled processes of heat and moisture transfer by a system of two nonlinear second-order partial differential equations which uses as driving potentials the temperature and the moisture content gradients. The system is nonlinear mainly since of the phase change term in the energy conservation equation and also due to the fact the transport coefficients are highly moisture content and/or temperature dependent. Although, the paramount importance of accurately predict moisture content and temperature fields in several fields of science,  detailed simulation of the heat and moisture transfer has only been performed with the improvement of the computer systems. The main reason for that is due to the complexity of the problem and to the high computational cost to solve the coupled heat and moisture transfer equations, which is at least two orders of magnitude greater than that needed to solve only the heat conduction equation \cite{Deru2003}.

The numerical models used to predict heat and moisture transfer have to deal with multi-scale effects in both space and time domains, with different characteristic times and lengths. For example, simulation of building behavior is generally analyzed on the scale of one year (or more). However, the phenomena and particularly the boundary conditions evolve on a time scale of minutes or even seconds. The geometric configurations of the buildings require three-dimensional modeling of lengthy elements such as facades and ground. Furthermore, when dealing with heat and moisture, the nonlinear behavior of the materials should be taken into account. The combination of all those factors implies the use of more robust and efficient numerical methods since no analytical solution does exist for those problems.

For modeling purposes, the problem previously described is generally solved by the finite-difference methods \cite{Gasparin2017, Fan2004}, the finite-volume methods \cite{DosSantos2009, Mendes2005, Manz2003} and the finite-element methods \cite{Thomas1980, Lu2002, Rouchier2013}, which are well-established methods in the field of heat and mass transfer. The basics of a numerical approach is the idea of making an approximation of the solution, which takes a solution for a finite number of degrees of freedom (DOFs). The greater the number of discrete points, the closer to the exact solution will be the approximated solution \cite{Maliska2004}.

However, when the solution of one problem requires long simulation periods (years), considering an entire building, with a very fine time step and mesh refinement, computation becomes too time-consuming. To solve this problem, model reduction techniques can be used as an alternative to approach the solution of the problem and reduce the computational cost.

The intent to construct reduced-order models (ROMs) is to provide an accurate description of the physical phenomena by decreasing the number of degrees of freedom, while retaining the model's fidelity, at a computational cost much lower than the large original model \cite{Reddy2017}. In recent years, reduced-order modeling techniques have proven to be powerful tools for solving various problems. Important efforts have been dedicated to developing reduced-order models that can provide accurate predictions while dramatically reducing computational time, for a wide range of applications, covering different fields such as fluid mechanics, heat transfer,  structural dynamics, among others \cite{Lucia2004, Herzet2018, Bai2002}. Reduced-order models -- such as POD (Proper Orthogonal Decomposition), MBR (Modal Basis Reduction) and PGD (Proper Generalized Decomposition) -- have shown a relevant reduction of the computational cost and have been successfully employed by the building physics community \cite{Berger2015}. In those works, they have applied reduced-order models to build accurate solutions with less computational effort than the complete original model. Reduced-order models can be classified as \emph{a priori} or \emph{a posteriori} methods. The \emph{a posteriori} approaches need a preliminary computed (or even experimental) solution data of the large original problem to build the reduced one. Whereas the \emph{a priori} ones do not need preliminary information on the studied problem. The reduced-order model is unknown \emph{a priori} and is directly built. A careful attention must be paid regarding the definition of ROMs since sometimes it is related to the degradation of the physical model \cite{Schilders2008}, which is not the case of the present work.

Another promising approach to solve the coupled transfer problem is the spectral method, which is a robust and highly accurate method that has been applied to solve partial differential equations since the $70$'s, but lost its spot due to the difficulties to treat nonlinearities, complex geometries, irregular domains, and non-periodic boundary conditions. However, the Spectral methods have overcome some of the mentioned difficulties and now they are successfully applied in studies of wave propagation, meteorology, computational fluid dynamics, quantum mechanics and other fields \cite{Canuto2006}. The main attractiveness of this method is the superior rate of convergence and the low dissipation and dispersion errors, making its use also attractive to the industry. Nonetheless, spectral methods still have some constraints. For example, complex geometries are one of their main drawbacks as they work better when the geometry of the problem is fairly smooth and regular \cite{Boyd2000}, while finite-element methods are particularly well suited to problems in very complex geometries. Moreover, spectral methods can offer higher accuracy in geometries like boxes and spheres, which can be combined into more complex shapes \cite{Fornberg1996, Canuto2007}. In applications where geometry-related disadvantages are not present, the classic finite-element, finite-volume and finite-difference methods do not come close in terms of efficiency.

Some works related to transport phenomena can be found in literature involving diffusive \cite{Guo2012a, Wang2016}, convective \cite{Chen2016, RamReddy2015} and radiative \cite{Li2008a, Chen2015b, Ma2014} heat transfer. The spectral techniques applied in these works are diversified, adopted according to the geometry, boundary conditions and field of application. In recent works, researchers have implemented spectral methods for solving heat and moisture transfer in food engineering \cite{Pasban2017} and on fluid flow \cite{Motsa2015}. Recently, in \cite{Gasparin2019}, the authors have studied the moisture transfer in porous building materials considering layered domains, and in \cite{Gasparin2017c}, they have compared the Spectral method to others ROMs, applied to parametric problems of the building physics field.

Therefore, the scope of this work is to continue the investigations presented in \cite{Gasparin2019} and \cite{Gasparin2017c}, extending it to the coupled heat and mass transfer. Here, the Spectral method is used to compute one-dimensional heat and moisture diffusion transfer in porous materials, which is validated against experimental data from the literature. The problems treated here involve irregular domains, high nonlinear transport coefficients and non-periodic boundary conditions. The objective is to significantly reduce the computational cost while maintaining high fidelity solutions. This technique assumes separated tensorial representation of the solution by a finite sum of function products. It fixes a set of spatial basis functions to be the \textsc{Chebyshev} polynomials and then, a system of ordinary differential equations is built to compute the temporal coefficients of the solution using the \textsc{Tau--Galerkin} method.

The efficiency of the Spectral approach will be analyzed for simple and multilayered domains with highly nonlinear properties and with sharp boundary conditions and profiles of solutions. For this purpose, the manuscript is organized as follows. First, the description of the physical phenomena is presented (Section~\ref{sec:HAM_transfer}). Then, the Spectral technique is described (Section~\ref{sec:spectral_description}). In the sequence, the proposed method is applied to two different cases: \textit{(i)} considering heat and moisture transfer through a single layer (Section~\ref{sec:case_1layer}) and  \textit{(ii)} focusing on the heat and moisture transfer through a multilayered domain (Section~\ref{sec:case_2layers}). Finally, the method and the model are compared with experimental data from literature in Section~\ref{sec:validation}, considering a single material with real boundary conditions. The main conclusions of the study are outlined in Section~\ref{sec:conclusion}.


\section{Mathematical model}
\label{sec:HAM_transfer}

The physical problem considers heat and moisture transfer through a porous material defined in the one-dimensional spatial domain $\Ox \egal [\, 0, \, L \,]$ and in the time horizon $\Ot \egal [\, 0, \, \tau \,]\,$. The moisture transfer occurs due to capillary migration and vapour diffusion. The heat transfer is governed by diffusion and latent mechanisms. The physical model of the problem can be formulated as \cite{Berger2015}:
\begin{subequations}\label{eq:HAM_equation}
\begin{align}
  \label{eq:M_equation}
  \frac{1}{\Ps} \pd{w}{\phi} \ \pd{\Pv}{t} &\moins \pd{}{x} \Biggl[ \, \Biggl(\kl \, \frac{R_{\,v} \, T\, \rhol}{\Pv} \plus \delta_{\,v} \Biggr)\,  \pd{\Pv}{x}  \,  \Biggr] \egal 0 \,, \\[3pt]
  \label{eq:H_equation}
  \bigl(\, \rhoz \ \cz \plus w \ \cw \, \bigr) \ \pd{T}{t} &\moins \pd{}{x} \Biggl[\, \lambda \  \nabla T \plus \Lv \ \delta_{\,v} \  \pd{\Pv}{x} \, \Biggr] \egal 0\,,
\end{align}
\end{subequations}
where $w\ [\mathsf{kg/m^3}]$ is the material volumetric moisture content, $\phi\ [-]$, the relative humidity, $\delta_{\,v}\ [\mathsf{s}]$ and $\kl\ [\mathsf{s}]$, the vapour and liquid permeabilities, $\Pv\ [\mathsf{Pa}]$, the vapour pressure, $\Ps\ [\mathsf{Pa}]$, the saturation pressure, $T\ [\mathsf{K}]$, the temperature, $\Rv\ [\mathsf{J/(kg\cdot K)}]\,$, the water vapour gas constant, $\cz\ [\mathsf{J/(kg\cdot K)}]$, the material specific heat, $\rhoz\ [\mathsf{kg/m^3}]$, the material specific mass, $\rhol\ [\mathsf{kg/m^3}]\,$, the water specific mass, $\cw\ [\mathsf{J/(kg\cdot K)}]\,$, the water specific heat, $\lambda\ [\mathsf{W/(m\cdot K)}]$, the thermal conductivity, and, $\Lv\ [\mathsf{J/kg}]\,$, the latent heat of vaporization. Table~\ref{table:properties_water} presents the values of the water properties considered in this work.

The relation between  the moisture content $w$ and the relative humidity $\phi$ is given by the sorption isotherm and the relation between the vapour pressure $\Pv$ and the relative humidity $\phi$ is given by $\phi \egalb \nicefrac{\Pv}{\Ps(T)}\,$. In addition, the following assumptions are adopted in this study: \textit{(i)} no hysteresis effect; \textit{(ii)} temperature within the range $[0,\,40]\gC\,$; \textit{(iii)} no temperature dependency on the mass balance equation and \textit{(iv)} properties are dependent only on the vapour pressure field.

Thus, considering the following notation:
\begin{align*}
  \kM \ &\eqdef \ \kl \, \dfrac{\rhol\, \Rv \, T }{\Pv} \plus \delta_{\,v}\,: && \text{the total moisture transfer coefficient} \\ 
  & && \text{under vapour pressure gradient} ,\\
  \kTM \ &\eqdef \ \Lv \ \delta_{\,v}\,: && \text{the heat coefficient due to a  vapour pressure gradient} ,\\
  \kT \ &\eqdef \ \lambda\,: && \text{the heat transfer coefficient under temperature gradient} , \\
  \cM \ &\eqdef \ \frac{1}{\Ps} \pd{w}{\phi} \,: && \text{the moisture storage coefficient}  ,\\
  \cT \ &\eqdef \ \rhoz \ \cz \plus w \ \cw\,: && \text{the energy storage coefficient} ,
\end{align*}
system~\eqref{eq:HAM_equation} can be rewritten in one-dimensional form as:
\begin{subequations}\label{eq:HAM_equation2}
\begin{align}
  \cM\,(\Pv) \ \pd{\Pv}{t} &\moins \pd{}{x} \Biggl[ \, \kM\,(\Pv) \, \pd{\Pv}{x} \,  \Biggr] \egal 0 \,,  \\[5pt]
  \cT\,(\Pv) \ \pd{T}{t}  &\moins \pd{}{x} \Biggl[\, \kT\,(\Pv) \ \pd{T}{x} \plus \kTM\,(\Pv) \ \pd{\Pv}{x} \, \Biggr]\egal 0 \,.
\end{align}
\end{subequations}

Finally, the problem of interest is a coupled system of two nonlinear parabolic partial differential equations, with vapour pressure $\Pv$ and temperature $T$ gradients as driving potentials. 

\bigskip
\paragraph*{Boundary conditions.}

The moisture exchange between the environment and the surface is driven by vapour exchange, including evaporation and condensation, and by driving rain:
\begin{align}\label{eq:BC_moist}
  \mathbf{n} \cdot \Biggl(\kM \, \pd{\Pv}{x}\Biggr) &\egal \hM \, \Bigl(\, \Pv \moins \Pvinf\,(t) \,\Bigr) \moins g_{\,\infty}\,(t)\,,
\end{align}
where $\Pvinf\ [\mathsf{Pa}]$ stands for the vapour pressure far from the surface and, $\hM\ [\mathsf{s/m}]$, is the convective moisture transfer coefficient. If the bounding surface is in contact with the outside air, $g_{\,\infty}\ [\mathsf{kg/(m^2\cdot s)}] $ is the liquid flow from wind driven rain. The normal $\mathbf{n}$ assumes $+\,1$ or $-\,1$ at the left or right boundary sides.

The heat balance at the boundary includes the convective exchange, the latent heat transfer due to vapour exchange, and the sensible heat transfer due to precipitation, which is expressed as:
\begin{multline}\label{eq:BC_heat}
  \mathbf{n} \cdot \Biggl(\kT \, \pd{T}{x} \plus \kTM \, \pd{\Pv}{x} \Biggr) \egal \hT \, \Bigl(\, T \moins \Tinf\,(t) \,\Bigr)\\ 
  \plus \Lv \, \hM \, \Bigl(\, \Pv \moins \Pvinf\,(t) \,\Bigr) \moins  H_{\,l}\,g_{\,\infty}\,(t)\,,
\end{multline}
where $\Tinf\ [\mathsf{K}]$ is the temperature of the air that varies over time and $\hT\ [\mathsf{W/(m^2\cdot K)}]$ is the convective heat transfer coefficient. Regarding to the moisture part, $\Lv\ [\mathsf{J/kg}]$ is the latent heat of vaporization of water and $H_{\,l}\,=\, \cw\,(\Tinf\ -\ T_{\,\text{ref}})\ [\mathsf{J/kg}]$ is the liquid water enthalpy, with $T_{\,\text{ref}} = 273\, \mathsf{K}\,$.

\bigskip
\paragraph*{Initial conditions.}

The initial conditions can either have a uniform distribution or a profile more appropriated to the boundary conditions to reduce a warm-up simulation period, which can be very significant depending on the material hygrothermal properties and on the thickness of the building component.
\begin{align*}
  \Pv\,(x\,,t \egalb 0) &\egal \Pvi\, (x)  \,, \\
  T\,(x\,,t \egalb 0)  &\egal \Ti\, (x)\,.
\end{align*}

\bigskip
\paragraph*{Interface.}

The configuration assumed at the interface between materials follows the hydraulic continuity \cite{DeFreitas1996}, which considers interpenetration of both porous structure layers. Consider two different materials, both of them are homogeneous and isotropic, and the coupled heat and moisture transfer are simulated, through a perfectly airtight structure. The hydraulic continuity assumes that there is a continuous moisture distribution of vapour content and temperature:
\begin{subequations}
\begin{align}
  P_{\,v,\,1}\,(x_{\,\text{int}},t) & \egal P_{\,v,\,2}\,(x_{\,\text{int}},t) \,, \\
  T_{\,1}\,(x_{\,\text{int}},t) & \egal T_{\,2}\,(x_{\,\text{int}},t)\,,
\end{align}\label{eq:cont_field}
\end{subequations}
a continuous moisture flow and a continuous heat flux across the interface verify:
\begin{subequations}
\begin{align} 
  \Biggl(\,k_{\,M,\,1} \ \pd{P_{\,v,\,1}}{x}\, \Biggr)\Bigg|_{x_{\,\text{int}}} & \egal\Biggl(\, k_{\,M,\,2}  \ \pd{P_{\,v,\,2}}{x}\, \Biggr)\Bigg|_{x_{\,\text{int}}} \,,\\
  \Biggl(\, k_{\,T,\,1} \ \pd{T_{\,1}}{x} \plus k_{\,TM,\,1} \ \pd{P_{\,v,\,1}}{x} \, \Biggr)\Bigg|_{x_{\,\text{int}}} & \egal \Biggl(\, k_{\,T,\,2} \ \pd{T_{\,2}}{x} \plus k_{\,TM,\,2} \ \pd{P_{\,v,\,2}}{x} \, \Biggr)\Bigg|_{x_{\,\text{int}}}\,,
\end{align}\label{eq:cont_flow}
\end{subequations}
where $x_{\,\text{int}}\ \in\ \Ox $ represents the location of the interface between materials. Therefore, we can split the spatial domain in two parts, $\Omega_{\,x,\,1} \egalb [\,0\,,x_{\,\text{int}}\,]$ and $\Omega_{\,x,\,2} \egalb (\,x_{\,\text{int}}\,,\,L\,]\,$, which represent the spatial domain of material $1$ and material $2\,$.

\bigskip
\paragraph*{Fluxes and flows.}

One of the interesting outputs in the building physics framework is the heat flux, divided into  sensible $q_{\,s}$ and latent $q_{\,l}$ heat fluxes $[\mathsf{W/m^2}]$, which are defined as:
\begin{align*}
  q_{\,s}\,(t)\ \eqdef\ \moins \kT \,\left. \pd{T}{x}\right\vert_{x_{\,0}}  & & \text{and} & &
  q_{\,l}\,(t)\ \eqdef\ \moins \kTM \,\left. \pd{\Pv}{x}\right\vert_{x_{\,0}} \,. 
\end{align*}
The moisture flow $g$ $[\mathsf{kg/(m^2\cdot s)}]$ is similarly computed:
\begin{align*}
  g\,(t)\ &\eqdef\ \moins \kM \,\left. \pd{\Pv}{x}\right\vert_{x_{\,0}} \,, 
\end{align*}
where $x_{\,0}\, \in\, [\,0\,,L\,]\,$.

\begin{table}
\centering
\caption{\small\em Hygrothermal properties of water.}
\bigskip
\def\arraystretch{1.2}
\begin{tabular}{lll}
\hline
\textit{Property} & \textit{Value} & \textit{Unit}\\
\hline
Heat capacity, $\cw$ & $4180$ & $[\unitfrac{J}{(kg \cdot K)}]$\\
Latent heat of evaporation, $\Lv$ & $2.5 \dix{6}$ & $[\unitfrac{J}{kg}]$\\
Water gas constant, $\Rv$ & $462$ & $[\mathsf{J/(kg\cdot K)}]$\\
Density, $\rhol$ & $1000$  & $[\unitfrac{kg}{m^3}]$\\
Saturation pressure, $\Ps\,(T)$ & $997.3\cdot\biggl(\dfrac{T-159.5}{120.6}\biggr)^{8.275}$ & $[\mathsf{Pa}] $\\
\hline                
\end{tabular}
\label{table:properties_water}
\end{table}


\subsection{Dimensionless formulation}

Before solving directly the problem, it is of capital importance to get a dimensionless formulation of the problem under consideration \cite{Kahan1979}. In this way, we define the following dimensionless quantities:
\begin{align*}
  u \ &\eqdef \ \frac{T}{\Tref} \,,
& v \ &\eqdef \ \frac{\Pv}{\Pvref} \,, 
& \xs  &\eqdef \ \frac{x}{\Lref} \,, \\[3pt]
  \ts  &\eqdef \ \frac{t}{\tref}  \,,  
& \cMs \ &\eqdef \ \frac{\cM \cdot \Lref^{\,2} }{\kMref \cdot \tref} \,, 
& \cTs \ &\eqdef \ \frac{\cT \cdot \Lref^{\,2} }{\kTref \cdot \tref} \,, \\[3pt]
  \kMs \ &\eqdef \ \frac{\kM }{\kMref} \,,
& \kTs \ &\eqdef \ \frac{\kT }{\kTref} \,,
& \kTMs \ &\eqdef \ \frac{\kTM \cdot \Pvref}{\kTref \cdot \Tref} \,,  \\[3pt]
  \BiT \ &\eqdef \ \frac{\hT \cdot \Lref}{\kTref} \,, 
& \BiM \ &\eqdef \ \frac{\hM \cdot \Lref}{\kMref} \,, 
& \BiTM \ &\eqdef \ \frac{\hM \cdot \Lv\cdot \Lref \cdot \Pvref }{\kTref\cdot \Tref} \,,  \\[3pt]
  \gsinf \ &\eqdef \ \frac{g_{\,\infty}\cdot \Lref}{\kMref \cdot \Pvref}  \,,
&  \Hls \ &\eqdef \ \frac{\Hl \cdot \kMref \cdot \Pvref}{\kTref \cdot \Tref }\,.
\end{align*}
where the subscript \textbf{ref} represents a reference value, chosen according to the application problem and the superscript $\star$ represents a dimensionless quantity of the same variable. Therefore, the governing system~\eqref{eq:HAM_equation2} can be written in a dimensionless form as:
\begin{subequations}\label{eq:HAM_dimless}
\begin{align}\label{eq:moist_dimless}
  \cMs \ \pd{v}{\ts} & \moins \pd{}{\xs} \Biggl( \, \kMs \, \pd{v}{\xs} \,  \Biggr) \egal 0 \,, \\[3pt]
  \label{eq:heat_dimless}
  \cTs \ \pd{u}{\ts} & \moins \pd{}{\xs} \Biggl(\, \kTs \ \pd{u}{\xs} \plus \kTMs \ \pd{v}{\xs} \Biggr) \egal 0 \,.  
\end{align}
\end{subequations}

The dimensionless formulation of the boundary conditions are:
\begin{align*}
  \mathbf{n} \cdot \Biggl( \kMs \, \pd{v}{\xs} \Biggr)  \egal & \BiM \, \Bigl(\, v \moins \vinf\,(\ts) \,\Bigr) \moins \gsinf\,(\ts) \,, \\[3pt]
  \mathbf{n} \cdot \Biggl( \kTs \, \pd{u}{\xs} \plus \kTMs \, \pd{v}{\xs} \Biggr) \egal & \BiT \, \Bigl(\, u \moins \uinf\,(\ts) \,\Bigr)\\ 
  & \plus \BiTM\, \Bigl(\, v \moins \vinf\,(\ts) \,\Bigr) \moins \Hls\, \gsinf\,(\ts) \,,
\end{align*}
and of the initial conditions are:
\begin{align*}
  u\, (\xs\,,\ts\,=\,0) &\egal u_{\,0}\, (\xs) \,, \\
  v\, (\xs\,,\ts\,=\,0) &\egal v_{\,0}\, (\xs) \,.
\end{align*}

Interface conditions in the dimensionless form are written as:
\begin{subequations}
\begin{align}
  v_{\,1}\,(\xs_{\,\text{int}},\ts) & \egal v_{\,2}\,(\xs_{\,\text{int}},\ts) \,, \\
  u_{\,1}\,(\xs_{\,\text{int}},\ts) & \egal u_{\,2}\,(\xs_{\,\text{int}},\ts) \,, \\
  k_{\,M,\,1}^{\,\star} \ \pd{v_{\,1}}{\xs}\Bigg|_{\xs_{\,\text{int}}} & \egal k_{\,M,\,2}^{\,\star}  \ \pd{v_{\,2}}{\xs}\Bigg|_{\xs_{\,\text{int}}} \,,\\
  \Biggl(\, k_{\,T,\,1}^{\,\star} \ \pd{u_{\,1}}{\xs} \plus k_{\,TM,\,1}^{\,\star} \ \pd{v_{\,1}}{\xs} \, \Biggr)\Bigg|_{\xs_{\,\text{int}}} & \egal \Biggl(\, k_{\,T,\,2}^{\,\star} \ \pd{u_{\,2}}{\xs} \plus k_{\,TM,\,2}^{\,\star} \ \pd{v_{\,2}}{\xs} \, \Biggr)\Bigg|_{\xs_{\,\text{int}}}\,.
\end{align}\label{eq:diml_interf}
\end{subequations}

In the following, we drop $\star$ for the sake of clarity.


\section{Spectral reduced-order model}
\label{sec:spectral_description}

Spectral methods consider a sum of polynomials that suit for the whole domain, providing a high approximation of the solution. The smoother a function is, the faster the convergence of its spectral series \cite{Boyd2000}. For considerably smooth problems, the error decreases exponentially, making the solution with the same order of accuracy of other methods but with a much lower number of degrees of freedom. As a result, this method has a low memory usage, allowing to store and operate a lower number of variables \cite{Trefethen1996}. The Spectral methods used in this work are the \textsc{Chebyshev} polynomials on the basis function and the \textsc{Tau--Galerkin} method to compute the temporal coefficients.

\subsection{Method description}
\label{sec:spectral_nonlinear}

Problem \eqref{eq:HAM_dimless} has an important difficulty in dealing with the nonlinearities of the storage $c$ and diffusion $k$ coefficients, all of them depending on the moisture content field. These coefficients are usually given by empirical functions from experimental data. For this reason, Eq.~\eqref{eq:moist_dimless} and \eqref{eq:heat_dimless} are recalled with a simplified notation:
\begin{subequations}
\begin{align}\label{eq:dimlss_moist_simpl}
  \cM \, (v) \, \pd{v}{t} &\moins \pd{}{x} \left[ \, \kM \, (v) \, \pd{v}{x} \, \right] \egal 0 \,, \\[3pt]
  \cT \, (v) \ \pd{u}{t} &\moins \pd{}{x} \Biggl[\, \kT\, (v) \ \pd{u}{x} \plus \kTM\, (v) \ \pd{v}{x} \Biggr] \egal 0 \label{eq:dimlss_temp_simpl}\,.
\end{align}
\end{subequations}
In addition, boundary conditions are simplified and also written with a shorter notation, just for the sake of explaining the method in a pedagogical way/manner:
\begin{subequations}
\begin{align}\label{eq:dimlss_BC_simpl_v}
  \mathbf{n}\cdot \Biggl( \kM\, (v) \, \pd{v}{x} \Biggr) \egal & \BiML \, \Bigl(\, v \moins \vinfL\,(t) \,\Bigr) \,, \\[3pt]
  \mathbf{n}\cdot \Biggl( \kT\, (v) \, \pd{u}{x} \plus \kTM\, (v) \, \pd{v}{x} \Biggr) \egal & \BiTR \, \Bigl(\, u \moins \uinfR \,(t) \,\Bigr)\\
  & \plus \BiTMR \, \Bigl(\, v \moins \vinfR\,(t) \,\Bigr) \label{eq:dimlss_BC_simpl_u} \,.
\end{align}
\end{subequations}
A special attention must be given to the spatial domain because the \textsc{Chebyshev} Spectral method is traditionally presented on the canonical interval $\big[ -1\,,  1 \, \big]\,$. Thus, if the dimensionless interval is not within $\big[ -1\,,  1 \, \big]\,$, a change of variables (domain transformation) must be performed for the computational domain.

In order to apply better the spectral method, Eqs.~\eqref{eq:dimlss_moist_simpl} and \eqref{eq:dimlss_temp_simpl} are written in the non-conservative form as:
\begin{subequations}\label{eq:rearang}
\begin{align}\label{eq:rearang_moist}
  \pd{v}{t} &\moins \nu \, (v) \, \pd{^{\,2}v}{x^{\,2}} \moins \lambda \, (v)\, \pd{v}{x} \egal 0\,, \\
  \pd{u}{t} &\moins \alpha \, (v) \, \pd{^{\,2}u}{x^{\,2}} \moins \beta \, (v)\, \pd{u}{x} \moins \gamma \, (v) \, \pd{^{\,2}v}{x^{\,2}} \moins \delta \, (v)\, \pd{v}{x} \egal 0 \label{eq:rearang_heat} \,,
\end{align}
\end{subequations}
where,
\begin{align*}
  \nu \, (v) &\eqdef \dfrac{\kM\, (v)}{\cM\, (v) }\,, &  \lambda \, (v) &\eqdef \dfrac{1}{\cM\, (v) }\cdot  \pd{\Bigl(\, \kM\, (v)\, \Bigr)}{x}\,, \\   
  \alpha \, (v) &\eqdef \dfrac{\kT\, (v)}{\cT\, (v) }\,, & \beta \, (v) &\eqdef \dfrac{1}{\cT\, (v) }\cdot  \pd{\Bigl(\, \kT\, (v)\, \Bigr)}{x}\,, \\ 
  \gamma \, (v) &\eqdef \dfrac{\kTM\, (v)}{\cT\, (v) }\,, & \delta \, (v) &\eqdef \dfrac{1}{\cT\, (v) }\cdot \pd{\Bigl(\, \kTM\, (v)\, \Bigr)}{x}\,.
\end{align*}

The unknowns $u\,(\,x,\,t\,)$ and $v\,(\,x,\,t\,)$ from Eq.~\eqref{eq:rearang} are accurately represented as a finite sum \cite[Chap.~6]{Mendes2017}:
\begin{subequations}
\begin{align}
  v\, (\,x,\, t\,)\ \approx\ v_{\, n}\, (\,x,\, t\,) &\egal \sum_{i\, =\, 0}^{n} \, a_{\,i}\, (t)\, \T_{\,i}\, (x)\label{eq:series_ap_v} \,, &  i \egal 0,1,2,\ldots,n\,,\\
  u\, (\,x,\, t\,)\ \approx\ u_{\, n}\, (\,x,\, t\,) &\egal \sum_{i\, =\, 0}^{n} \, b_{\,i}\, (t)\, \T_{\,i}\, (x)\label{eq:series_ap_u} \,, &  i \egal 0,1,2,\ldots,n\,. 
\end{align}\label{eq:series}
\end{subequations}
Here, $\{\T_{\,i}\, (x)\}_{\,i\, =\, 0}^{\,n}$ are the \textsc{Chebyshev} polynomials, $\{a_{\,i}\, (t)\}_{\,i\, =\, 0}^{\,n}$ are the corresponding time-dependent spectral coefficients and $n$ represents the number of degrees of freedom of the solution component. Eqs.~\eqref{eq:series_ap_v} and \eqref{eq:series_ap_u} can be seen as a series truncation from $N \, = \, n\, +\, 1$ modes. The \textsc{Chebyshev} polynomials are chosen as the basis functions since they are optimal in $\mathcal{L}_{\,\infty}$ approximation norm \cite{Gautschi2004}. Therefore, the expression of the derivatives in the \textsc{Chebyshev} basis are:
\begin{subequations}\label{eq:derivatives}
\begin{align}
  \pd{v_{\,n}}{x} &\egal \sum_{i\, =\, 0}^n \, a_{\,i}\,(t)\, \pd{\T_{\,i}}{x}\,(x)\egal \sum_{i\, =\, 0}^n \tilde{a}_{\,i}\,(t)\, \T_{\,i}\,(x)\,,\label{eq:derivative1}\\
  \pd{^{\,2} v_{\,n}}{x^{\,2}} &\egal \sum_{i\, =\, 0}^n \, a_{\,i}\,(t)\, \pd{^{\,2} \T_{\,i}}{x^{\,2}}\,(x)\egal \sum_{i\, =\, 0}^n \Tilde{\Tilde{a}}_{\,i}\,(t)\, \T_{\,i}\,(x) \,, \label{eq:derivative2}\\
  \pd{v_{\,n}}{t} &\egal \sum_{i\, =\, 0}^n \, \dot{a}_{\,i}\,(t)\, \T_{\,i}\,(x)\,,\label{eq:derivative3} 
\end{align}
\end{subequations}
where the dot denotes $\dot{a}_{\,i}\, (t) \eqdef \dfrac{\mathrm{d}\,a\, (t) }{\mathrm{d}t} $ according to \textsc{Newton} notation. Note that the derivatives are re-expanded in the same basis function. As a result, coefficients $\{\tilde{a}_{\,i}\, (t)\}$ and $\{\Tilde{\Tilde{a}}_{\,i}\, (t)\}$ must be re-expressed in terms of coefficients $\{a_{\,i}\, (t)\}\,$. The connection is given explicitly from the recurrence relation of the \textsc{Chebyshev} polynomial derivatives  \cite{Peyret2002}:
\begin{align*}
  & \tilde{a}_{\,i} \egal \dfrac{2}{c_{\,i}} \sum_{\substack{p\, =\,i\,+\,1 \\ p\,+\,i\; \text{odd}}}^{n-1} \, p \, a_{\,p}\, , & i \egal 0,1,\ldots,n-1, \\[3pt]
  & \Tilde{a}_{\,n} \ \equiv \ 0 \,, \\[3pt]
  & \Tilde{\Tilde{a}}_{\,i} \egal \dfrac{1}{c_{\,i}} \sum_{\substack{p\, =\, i\,+\,2 \\ p\,+\,i\; \text{even}}}^{n-2}\, p\,\Bigl(\,p^{\,2} \moins i^{\,2}\,\Bigr)\, a_{\,p}\, , & i \egal 0,1,\ldots,n-2,  \\[3pt]
  & \Tilde{\Tilde{a}}_{\,n-1} \ \equiv \  \Tilde{\Tilde{a}}_{\,n} \ \equiv \ 0 \,,
\end{align*}
with,
\begin{align*}
  c_{\,i} \egal \left\lbrace 
  \begin{matrix}
  \ 2 \,, & \text{if} & i \egal 0\,,\\
  \ 1 \,, & \text{if} & i \ >\ 0\,.
  \end{matrix} \right.
\end{align*}
The derivatives of $u\,(x,\,t)$ are written in a similar way by just replacing $v$ by $u$ and $a$ by $b\,$. Thus, the derivatives defined by Eqs.~\eqref{eq:derivative1}, \eqref{eq:derivative2} and \eqref{eq:derivative3} are replaced into Eqs.~\eqref{eq:rearang_moist} and \eqref{eq:rearang_heat} to provide the residuals:
\begin{subequations}
{\footnotesize
\begin{align}
R_{\,1}\egal & \sum_{i\, = \, 0}^{n} \dot{a}_{\,i}\, (t)\, \T_{\,i}\, (x) \moins \nu \biggl(\, \sum_{i\, = \, 0}^{n} a_{\,i} \,(t)\, \T_{\,i}\,(x) \biggr) \sum_{i\, =\, 0}^{n} \Tilde{\Tilde{a}}_{\,i}\, (t)\, \T_{\,i}\, (x) \moins \lambda \biggl( \, \sum_{i\, = \, 0}^{n} \, a_{\,i}\, (t)\, \T_{\,i}\, (x) \biggr) \sum_{i\, =\, 0}^{n} \tilde{a}_{\,i}\,(t)\, \T_{\,i}\,(x) \,, \label{eq:residual_v} \\[5pt] 
R_{\,2} \egal &\sum_{i\, = \, 0}^{n} \dot{b}_{\,i}\, (t)\, \T_{\,i}\, (x) \moins \alpha \biggl(\, \sum_{i\, = \, 0}^{n} a_{\,i} \,(t)\, \T_{\,i}\,(x) \biggr) \sum_{i\, =\, 0}^{n} \Tilde{\Tilde{b}}_{\,i}\, (t)\, \T_{\,i}\, (x) \moins \beta \biggl( \, \sum_{i\, = \, 0}^{n} \, a_{\,i}\, (t)\, \T_{\,i}\, (x) \biggr) \sum_{i\, =\, 0}^{n} \tilde{b}_{\,i}\,(t)\, \T_{\,i}\,(x) \nonumber \\ 
&\moins \gamma \biggl(\, \sum_{i\, = \, 0}^{n} a_{\,i} \,(t)\, \T_{\,i}\,(x) \biggr) \sum_{i\, =\, 0}^{n} \Tilde{\Tilde{a}}_{\,i}\, (t)\, \T_{\,i}\, (x) \moins \delta \biggl( \, \sum_{i\, = \, 0}^{n} \, a_{\,i}\, (t)\, \T_{\,i}\, (x) \biggr) \sum_{i\, =\, 0}^{n} \tilde{a}_{\,i}\,(t)\, \T_{\,i}\,(x) \,, \label{eq:residual_u}
\end{align}}
\end{subequations}

which are considered a misfit of the approximate solution. The purpose is to minimize the residual so the solution satisfies the governing equations. To this end, the residual is minimized via the \textsc{Tau--Galerkin} method, which requires that Eqs.~\eqref{eq:residual_v} and \eqref{eq:residual_u} be orthogonal to the \textsc{Chebyshev} basis functions $\langle\,R\,,\T_{\,j} \,\rangle \,=\, 0\,$:
\begin{align}\label{eq:inner_prodc}
  \langle\,R\,,\T_{\,j} \,\rangle \egal \int_{-1}^{1}\, \dfrac{R\,(\,x\,,t\,)\,\T_{\,j}\,(\,x\,)}{\sqrt{1 \moins x^{\,2}}}\, \mathrm{d} x \egal 0  \,, & & j \egal 0,1,2,\ldots,n-2\,.
\end{align}
As a result, the project residuals are:
\begin{subequations}
\begin{align} \label{eq:res_simpl_v}
  \M \cdot \dot{a}_{\,i}\, (t) &\egal \mathrm{G}_{\, i,\, j} \cdot \Tilde{\Tilde{a}}_{\,i}\, (t) \plus \Lambda_{\,i,\,j} \cdot \Tilde{a}_{\,i}\, (t) \,,\\ \label{eq:res_simpl_u}
  \M \cdot \dot{b}_{\,i}\, (t) &\egal \mathrm{M}_{\, i,\, j} \cdot \Tilde{\Tilde{b}}_{\,i}\, (t) \plus \mathrm{N}_{\, i,\, j} \cdot \Tilde{b}_{\,i}\, (t) \plus \mathrm{F}_{\, i,\, j} \cdot \Tilde{\Tilde{a}}_{\,i}\, (t) \plus \mathrm{J}_{\, i,\, j}\cdot \Tilde{a}_{\,i}\, (t)\,,
\end{align}\label{eq:res_simpl}
\end{subequations}
where, $\M$ is a diagonal and the singular matrix ($\mathrm{rank}\,(\,\M\,)\,=\,N\,-\,2$) which contains the coefficients of the \textsc{Chebyshev} weighted orthogonal system. The matrix $\M$ has the following form:
\begin{align*}
  \M \egal \left[
  \begin{array}{cccccc}
  \pi &   &   &  &  &  \\
      & \frac{\pi}{2} &   &  & \mathbf{0} & \\
   &  & \ddots &  & & \\
   &  &  & \frac{\pi}{2} & & \\
   &\mathbf{0}  &  &  & 0 &  \\
   &  &  &  & & 0
  \end{array} \right]\,,
\end{align*}
and, the matrices with indices ($i\,,$ $j$) written as:
{\footnotesize
\begin{align*}
\mathrm{G}_{\,i,\, j} &\egal \int_{-1}^{1} \, \dfrac{\nu\, \Bigl(\, \sum_{i=0}^{n} a_{\,i} \,(t)\, \T_{\,i}\,(x)\, \Bigr)\, \T_{\,i}\, (x) \, \T_{\,j}\, (x) }{\sqrt{1 \moins x^{\,2}}} \, \mathrm{d}x \,,  &
\Lambda_{\,i,\, j} &\egal \int_{-1}^{1} \, \dfrac{\lambda\, \Bigl(\, \sum_{i=0}^{n} a_{\,i} \,(t)\, \T_{\,i}\,(x)\, \Bigr)\, \T_{\,i}\, (x) \, \T_{\,j}\, (x) }{\sqrt{1 \moins x^{\,2}}} \,  \mathrm{d}x \,, \\
\mathrm{M}_{\,i,\, j} &\egal \int_{-1}^{1} \, \dfrac{\alpha\, \Bigl(\, \sum_{i=0}^{n} a_{\,i} \,(t)\, \T_{\,i}\,(x)\, \Bigr)\, \T_{\,i}\, (x) \, \T_{\,j}\, (x) }{\sqrt{1 \moins x^{\,2}}} \, \mathrm{d}x \,, &
\mathrm{N}_{\,i,\, j} &\egal \int_{-1}^{1} \, \dfrac{\beta\, \Bigl(\, \sum_{i=0}^{n} a_{\,i} \,(t)\, \T_{\,i}\,(x)\, \Bigr)\, \T_{\,i}\, (x) \, \T_{\,j}\, (x) }{\sqrt{1 \moins x^{\,2}}} \,  \mathrm{d}x \,, \\
\mathrm{F}_{\,i,\, j} &\egal \int_{-1}^{1} \, \dfrac{\gamma\, \Bigl(\, \sum_{i=0}^{n} a_{\,i} \,(t)\, \T_{\,i}\,(x)\, \Bigr)\, \T_{\,i}\, (x) \, \T_{\,j}\, (x) }{\sqrt{1 \moins x^{\,2}}} \, \mathrm{d}x \,, &
\mathrm{J}_{\,i,\, j} &\egal \int_{-1}^{1} \, \dfrac{\delta\, \Bigl(\, \sum_{i=0}^{n} a_{\,i} \,(t)\, \T_{\,i}\,(x)\, \Bigr)\, \T_{\,i}\, (x) \, \T_{\,j}\, (x) }{\sqrt{1 \moins x^{\,2}}} \,  \mathrm{d}x \,. 
\end{align*}}
with indices $i\,$, $j$ are the ones defined in Eqs.~\eqref{eq:series} and \eqref{eq:inner_prodc}.

By using the \textsc{Chebyshev--Gau}\ss{} quadrature, the integrals are also approximated by a finite sum:
\begin{align*}
\mathrm{G}_{\,i,\, j} &\ \approx\ \dfrac{\pi}{m} \ \sum_{k\, =\, 1}^{m}\, \nu_{\,k} \ \T_{\,i}\, (x_{\,k})\, \T_{\,j}\, (x_{\,k})\,, &
\Lambda_{\,i,\, j} &\ \approx\ \dfrac{\pi}{m} \ \sum_{k\, =\, 1}^{m}\, \lambda_{\,k} \ \T_{\,i}\, (x_{\,k})\, \T_{\,j}\, (x_{\,k})\,, \\
\mathrm{M}_{\,i,\, j} &\ \approx\ \dfrac{\pi}{m} \ \sum_{k\, =\, 1}^{m}\, \alpha_{\,k} \ \T_{\,i}\, (x_{\,k})\, \T_{\,j}\, (x_{\,k})\,, &
\mathrm{N}_{\,i,\, j} &\ \approx\ \dfrac{\pi}{m} \ \sum_{k\, =\, 1}^{m}\, \beta_{\,k} \ \T_{\,i}\, (x_{\,k})\, \T_{\,j}\, (x_{\,k})\,, \\
\mathrm{F}_{\,i,\, j} &\ \approx\ \dfrac{\pi}{m} \ \sum_{k\, =\, 1}^{m}\, \gamma_{\,k} \ \T_{\,i}\, (x_{\,k})\, \T_{\,j}\, (x_{\,k})\,, &
\mathrm{J}_{\,i,\, j} &\ \approx\ \dfrac{\pi}{m} \ \sum_{k\, =\, 1}^{m}\, \delta_{\,k} \ \T_{\,i}\, (x_{\,k})\, \T_{\,j}\, (x_{\,k})\,,
\end{align*}
where, 
\begin{align*}
\nu_{\,k}\ &\eqdef\ \nu \,\Biggl( \, \sum_{i\, =\, 0}^n \,a_{\,i}\,(t)\, \T_{\,i}\,(x_{\,k}) \, \Biggr) \,,&
\lambda_{\,k}\ &\eqdef\ \lambda \,\Biggl( \, \sum_{i\, =\, 0}^n \,a_{\,i}\,(t)\, \T_{\,i}\,(x_{\,k}) \, \Biggr) \,, \\
\alpha_{\,k}\ &\eqdef\ \alpha \,\Biggl( \, \sum_{i\, =\, 0}^n \,a_{\,i}\,(t)\, \T_{\,i}\,(x_{\,k}) \, \Biggr) \,,&
\beta_{\,k}\ &\eqdef\ \beta \,\Biggl( \, \sum_{i\, =\, 0}^n \,a_{\,i}\,(t)\, \T_{\,i}\,(x_{\,k}) \, \Biggr) \,, \\
\gamma_{\,k}\ &\eqdef\ \gamma \,\Biggl( \, \sum_{i\, =\, 0}^n \,a_{\,i}\,(t)\, \T_{\,i}\,(x_{\,k}) \, \Biggr) \,,&
\delta_{\,k}\ &\eqdef\ \delta \,\Biggl( \, \sum_{i\, =\, 0}^n \,a_{\,i}\,(t)\, \T_{\,i}\,(x_{\,k}) \, \Biggr) \,,
\end{align*}
and $x_{\,k}$ are the \textsc{Chebyshev} nodes:
\begin{align*}
x_{\,k} \egal \cos\, \Biggl( \, \dfrac{2\, k \moins 1}{2\, m}\, \pi \, \Biggr) \,, & & k \egal 1, \,2, \,\ldots,\, m \,.
\end{align*}
The value of $m$ is approximately the same as the number of modes, as discussed in \cite{Gasparin2019}.

To complete the problem, the boundary conditions are also written in the form of the residuals. First, the boundary conditions associated to the moisture transport --- Eq.~\eqref{eq:dimlss_BC_simpl_v}:
\begin{align*}
\omega_{\,1} &\egal \kM \, \biggl(\, \sum_{i\,=\,0}^{n} \,a_{\,i}\,(t)\, \T_{\,i}\, (-1) \biggr)\, \sum_{i\, =\, 0}^{n}\, \tilde{a}_{\,i}\, (t)\, \T_{\,i}\, (-1) \moins \BiML \biggl( \, \sum_{i\,=\,0}^{n}\, a_{\,i}\,(t)\, \T_{\,i}\, (-1) \moins \vinfL \biggr)\,, \\
\omega_{\,2} &\egal \kM \, \biggl(\, \sum_{i\,=\,0}^{n}\, a_{\,i}\,(t)\,\T_{\,i}\,(1)  \biggr)\,  \sum^n_{i\, =\, 0}\, \tilde{a}_{\,i}\, (t)\,\T_{\,i}\,(1)  \plus \BiMR \biggl(\, \sum^n_{i\, =\, 0}\, a_{\,i}\,(t)\,\T_{\,i}\,(1) \moins \vinfR \biggr)\, \,.
\end{align*}
Then, the boundary conditions for Eq.~\eqref{eq:dimlss_BC_simpl_u}, regarding the heat transport:
\begin{align*}
\kappa_{\,1} \egal & \kT \, \biggl(\, \sum_{i\,=\,0}^{n} \,a_{\,i}\,(t)\, \T_{\,i}\, (-1) \biggr)\, \sum_{i\, =\, 0}^{n}\, \tilde{b}_{\,i}\, (t)\, \T_{\,i}\, (-1) \plus 
\kTM \, \biggl(\, \sum_{i\,=\,0}^{n} \,a_{\,i}\,(t)\, \T_{\,i}\, (-1) \biggr)\, \sum_{i\, =\, 0}^{n}\, \tilde{a}_{\,i}\, (t)\, \T_{\,i}\, (-1)\nonumber  \\
&\moins \BiTL \biggl( \, \sum_{i\,=\,0}^{n}\, b_{\,i}\,(t)\, \T_{\,i}\, (-1) \moins \uinfL \biggr)
\moins \BiTML \biggl( \, \sum_{i\,=\,0}^{n}\, a_{\,i}\,(t)\, \T_{\,i}\, (-1) \moins \vinfL \biggr) \,,\\[5pt]
\kappa_{\,2} \egal  &\kT \, \biggl(\, \sum_{i\,=\,0}^{n} \,a_{\,i}\,(t)\, \T_{\,i}\,(1) \biggr)\, \sum_{i\, =\, 0}^{n}\, \tilde{b}_{\,i}\, (t)\, \T_{\,i}\, (1) \plus 
\kTM \, \biggl(\, \sum_{i\,=\,0}^{n} \,a_{\,i}\,(t)\, \T_{\,i}\,(1) \biggr)\, \sum_{i\, =\, 0}^{n}\, \tilde{a}_{\,i}\, (t)\, \T_{\,i}\,(1)\nonumber  \\
&\plus \BiTR \biggl( \, \sum_{i\,=\,0}^{n}\, b_{\,i}\,(t)\, \T_{\,i}\, (1) \moins \uinfR \biggr)
\plus \BiTMR \biggl( \, \sum_{i\,=\,0}^{n}\, a_{\,i}\,(t)\, \T_{\,i}\, (1) \moins \vinfR \biggr) \,, 
\end{align*}
with $T_{\,i}\,(\,-\,1)\, =\, (\,-\,1)^{\,i}$ and $T_{\,i}\,(\, 1\,)\, \equiv\, 1$ (see \cite{Peyret2002} for more details).

In this way, it is possible to compose the system of ODEs to be solved, plus the four additional algebraic expressions regarding the boundary conditions. Finally, the system of differential--algebraic equations (DAEs) has the following form:
\begin{align}\label{eq:system_DAE}
\left(\begin{array}{cc}
\M & 0 \\
0 & \M 
\end{array}\right) & 
\left[\begin{array}{c}
\dot{a}_{\,n} \\
\dot{b}_{\,n}  
\end{array}\right] \egal 
\left(\begin{array}{cc}
\A & 0 \\
\B & \C 
\end{array}\right) \cdot
\left[\begin{array}{c}
a_{\,n}\,(t) \\
b_{\,n}\,(t)  
\end{array}\right] \plus
\left[\begin{array}{c}
\b_{\,1}\,(t) \\
\b_{\,2}\,(t)  
\end{array}\right]  \,,
\end{align}
where $\b_{\,1}\,(t)$ and $\b_{\,2}\,(t)$ are vectors containing the boundary conditions, previously defined by $\omega_{\,1}\,$, $\omega_{\,2}\,$, $\kappa_{\,1}$ and $\kappa_{\,2}\,$:
\begin{align*}
\b_{1}\,(t)\egal \left[
\begin{array}{c}
0  \\
0  \\
\vdots   \\
0   \\
\omega_{\,1}  \\
\omega_{\,2}
\end{array} \right]
& & \text{and} & & \b_{2}\,(t)\egal \left[
\begin{array}{c}
0  \\
0  \\
\vdots   \\
0   \\
\kappa_{\,1}   \\
\kappa_{\,2} 
\end{array} \right]\,.
\end{align*}
Matrix $\A$ is written from the right member of Eq.~\eqref{eq:res_simpl_v}, and, matrices $\B$ and $\C$ are written from the right member of Eq.~\eqref{eq:res_simpl_u}.

Initial values of the coefficients $\{a_{\,i}\,(t\, =\, 0)\}$ and $\{b_{\,i}\,(t\, =\, 0) \}$ are calculated by the orthogonal projection of the initial condition \cite{Canuto2006}:
\begin{subequations}
\begin{align}
a_{\,0,\,i}\ \equiv\ a_{\,i}\,( 0) \egal \dfrac{2}{\pi\, c_{\,i}}\, \int_{-1}^{\,1}\, \dfrac{v_{\, 0}\,(x)\, \T_{\,i}\,(x)}{\sqrt{1 \moins x^{\,2}}}\, \mathrm{d}x\,, & &
i \egal 0, \,1, \,\ldots,\, n \,, \\
b_{\,0,\,i}\ \equiv\ b_{\,i}\,( 0) \egal \dfrac{2}{\pi\, c_{\,i}}\, \int_{-1}^{\,1}\, \dfrac{u_{\, 0}\,(x)\, \T_{\,i}\,(x)}{\sqrt{1 \moins x^{\,2}}}\, \mathrm{d}x\,, & &
i \egal 0, \,1, \,\ldots,\, n \,,
\end{align}
\label{eq:spect_initial}
\end{subequations}
where, $v_{\, 0}\,(x)$ and $u_{\, 0}\,(x)$, are the dimensionless initial condition.

Therefore, the \emph{reduced} system of ODEs composed from Eqs.~\eqref{eq:system_DAE} and \eqref{eq:spect_initial} can be solved. Different approaches can be used to solve the system of ODEs~\eqref{eq:system_DAE}. The most straightforward solution is to apply a numerical integration scheme, with moderate accuracy. So, with an embedded error control and not so stringent tolerances, it can be done very efficiently. In this work, the \texttt{Matlab}\textsuperscript{\texttrademark} environment was used to perform simulations, and the solvers \texttt{ODE15s} or \texttt{ODE23t} were used to solve the differential-algebraic system of equations (DAEs). The output are the vectors of spectral coefficients $\{a_{\,i}\,(t)\}_{\,i\,=\,0}^{\,n}$ and $\{b_{\,i}\,(t) \}_{\,i\,=\,0}^{\,n}\,$. Then, it enables to reconstruct the solution thanks to spectral representations.

\subsection{Extension to multilayered domains}

\begin{figure}
\begin{center}
\includegraphics[width=0.75\textwidth]{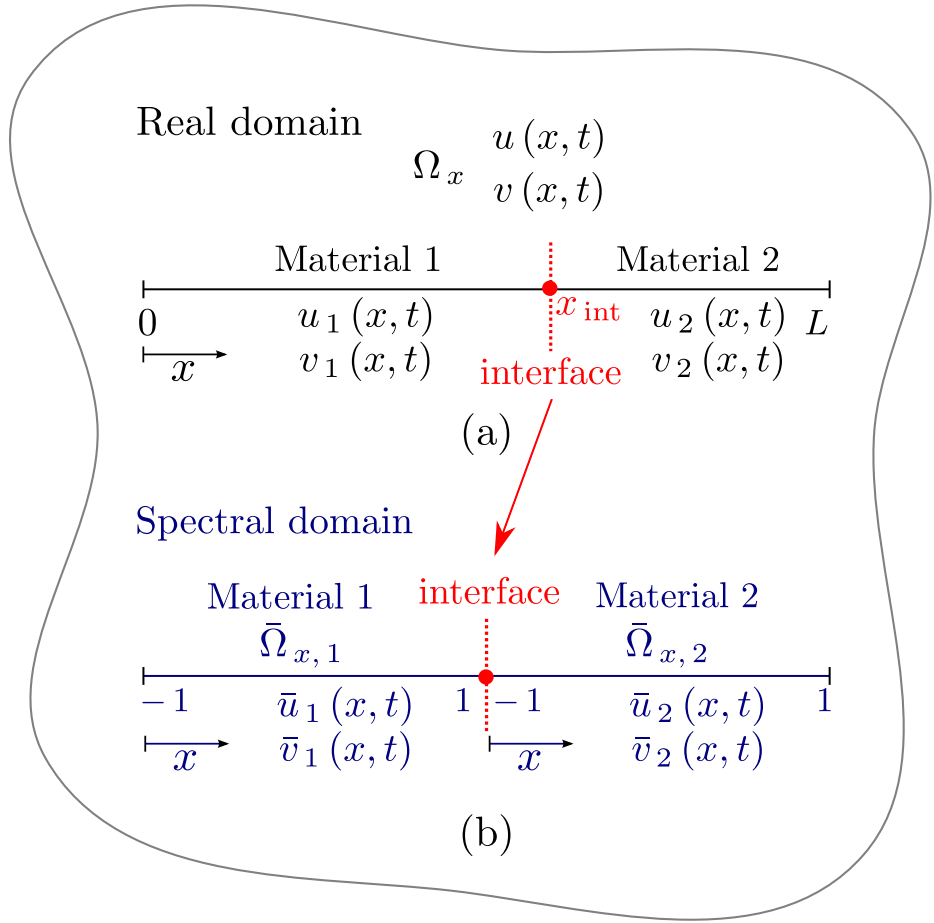}
\caption{\small\em Schematic representation of the domain division with the real domain (a) transformed linearly to obtain the spectral domain (b).}
\label{fig_AN4:layred_spectral}
\end{center}
\end{figure}

We present the idea for two materials, but it can be generalized to any finite number of subdomains straightforwardly. Consider that the original spatial domain $\Omega_{\,x} \egalb [\,0,\,L\,]$ is decomposed into two subdomains $\Omega_{\,x,\,1} \egalb [\,0,\,x_{\,\text{int}}\,]$ and $\Omega_{\,x,\,2} \egalb (\,x_{\,\text{int}},\,L\,]\,$, which represent each material surface and $x_{\,\text{int}}$ represents the location of the interface, as previously defined. From this, the unknowns $v\,(x,\,t)$ and $u\,(x,\,t)$ are then written as:
\begin{align*}
v\, (x,\, t) &\egal \left\{ \begin{array}{cc} 
v_{\,1}\, (x,\, t)\,, & x\in \Omega_{\,x,\,1}\,,\\
v_{\,2}\, (x,\, t), & x\in \Omega_{\,x,\,2}\,, \end{array} \right. \\
u\, (x,\, t) &\egal \left\{ \begin{array}{cc} 
u_{\,1}\, (x,\, t), & x\in \Omega_{\,x,\,1}\,,\\
u_{\,2}\, (x,\, t), & x\in \Omega_{\,x,\,2}\,, \end{array}\right. 
\end{align*}
for $t \geqslant 0\,$. Thus, $v_{\,1}\,$, $v_{\,2}\,$, $u_{\,1}$ and $u_{\,2}$ are written respectively as:
\begin{align*}
v_{\,1}\, (x,\, t) &\egal \sum_{i\, =\, 0}^{n} \, a_{\,i,\,1}\, (t)\, \T_{\,i}\, (x)\,,  &  
v_{\,2}\, (x,\, t) &\egal \sum_{i\, =\, 0}^{n} \, a_{\,i,\,2}\, (t)\, \T_{\,i}\, (x)\,, \\
u_{\,1}\, (x,\, t) &\egal \sum_{i\, =\, 0}^{n} \, b_{\,i,\,1}\, (t)\, \T_{\,i}\, (x)\,,  &  
u_{\,2}\, (x,\, t) &\egal \sum_{i\, =\, 0}^{n} \, b_{\,i,\,2}\, (t)\, \T_{\,i}\, (x)\,.
\end{align*}
Note that the \textsc{Chebyshev} polynomials $\T\,(x)$ are always the same and $x$ must always be in the closed set $[\,-1,\,1\,]\,$. To assure this, the subdomains $\Omega_{\,x,\,1} $ and $ \Omega_{\,x,\,2}$ are linearly transformed to the spectral domains $\bar{\Omega}_{\,x,\,1}$ and $\bar{\Omega}_{\,x,\,2\,}$, as illustrated in Figure~\ref{fig_AN4:layred_spectral}, so they can fit within the interval of interest, which should be seen as a horizontal coordinate in each material. In addition, by separating the real domain in two spectral domains, the smoothness of the solution is assured.

The condition at the interface between the two materials states the continuity of the fields and the flows as in Eq.~\eqref{eq:diml_interf}. It implies that the derivatives of the fields $u$ and $v$ are not continuous at the interface. This important remark has to be taken into account in the construction of the spectral reduced-order model. Indeed, the domain is decomposed in subdomains to maintain a smooth solution and particularly a continuous derivative on each sub-domain. In this way, the model order reduction is optimal and ensures the error of the Spectral-ROM to decrease exponentially. It is possible to build the reduced order model considering the whole domain (without decomposition). However, the convergence is undermined since the solution and its derivatives are not smooth at the interface between two materials. Many more modes would be necessary to reach the same accuracy, as detailed in Theorem~$1$ of \cite[Chap.~4]{Trefethen2000}.

By considering the two materials, the DAE system represented in Eq.~\eqref{eq:system_DAE} becomes:
{\footnotesize
\begin{align*}
\left( 
\begin{array}{cccc}
\M &  & & \mathbf{0}\\
 & \M & & \\ 
 & & \M & \\
\mathbf{0} & & & \M 
\end{array}
\right) \cdot
\left[ 
\begin{array}{c}
\dot{a}_{\,i,\,1} \\[5pt]
\dot{b}_{\,i,\,1} \\[5pt]
\dot{a}_{\,i,\,2} \\[5pt]
\dot{b}_{\,i,\,2} 
\end{array}
\right] \egal
\left(\begin{array}{cccc}
\A_{\,1} & \mathbf{0}  &  & \mathbf{0}\\[5pt]
\B_{\,1} & \C_{\,1} &   &  \\[5pt]
 &  & \A_{\,2} &   \mathbf{0} \\[5pt]
\mathbf{0} &  & \B_{\,2} & \C_{\,2}  \\[5pt]
\end{array}\right) \cdot
\left[\begin{array}{c}
a_{\,n,\,1}\,(t) \\[5pt]
b_{\,n,\,1}\,(t) \\[5pt]
a_{\,n,\,2}\,(t) \\[5pt]
b_{\,n,\,2}\,(t)   
\end{array}\right] +
\left[ 
\begin{array}{c}
b_{\,1}\, (t) \\[5pt]
b_{\,2}\, (t) \\[5pt]
b_{\,3}\, (t) \\[5pt]
b_{\,4}\, (t) 
\end{array} 
\right]\,.
\end{align*}}
The interface conditions -- Eq.~\eqref{eq:diml_interf} -- are then written in the spectral form as:
\begin{align*}
\vartheta_{\,1} &\egal \sum^n_{i\, =\, 0}\, a_{\,i,\,1}\, (t)\,  \moins \sum^n_{i\, =\, 0}\, a_{\,i,\,2}\,(t)\, (-1)^{\,i}   \,, \\
\vartheta_{\,2} &\egal k_{\,M,\,1} \Biggl(\, \sum_{i\,=\,0}^{n} \,a_{\,i,\,1}(t)\Biggr)\, \sum_{i\, =\, 0}^{n}\, \tilde{a}_{\,i,\,1}(t)\, \moins k_{\,M,\,2} \Biggl(\, \sum_{i\,=\,0}^{n} \,a_{\,i,\,2}(t)\, (-1)^{\,i}\Biggr)\, \sum_{i\, =\, 0}^{n}\, \tilde{a}_{\,i,\,2}(t)\, (-1)^{\,i} \,, \\
\vartheta_{\,3} &\egal \sum^n_{i\, =\, 0}\, b_{\,i,\,1}\, (t)\,  \moins \sum^n_{i\, =\, 0}\, b_{\,i,\,2}\,(t)\, (-1)^{\,i}  \,,  \\
\vartheta_{\,4} &\egal k_{\,T,\,1} \Biggl(\, \sum_{i\,=\,0}^{n} \,a_{\,i,\,1}(t)\Biggr)\, \sum_{i\, =\, 0}^{n}\, \tilde{b}_{\,i,\,1}(t)\, \plus k_{\,TM,\,1} \Biggl(\, \sum_{i\,=\,0}^{n} \,a_{\,i,\,1}(t)\Biggr)\, \sum_{i\, =\, 0}^{n}\, \tilde{a}_{\,i,\,1}(t)\,\\
&\moins k_{\,T,\,2} \Biggl(\, \sum_{i\,=\,0}^{n} \,a_{\,i,\,2}(t)\, (-1)^{\,i}\Biggr)\, \sum_{i\, =\, 0}^{n}\, \tilde{b}_{\,i,\,2}(t)\, (-1)^{\,i}  \moins k_{\,TM,\,2} \Biggl(\, \sum_{i\,=\,0}^{n} \,a_{\,i,\,2}(t)\, (-1)^{\,i}\Biggr)\, \sum_{i\, =\, 0}^{n}\, \tilde{a}_{\,i,\,2}(t)\, (-1)^{\,i} \,, 
\end{align*}
which are included in vectors $b_{\,1}$ and $b_{\,2}$ and set equal to zero. In the same way, the boundary conditions are written in the spectral form as:
\begin{align*}
\sigma_{\,1} &\egal k_{\,M,\,1} \, \biggl(\, \sum_{i\,=\,0}^{n} \,a_{\,i}\,(t)\, \T_{\,i}\, (-1) \biggr)\, \sum_{i\, =\, 0}^{n}\, \tilde{a}_{\,i}\, (t)\, \T_{\,i}\, (-1) \moins \BiML \biggl( \, \sum_{i\,=\,0}^{n}\, a_{\,i}\,(t)\, \T_{\,i}\, (-1) \moins \vinfL \biggr)\,, \\
\sigma_{\,2} &\egal k_{\,M,\,2} \, \biggl(\, \sum_{i\,=\,0}^{n}\, a_{\,i}\,(t)\,\T_{\,i}\,(1)  \biggr)\,  \sum^n_{i\, =\, 0}\, \tilde{a}_{\,i}\, (t)\,\T_{\,i}\,(1)  \plus \BiMR \biggl(\, \sum^n_{i\, =\, 0}\, a_{\,i}\,(t)\,\T_{\,i}\,(1) \moins \vinfR \biggr)\,, \\
\sigma_{\,3} &\egal k_{\,T,\,1} \, \biggl(\, \sum_{i\,=\,0}^{n} \,a_{\,i}\,(t)\, \T_{\,i}\, (-1) \biggr)\, \sum_{i\, =\, 0}^{n}\, \tilde{b}_{\,i}\, (t)\, \T_{\,i}\, (-1) \plus 
k_{\,TM,\,1} \, \biggl(\, \sum_{i\,=\,0}^{n} \,a_{\,i}\,(t)\, \T_{\,i}\, (-1) \biggr)\, \sum_{i\, =\, 0}^{n}\, \tilde{a}_{\,i}\, (t)\, \T_{\,i}\, (-1)\nonumber  \\
&\moins \BiTL \biggl( \, \sum_{i\,=\,0}^{n}\, b_{\,i}\,(t)\, \T_{\,i}\, (-1) \moins \uinfL \biggr)
\moins \BiTML \biggl( \, \sum_{i\,=\,0}^{n}\, a_{\,i}\,(t)\, \T_{\,i}\, (-1) \moins \vinfL \biggr) \,,\\[5pt]
\sigma_{\,4} &\egal k_{\,T,\,2} \, \biggl(\, \sum_{i\,=\,0}^{n} \,a_{\,i}\,(t)\, \T_{\,i}\,(1) \biggr)\, \sum_{i\, =\, 0}^{n}\, \tilde{b}_{\,i}\, (t)\, \T_{\,i}\, (1) \plus 
k_{\,TM,\,2} \, \biggl(\, \sum_{i\,=\,0}^{n} \,a_{\,i}\,(t)\, \T_{\,i}\,(1) \biggr)\, \sum_{i\, =\, 0}^{n}\, \tilde{a}_{\,i}\, (t)\, \T_{\,i}\,(1)\nonumber  \\
&\plus \BiTR \biggl( \, \sum_{i\,=\,0}^{n}\, b_{\,i}\,(t)\, \T_{\,i}\, (1) \moins \uinfR \biggr)
\plus \BiTMR \biggl( \, \sum_{i\,=\,0}^{n}\, a_{\,i}\,(t)\, \T_{\,i}\, (1) \moins \vinfR \biggr) \,, 
\end{align*}
which are included in $b_{\,3}$ and $b_{\,4}$ and set equal to zero. Vectors $b_{\,1}$, $b_{\,2}$ , $b_{\,3}$ and $b_{\,4}$ are $N \times 1$ column vectors with the form:
\begin{align*}
b_{\,1} = \left[
\begin{array}{c}
0\\
0\\
\vdots \\
0\\
\vartheta_{\,1}\\
\vartheta_{\,2}
\end{array} \right]\,,
& &
b_{\,2} = \left[
\begin{array}{c}
0\\
0\\
\vdots \\
0\\
\vartheta_{\,3}\\
\vartheta_{\,4}
\end{array} \right]\,,
& &
b_{\,3} = \left[
\begin{array}{c}
0\\
0\\
\vdots \\
0\\
\sigma_{\,1}\\
\sigma_{\,2}
\end{array} \right]
& & \text{and} &  &
b_{\,4} = \left[
\begin{array}{c}
0\\
0\\
\vdots \\
0\\
\sigma_{\,3}\\
\sigma_{\,4}
\end{array} \right] \,.
\end{align*}

With all elements listed before, it is possible to set the system to be solved. The system of ODEs for solving the coupled heat and moisture transfer with two layers has the size of $4\cdot N$ with \textbf{four} additional algebraic expressions for the boundary and interface conditions. The initial condition is also given by Eq.~\eqref{eq:spect_initial} and the DAE system is solved by the \texttt{Matlab}\textsuperscript{\texttrademark} solver \texttt{ODE15s}. In this work, the approach was presented for a two-layered wall for the sake of clarity, knowing it can be easily extended to any number of layers and any other boundary conditions at their interfaces.


\section{Numerical benchmark}

To analyze the accuracy of the proposed method, the error between the solution $Y^{\mathrm{num}}$, obtained by the Spectral or IMEX methods, and the reference solutions $Y^{\mathrm{ref}}$, are computed as functions of $x$ using the following \textsc{Euclidean} norms:
\begin{align*}
  \varepsilon_{\,2}^{\,Y}\, (\,x\,)\ &\eqdef\ \sqrt{\,\frac{1}{N_{\,t}} \, \sum_{j\, =\, 1}^{N_{\,t}} \, \Bigl( \, Y_{\, j}^{\, \mathrm{num}}\, (\, x\,, t_{\,j}\,) \moins Y_{\, j}^{\mathrm{\, ref}}\, (\, x\,, t_{\,j}\,) \, \Bigr)^{\,2}}\,, 
\end{align*}
where $N_{\,t}$ is the number of temporal steps and $Y$ can be the temperature $T$ or the vapour pressure $\Pv\,$. The reference solution $Y^{\mathrm{ref}}\,(\,x\,, t\,)$ is computed using the \texttt{Matlab}\textsuperscript{\texttrademark} open source toolbox \texttt{Chebfun} \cite{Driscoll2014}. Moreover, the uniform norm error -- $\varepsilon_{\, \infty}^{\,Y}$ -- is given by the maximal values of $\varepsilon_{\,2}^{\,Y}\, (\,x\,)$\,:
\begin{align*}
  \varepsilon_{\, \infty}^{\,Y}\ &\eqdef\ \sup_{x \ \in \ \bigl[\, 0 \,,\, L \,\bigr]} \, \varepsilon_{\,2}^{\,Y}\, (\,x\,) \,.
\end{align*}


\subsection{Single-layered domain}
\label{sec:case_1layer}

Simulations of one-dimensional coupled heat and moisture transport are carried out with the spectral method, the IMplicit-EXplicit (IMEX) method and a reference solution to verify the applicability of the spectral method. The IMEX scheme approximates the continuous operator to order $\O\,(\dx^{\,2}\ +\ \dt)\,$. The advantage of this semi-implicit scheme over the fully implicit one is to avoid sub-iterations in the solution procedure and, at the same time, being stable and consistent.

\begin{figure}
\centering
\subfigure[a][\label{fig_AN1:BC_T}]{\includegraphics[width=0.48\textwidth]{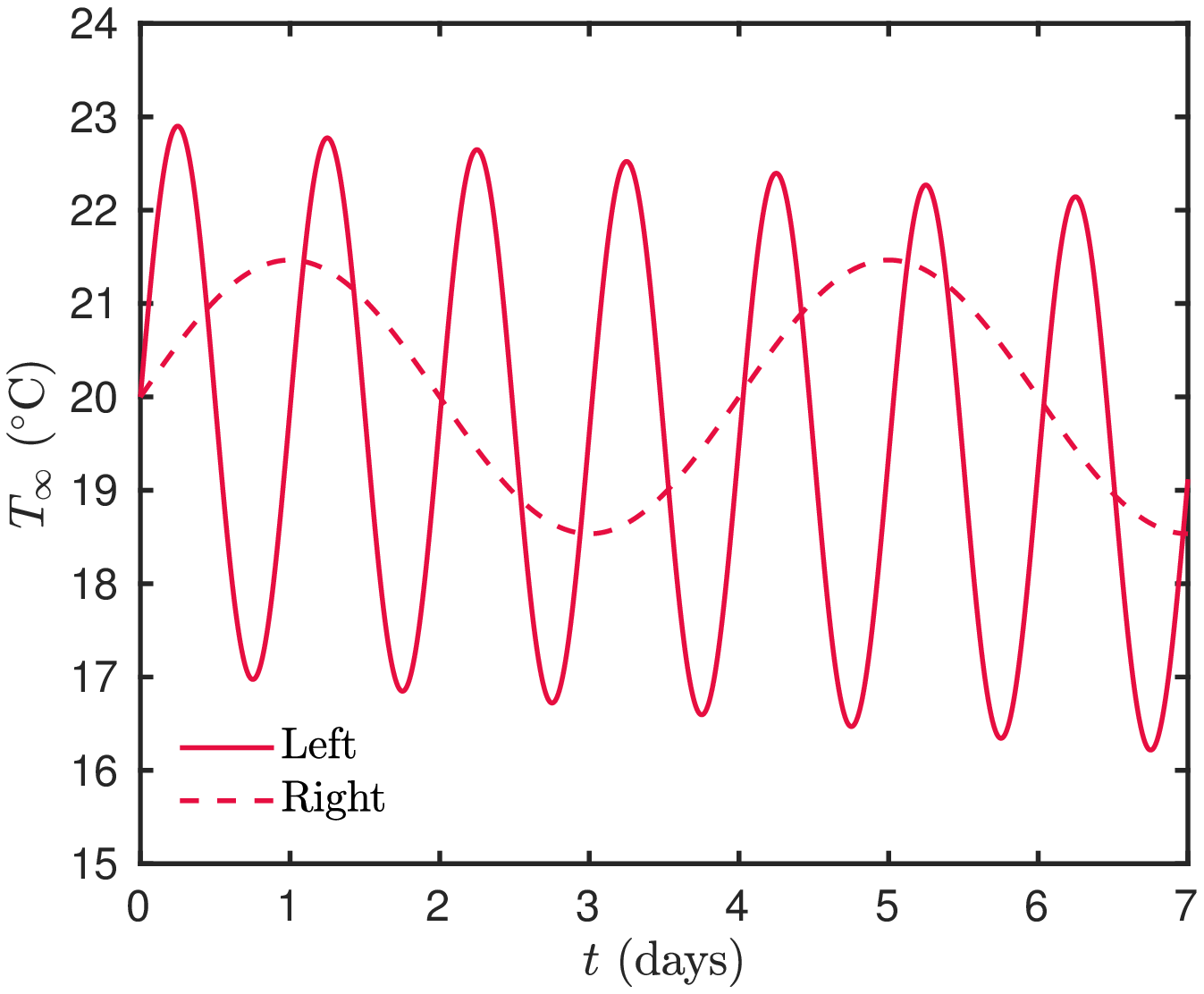}} \hspace{0.3cm}
\subfigure[b][\label{fig_AN1:BC_RH}]{\includegraphics[width=0.48\textwidth]{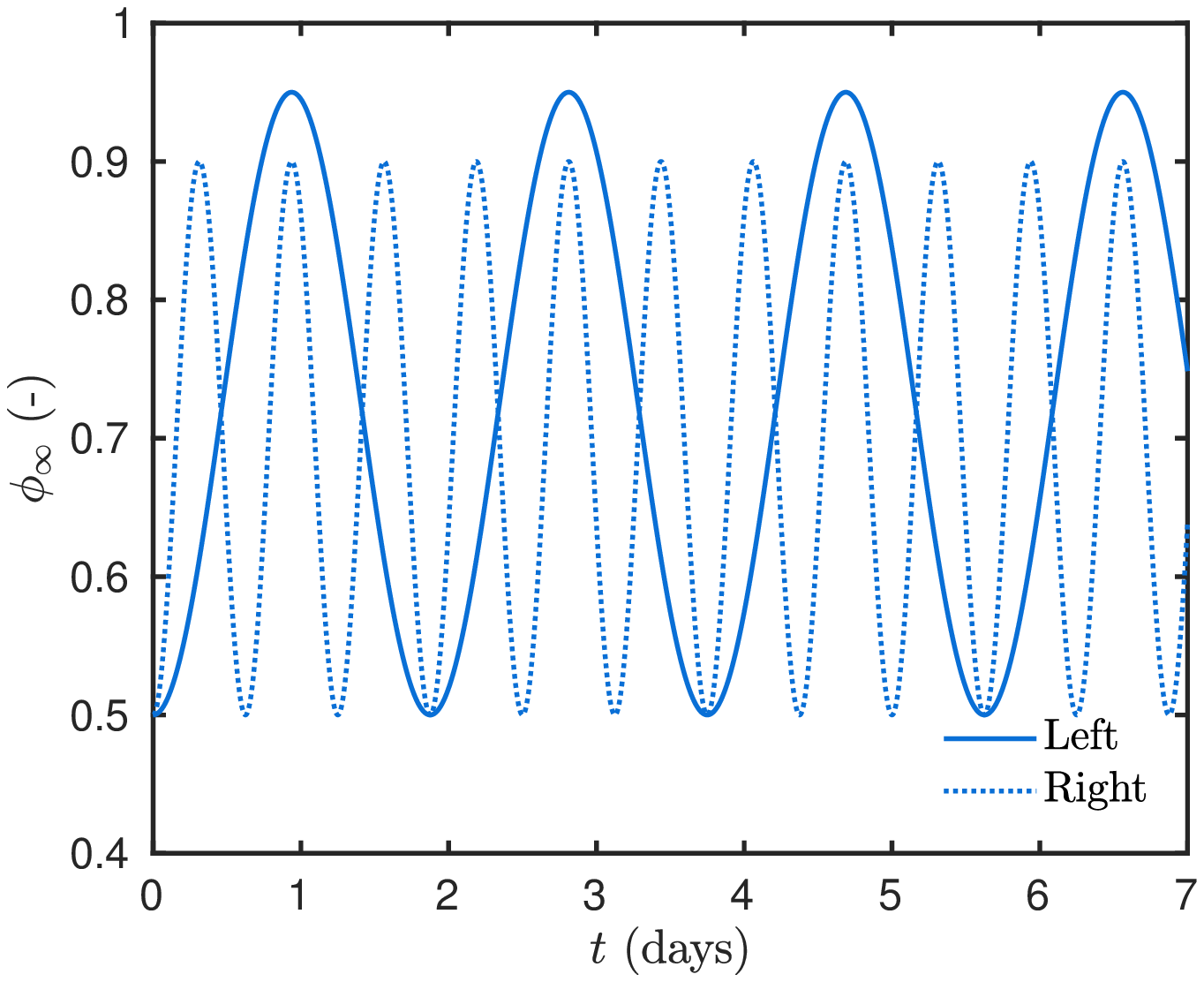}}
\caption{\small\em Ambient temperature (a) and relative humidity (b).}
\label{fig_AN1:BC}
\end{figure}

This case study considers moisture-dependent coefficients $\cM\,$, $\kM\,$, $\cT\,$, $\kT$ and $\kM$ as illustrated in Figure~\ref{fig_AN:properties_coeff} of Appendix~\ref{annexe:material_coefficients}. Their variations are similar to the load bearing material from \textsc{Hagentoft} \cite{Hagentoft2004}, which are presented in Table~\ref{table:properties_loadbearing} of Appendix~\ref{annexe:material_properties}. The material of $0.1\ \mathsf{m}$ of length has uniform vapour pressure and temperature initial conditions, in which $\Pvi \, =\, 1.16 \dix{\,3}\,\mathsf{Pa}$ and $\Ti \egalb 293\, \mathsf{K}$. The ambient temperature and relative humidity at the boundaries vary sinusoidally as illustrated in Figures~\ref{fig_AN1:BC_T} and \ref{fig_AN1:BC_RH}. The convective vapour transfer coefficients are $\hML \, =\, 2 \cdot 10^{\,-7}\, \mathsf{s/m}$ and $\hMR \, =\, 3 \cdot 10^{\,-8}\, \mathsf{s/m}$, for the left and right boundaries, while the convective heat transfer coefficients are set to $\hTL \egalb 25\, \mathsf{W/(m^2\cdot K)}$ and $\hTR \egalb 8\, \mathsf{W/(m^2\cdot K)}$. At the boundaries, only the convective exchange is considered. The simulation is performed for $7\, \mathsf{days}\,$. The dimensionless values can be found in Appendix~\ref{annexe:dimensionless}.

The Spectral method is composed by $N \, = \, 10$ modes with $m \, = \, 15\,$. The \texttt{ODE15s} was used to solve System~\eqref{eq:system_DAE}, with a tolerance set to $10^{\,-\,5}\,$. This solver is adaptive in time, although it can provide the integration in time for the instants required. For this case, the Spectral method was compared to the IMEX and to a reference solution computed using the \texttt{Chebfun} \texttt{Matlab}\textsuperscript{\texttrademark} toolbox \cite{Driscoll2014}. All solutions have been computed with $\Delta \xs \,=\, 10^{\,-\,2}\,$. For the time discretization, the IMEX solution was computed with $\Delta \ts \,=\, 10^{\,-\,2}$ and the Spectral with $\Delta \ts \,=\, 10^{\,-\,1}\,$, to have the same order of error of the solution.

\begin{figure}
\centering
\subfigure[a][\label{fig_AN1:Evolution_Pv}]{\includegraphics[width=0.48\textwidth]{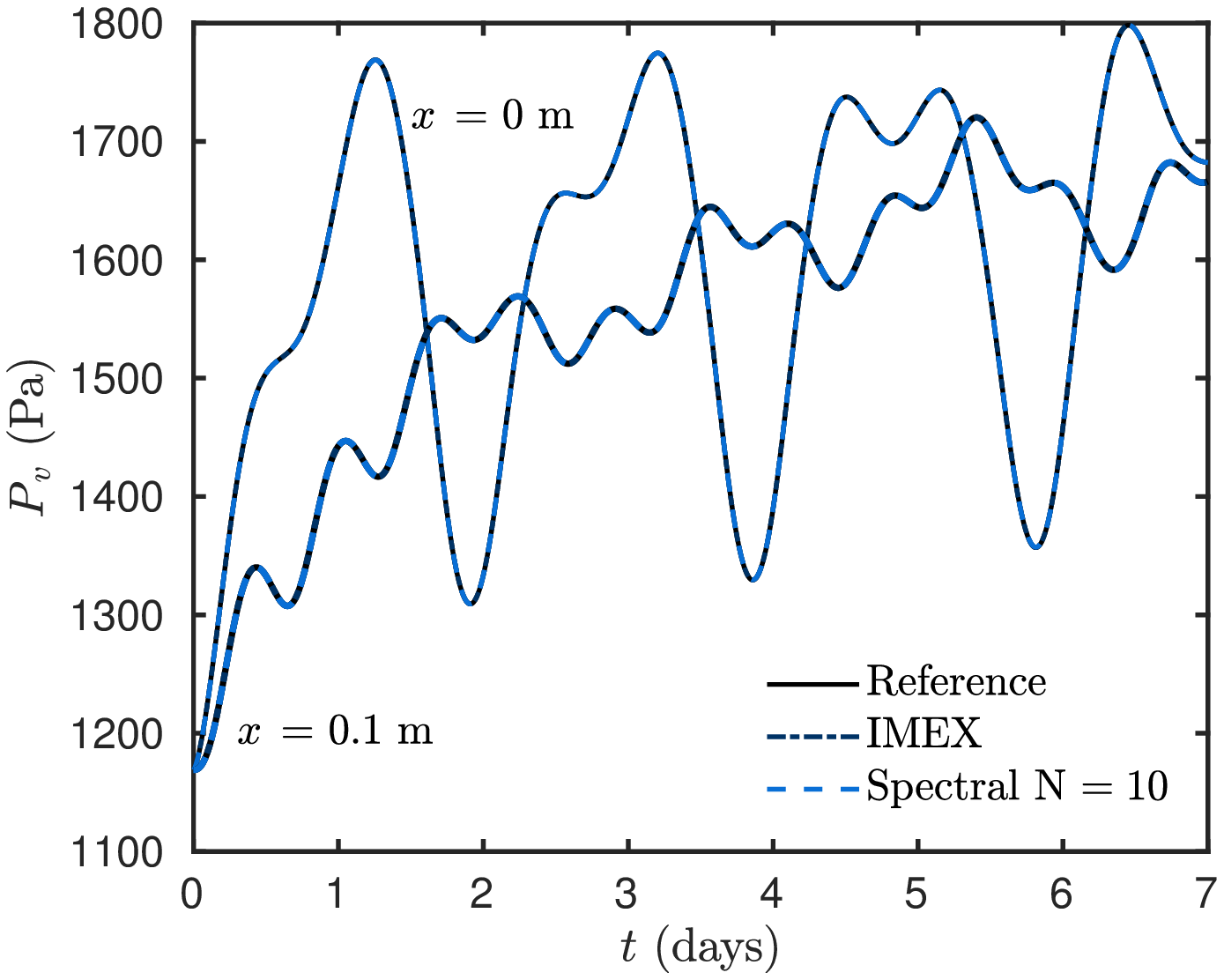}} \hspace{0.3cm}
\subfigure[b][\label{fig_AN1:Evolution_T}]{\includegraphics[width=0.47\textwidth]{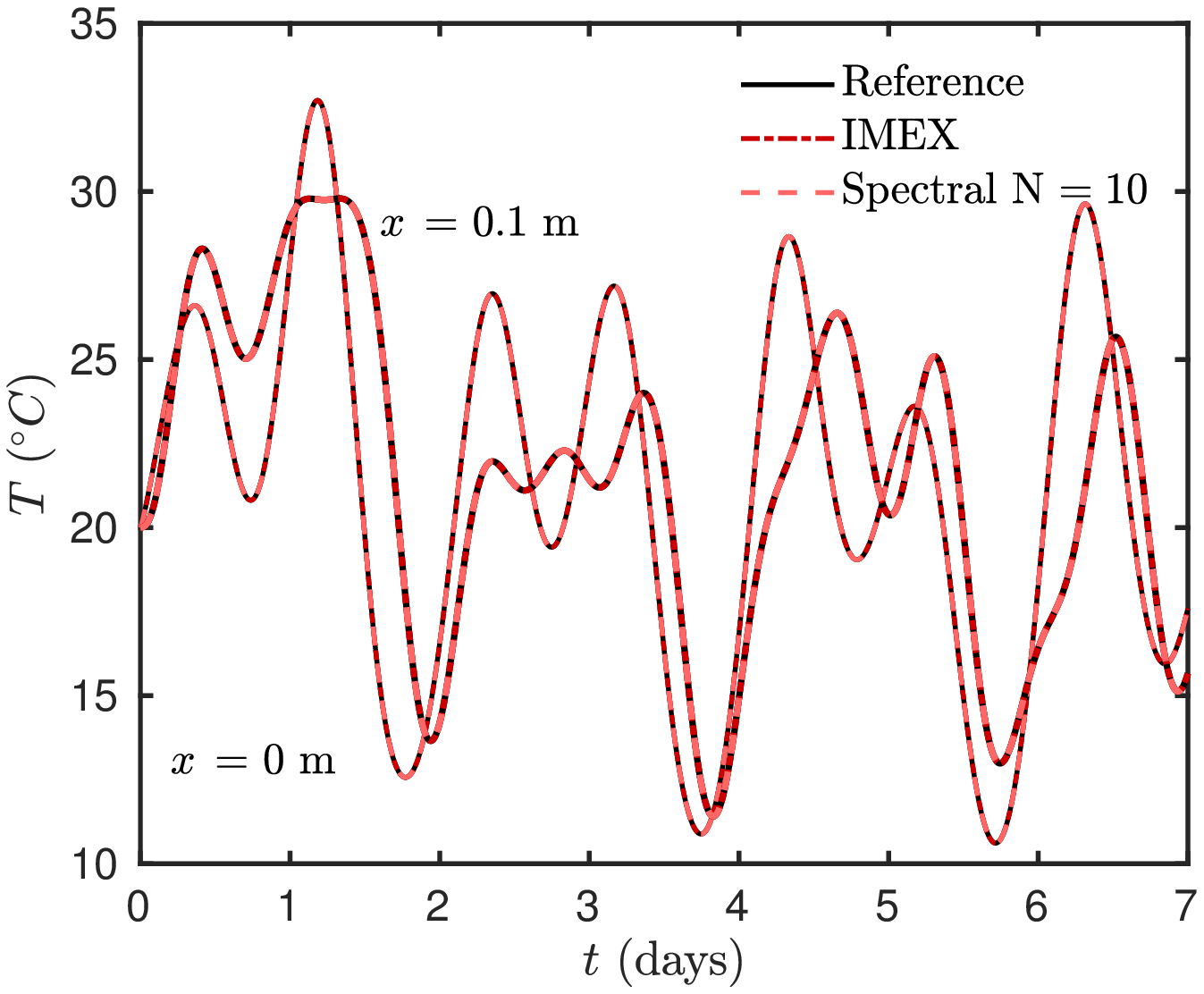}}
\caption{\small\em Vapour pressure evolution at the boundaries (a) and vapour pressure profiles inside the material (b).}
\end{figure}

Vapour pressure variations at the boundaries are shown in Figure~\ref{fig_AN1:Evolution_Pv}. In the first four days, it is possible to observe the influence of the initial condition in which the vapour pressure rises significantly, meaning that when the simulation started the material was not in balance with the ambient environment. The vapour pressure at $x \egalb 0\, \mathsf{m}$ and $x \egalb 0.1\, \mathsf{m}$ oscillates according to the boundary conditions while also retaining the moisture. As the convective vapour transfer coefficient is higher at the left side, the material will exchange more with the external environment making variations more visible.

Figure~\ref{fig_AN1:Evolution_T} shows the temperature variations at the boundaries of the material. The temperature slowly oscillates according to the temperature ambient conditions and also with the vapour pressure variations. As one could expect, vapor pressure and temperature values inversely oscillate.

\begin{figure}
\centering
\subfigure[a][\label{fig_AN1:Error_fx}]{\includegraphics[width=0.48\textwidth]{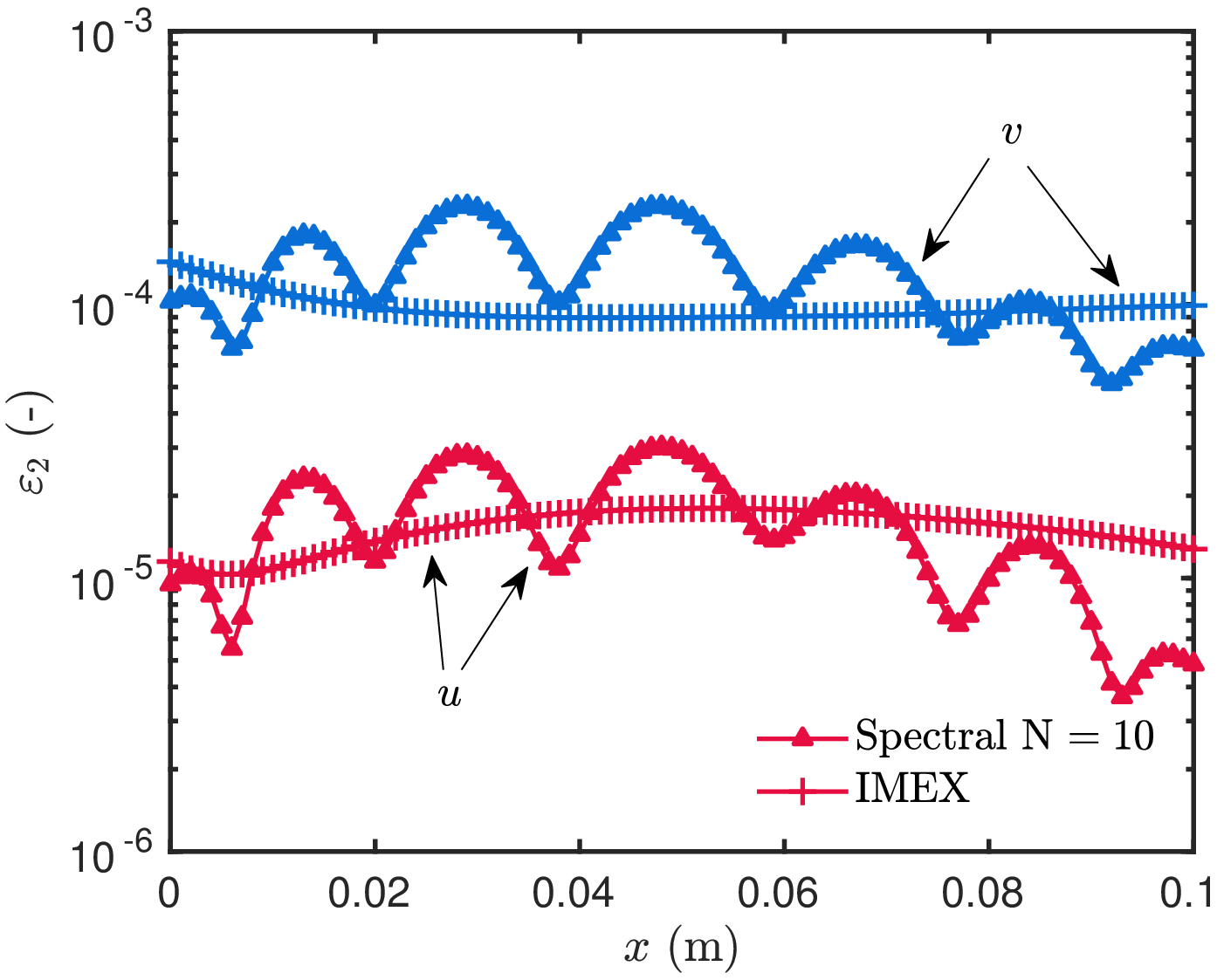}} \hspace{0.3cm}
\subfigure[b][\label{fig_AN1:coefA_spc_last}]{\includegraphics[width=0.46\textwidth]{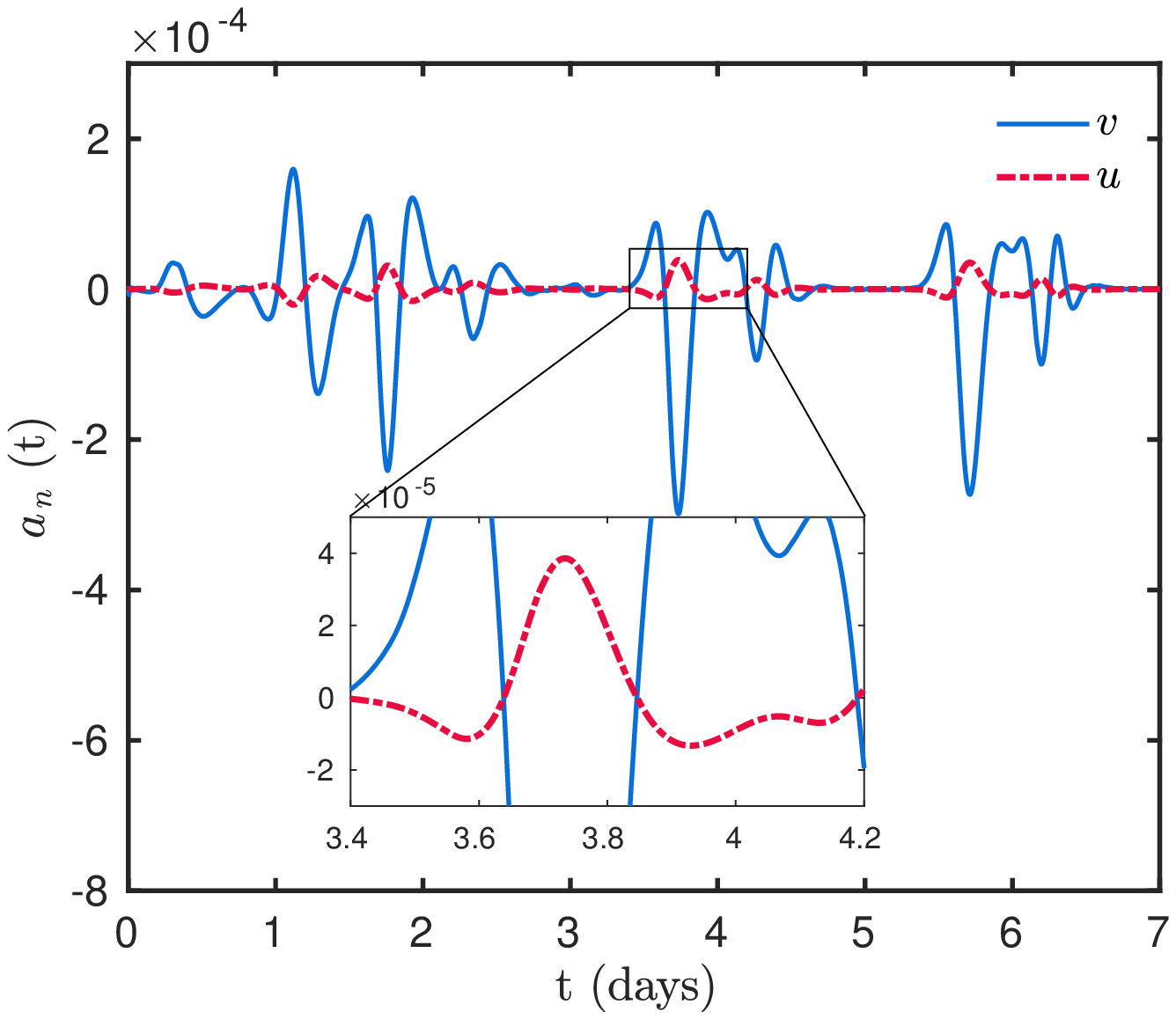}}
\caption{\small\em Error of both solutions (a) and last spectral coefficient (b).}
\end{figure}

Results of the error $\varepsilon_{\,2}$ as a function of $x$ are shown in Figure~\ref{fig_AN1:Error_fx}. The errors associated to the IMEX scheme and to the Spectral method are of order of $\O\,(10^{-5}\,)$, for temperature, and of order of $\O (10^{-4}\,)$ for  moisture. The computer run time used to perform this case and the error $\varepsilon_{\,\infty}$ are given in Table~\ref{table:AN1_cputime}, which shows the spectral method is more efficient, even compared to the IMEX, which is an improved method based on the Euler scheme as it does not need sub-iterations. The number of degrees of freedom (DOF) of the Spectral solution is considerably lower if compared to the IMEX solution, which impacts directly on the computational time, making the Spectral approach $7$ times more efficient for this case.

Figure~\ref{fig_AN1:coefA_spc_last} displays the last coefficient $a_{\, n}$ of the Spectral-ROM solution for the temperature $u$ and vapour pressure $v$ solutions. The last spectral coefficient is the smallest one and it gives the order of approximation of the residual \cite[Page~51]{Boyd2000}. It works as an upper bound on the error. For example, the error of the $u$ solution in Figure~\ref{fig_AN1:Error_fx} is of order of $\O\,(10^{\,-5})$, which has the same magnitude of the last coefficient, as can be seen in Figure~\ref{fig_AN1:coefA_spc_last}.

\begin{figure}
\centering
\subfigure[a][\label{fig_AN1:Modes}]{\includegraphics[width=0.485\textwidth]{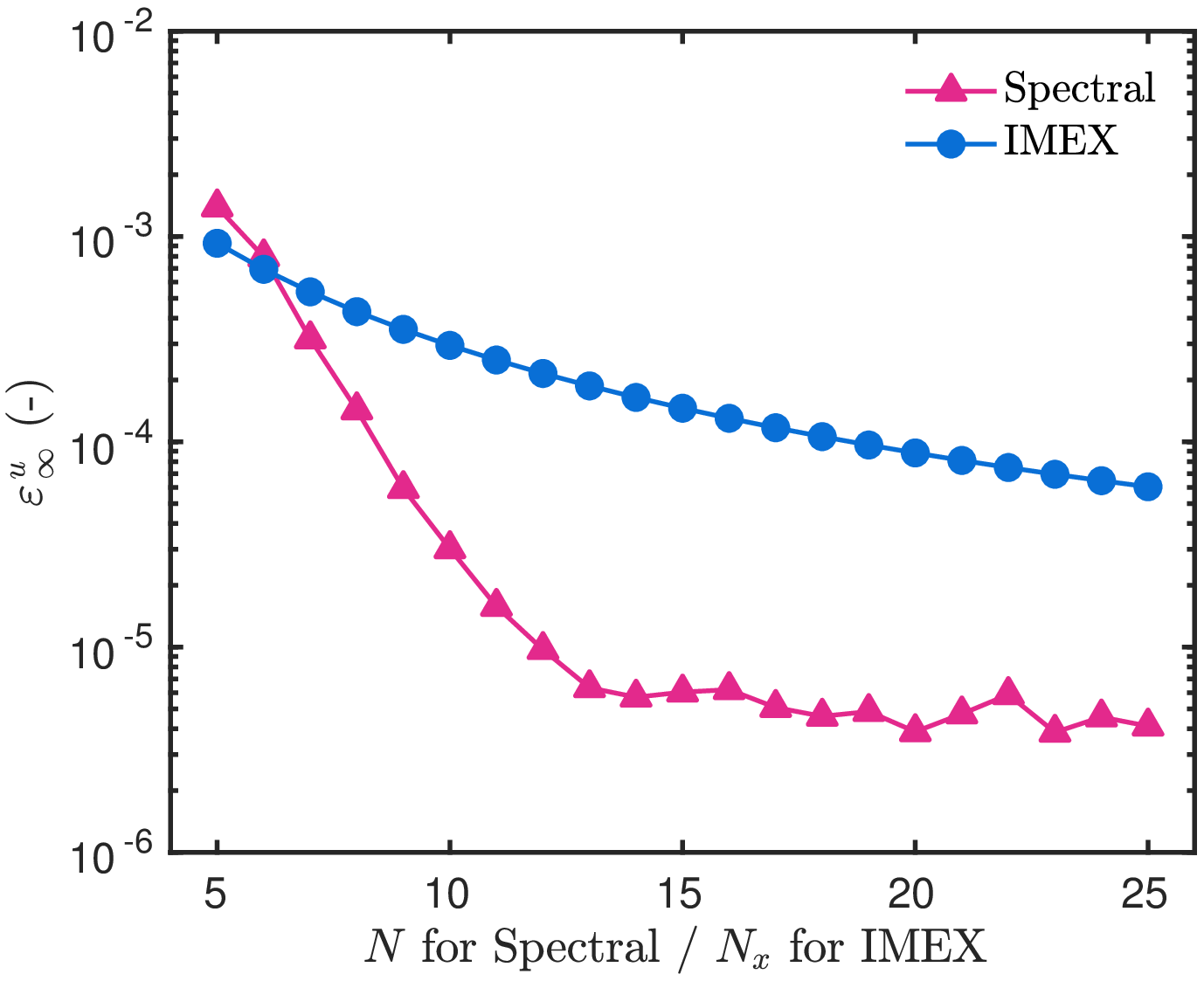}} \hspace{0.3cm}
\subfigure[b][\label{fig_AN1:CPU_time}]{\includegraphics[width=0.47\textwidth]{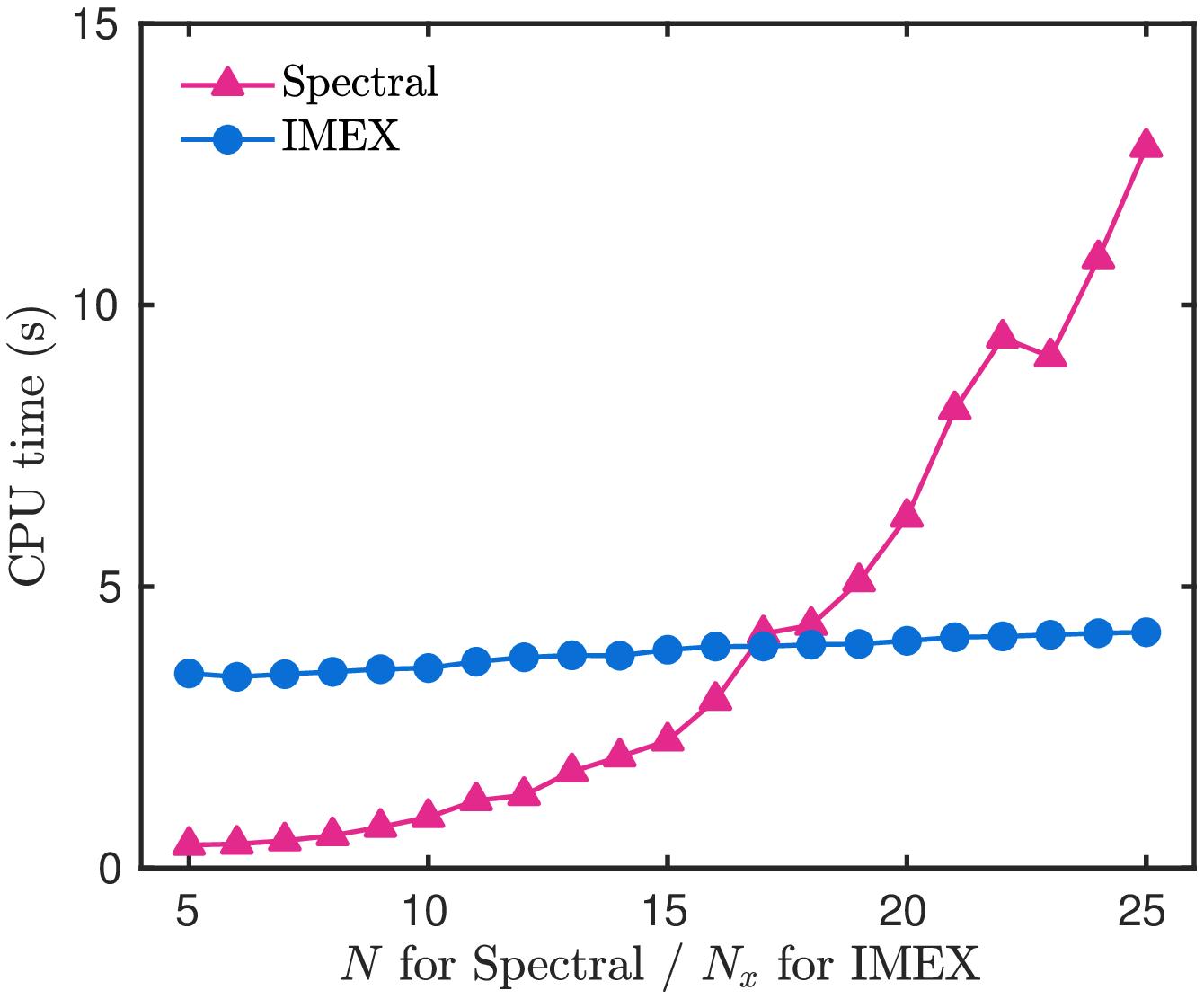}}
\caption{\small\em Error $\varepsilon_{\,\infty}$ (a)  and the equivalent CPU time (b) as a function of the number of modes $N$ for the Spectral solution and as a function of the number of spatial nodes $N_{\,x}$ for the IMEX solution.}
\end{figure}

To illustrate the rate of convergence of the methods, Figure~\ref{fig_AN1:Modes} presents the error $\varepsilon_{\,\infty}$ as a function of the number of modes $N$ for the Spectral solution, and, as a function of the number of spatial nodes $N_{\,x}$ for the IMEX solution. As the number of modes increases, the solution of the Spectral method converges exponentially and stabilizes with approximatively $13$ modes, reaching an accuracy of order of $\O\,(10^{-5})\,$, which is equivalent to the tolerance set on the solver \texttt{ode15s}. On the other hand, the IMEX method converges slower than the Spectral method. It needs at least $100$ spatial nodes to reach the same accuracy of the Spectral method: the CPU time of each simulation was measured and it is presented in Figure~\ref{fig_AN1:CPU_time}. The computational time increases faster for the Spectral method. However, the spectral solution does not need many modes to converge to an acceptable accuracy ($N\, \simeq\, 10$).

\begin{table}
\centering
\caption{\small\em Some features of the one-layer heat and moisture transfer case.}
\def\arraystretch{1.2}
\begin{tabular}{ccc}
\hline
  & \textit{IMEX} & \textit{Spectral $N=10$} \\
\hline
$\Delta \ts$  & $1.00\dix{-2}$  & $1.00\dix{-1}$  \\
$\varepsilon_{\,\infty}^{\,u}$ & $1.48\dix{-5}$ & $3.30\dix{-5}$   \\
$\varepsilon_{\,\infty}^{\,v}$ & $1.43\dix{-4}$ & $2.31\dix{-4}$    \\
DOF & $200$ & $20$ \\
\hline
\rowcolor[HTML]{C0C0C0} 
CPU time $(\mathsf{s})$ & $\mathbf{6.46}$  & $\mathbf{0.94}$ \\
\rowcolor[HTML]{C0C0C0} 
CPU time $(\%)$ & $\mathbf{100}$  & $\mathbf{15}$ \\
\hline 
\end{tabular}
\label{table:AN1_cputime}
\end{table}

The Spectral method has demonstrated good agreement to represent the physical model and the fidelity of the model is not deteriorated with the use of this approach.


\subsection{Multi-layered domain}
\label{sec:case_2layers}

\begin{figure}
\begin{center}
\includegraphics[width=0.75\textwidth]{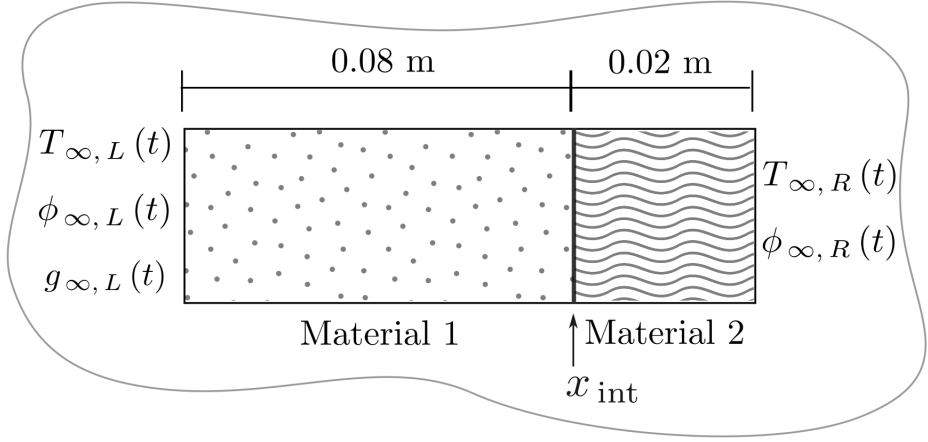}
\caption{\small\em Schematic representation of the two-layers wall.}
\label{fig_AN2:layred_configuration}
\end{center}
\end{figure}

This case study considers a porous wall formed by $2$ layers: $8$-$\mathsf{cm}$ load bearing material and $2$-$\mathsf{cm}$ finishing material. Figure~\ref{fig_AN2:layred_configuration} shows its schematic representation. The first layer has a faster liquid transfer, while the second layer acts as a hygroscopic finish. The properties of these materials are given in Tables~\ref{table:properties_loadbearing} and \ref{table:properties_finishing_mat}. Considering the temperature range of interest in building applications, temperature dependence was neglected when compared to their dependence on moisture content, with the transport coefficients calculated as a function of the moisture content.

\begin{figure}
\begin{center}
  \subfigure[][\label{fig_AN2:Rain_flux}]{\includegraphics[width=.48\textwidth]{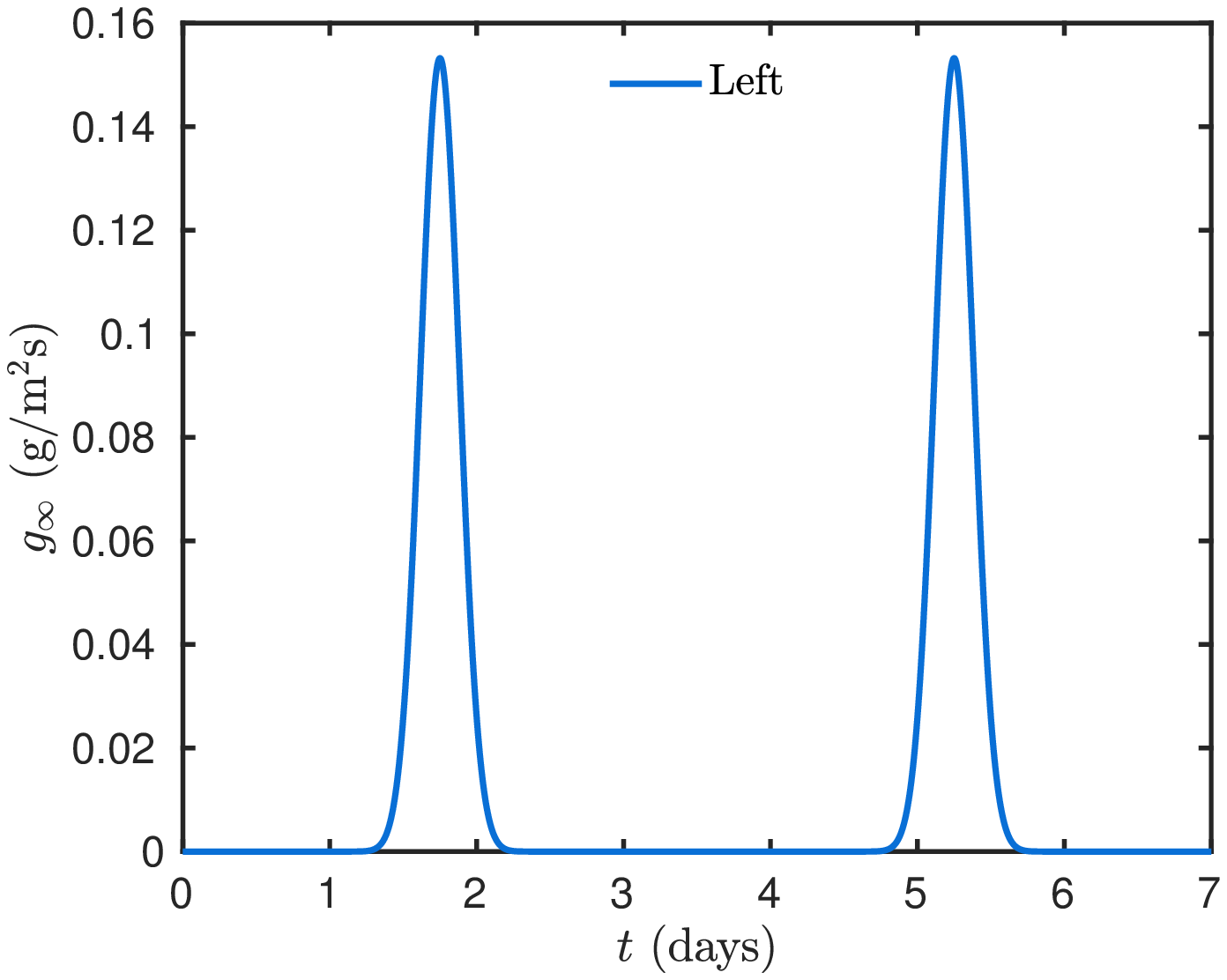}}  \hspace{0.3cm}
  \subfigure[][\label{fig_AN2:sensible_heat_rain}]{\includegraphics[width=.46\textwidth]{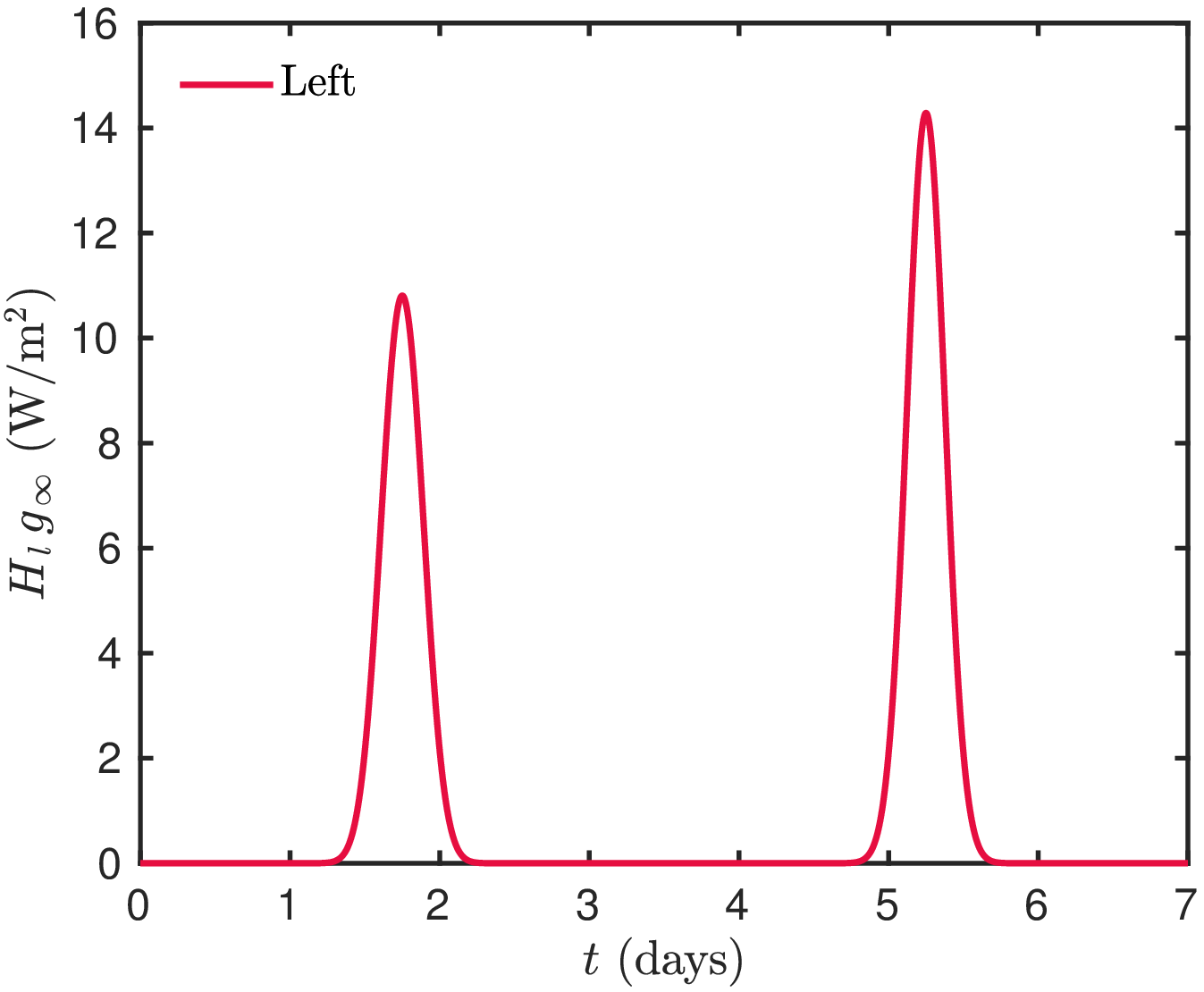}}
  \caption{\small\em Liquid flow $g_{\infty}$ at the left boundary (a) and associated sensible heat flux (b).}
\end{center}
\end{figure}

Initial conditions are considered to be uniform over the spatial domain, with the initial temperature of $\Ti \egalb 293.15\ \mathsf{K}$ and the initial vapour pressure of $\Pvi \egalb 1.16\cdot 10^{\,3}\  \mathsf{Pa}$, regarding to a relative humidity of $50\, \%\,$. The boundary conditions oscillate sinusoidally during $7$ days of simulation, which are represented in Figures~\ref{fig_AN1:BC_T} and \ref{fig_AN1:BC_RH}. The convective mass and heat transfer coefficients are set to $\hM \egalb 2 \cdot 10^{\,-7}\, \mathsf{s/m}$, $\hMR \, =\, 3 \cdot 10^{\,-8}\, \mathsf{s/m}$, $\hTL \egalb 25\, \mathsf{W/(m^2\cdot K)}$ and $\hTL \egalb 8\, \mathsf{W/(m^2\cdot K)}$. The liquid water flow (rain) has two peaks, one at $42\, \mathsf{h}$ and the other one at $126\, \mathsf{h}$, reaching a maximum value of $1.53\cdot 10^{\,-4}\ \unitfrac{kg}{(m^2\cdot s)}$ as shown in Figure~\ref{fig_AN2:Rain_flux}, which generates a sensible heat flux of $11\, \unitfrac{W}{m^2} $ and $14\, \unitfrac{W}{m^2}$ as displayed in Figure~\ref{fig_AN2:sensible_heat_rain}.

Simulations with the Spectral-ROM were performed using the \texttt{ode15s}, with a tolerance set to $\mathsf{tol} \egalb 10^{\,-\,5}\,$, with $N\egalb 8$ modes and $m\egalb 13\,$. The time is incremented with a discretization of $\Delta \ts \egalb 10^{\,-\,1}$ and the Spectral solution is project to $N_{\,x} \egalb 101$ spatial nodes. The IMEX solution is computed for a discretization parameter of $\Delta \ts \egalb 10^{\,-2}$ and $\Delta \xs \egalb 10^{\,-\,2}\,$.

\begin{figure}
  \centering
  \subfigure[a][\label{fig_AN2:Profiles_Pv}]{\includegraphics[width=0.48\textwidth]{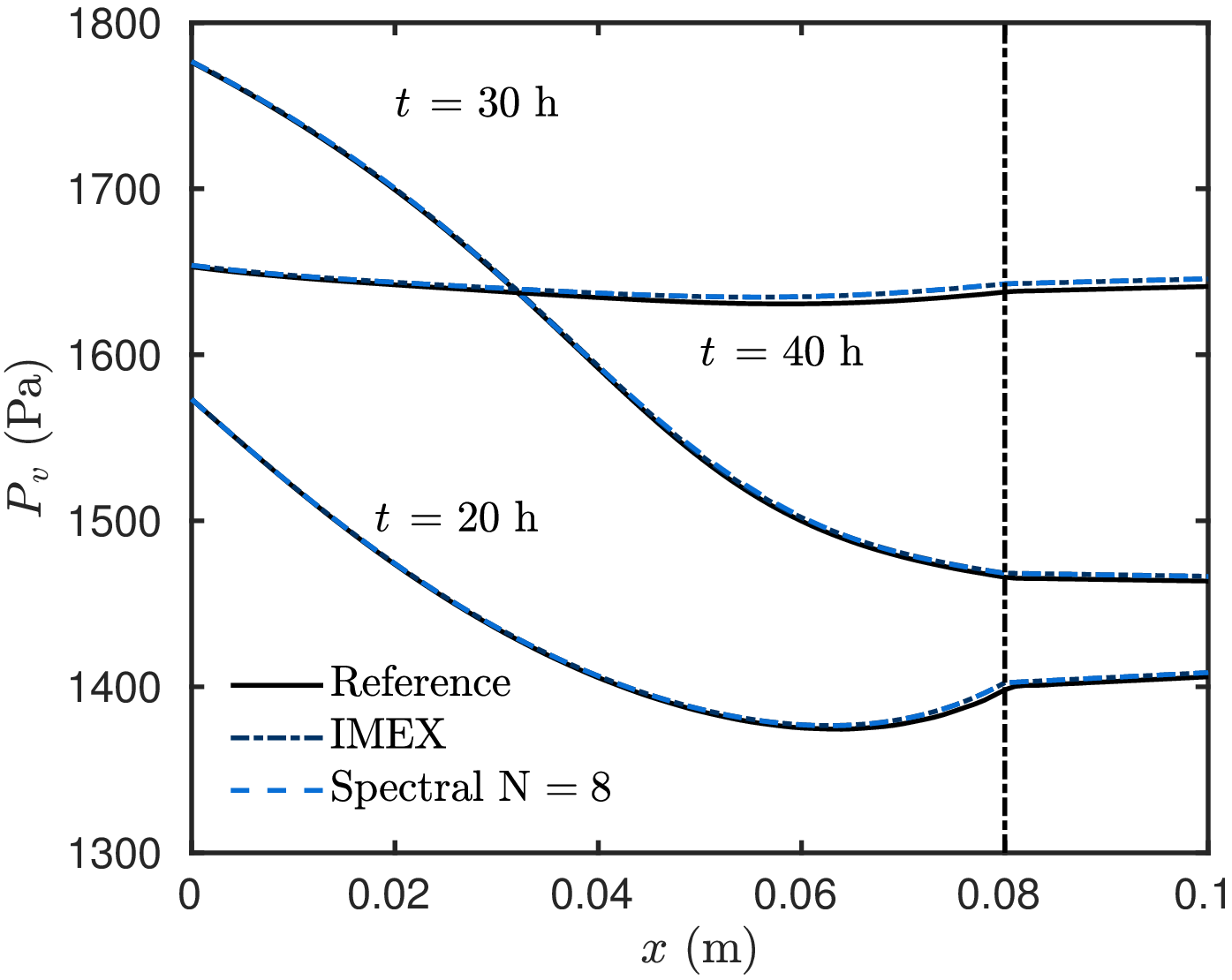}} \hspace{0.3cm}
  \subfigure[b][\label{fig_AN2:Profiles_T}]{\includegraphics[width=0.47\textwidth]{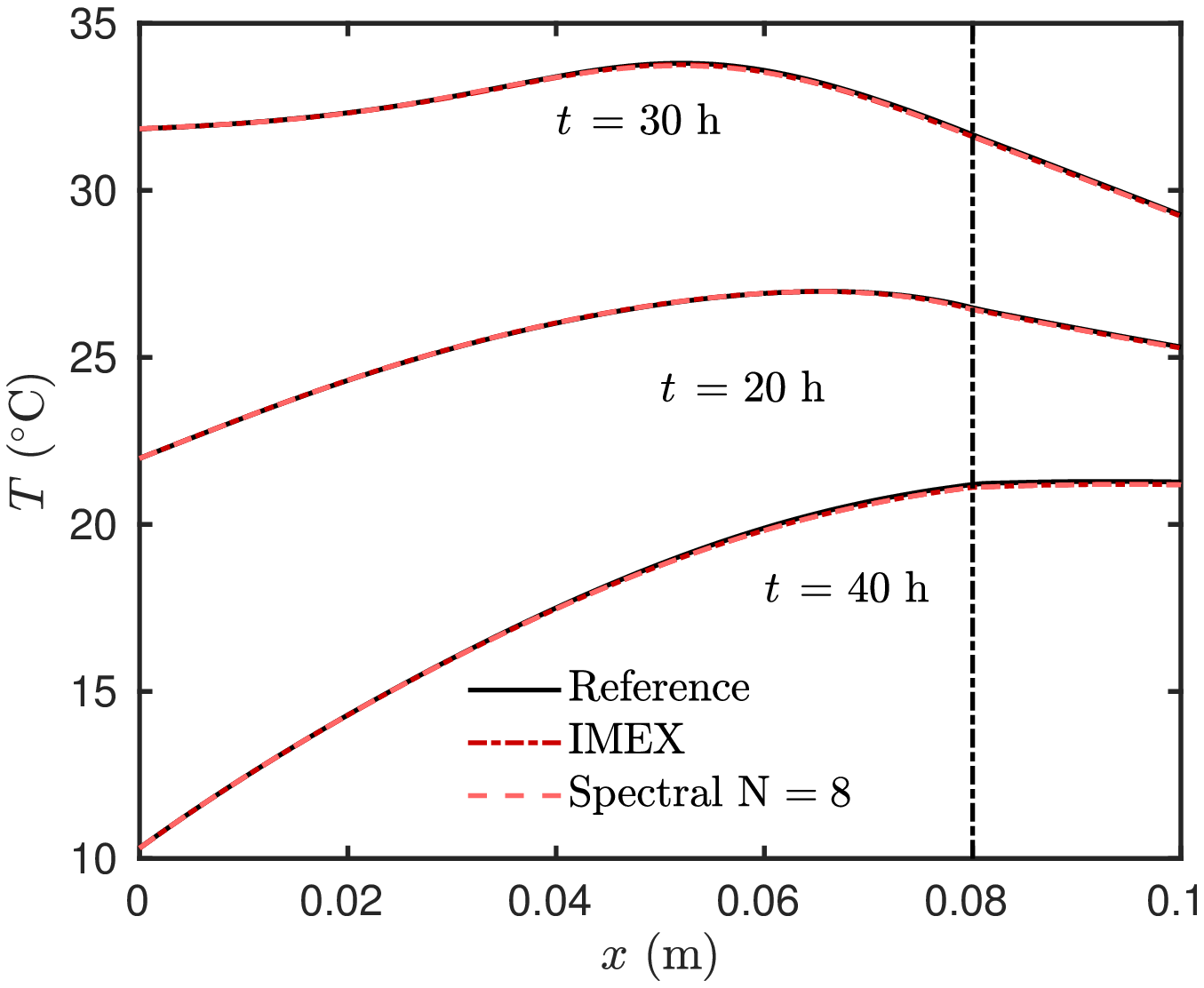}}
  \caption{\small\em Vapour pressure and (a) temperature (b) profiles at  $t \egalb \{20,\,30,\,40\}\, \mathsf{h}$.}
\end{figure}

Figure~\ref{fig_AN2:Profiles_Pv} presents three profiles of vapour pressure field, at the time instants $t \egalb \{20,\,30,\,40\}\, \mathsf{h}\,$, which are included on the interval of the first incoming raining flow. Variations of vapour pressure are more significant on the first layer because this material is more hygroscopic than the second one. At the interface, one can notice a discontinuity on the derivative. However, the continuity of the field and the one of the flow are assured. Additionally, Figure~\ref{fig_AN2:Profiles_T} presents the temperature profiles for the same time instants. Variations on the boundaries occur mainly due to the variations of $\Pvinf$ in both fields, and not only due to the rain flow, which has an opposite effect on the temperature.

\begin{figure}
  \centering
  \subfigure[a][\label{fig_AN2:Evolution_Pv}]{\includegraphics[width=0.48\textwidth]{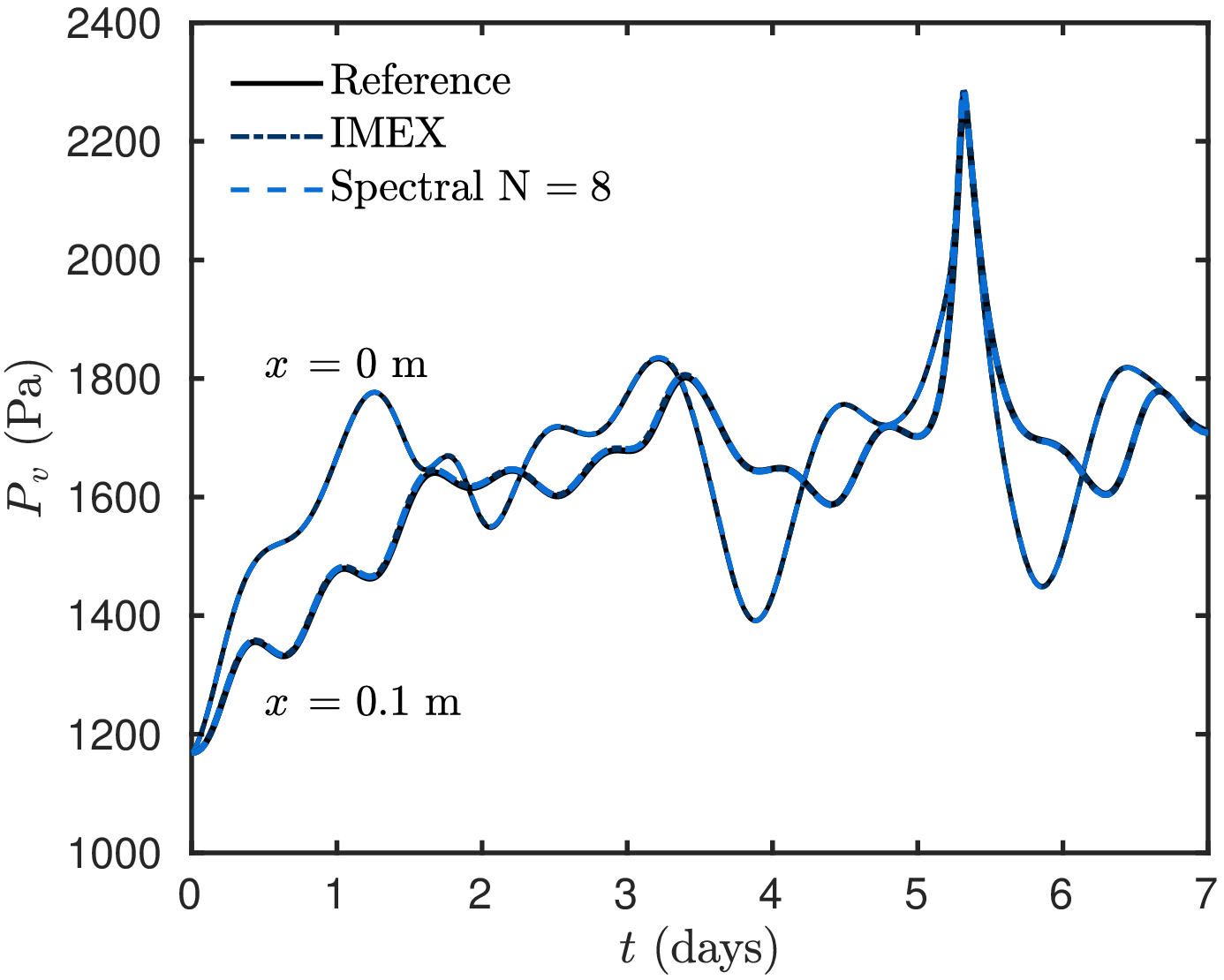}} \hspace{0.3cm}
  \subfigure[b][\label{fig_AN2:Evolution_T}]{\includegraphics[width=0.47\textwidth]{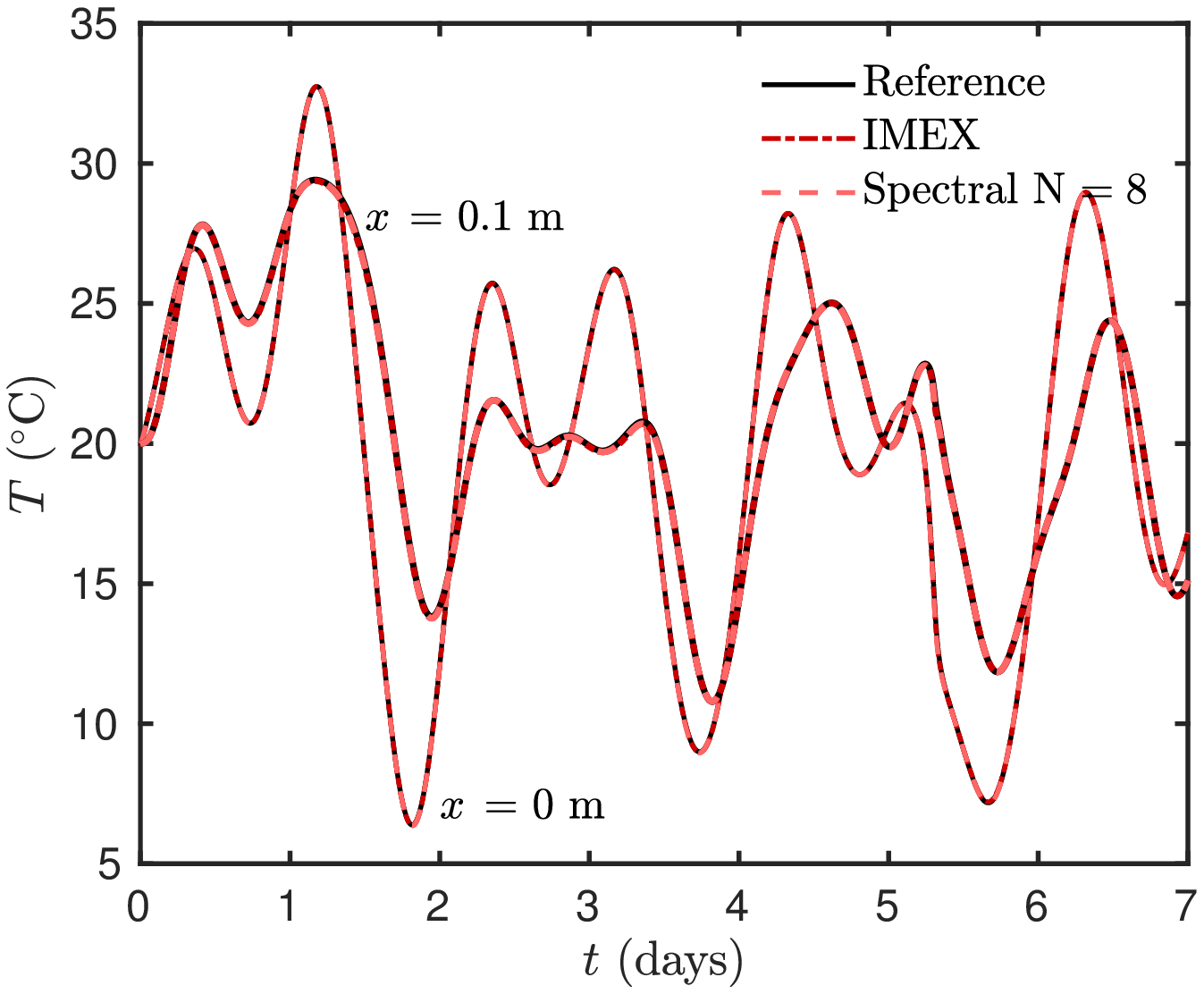}}
  \caption{\small\em Evolution of vapour pressure (a) and temperature (b) at the boundaries of the composite wall.}
\end{figure}

The evolution of the temperature and the vapour pressure at the boundary surfaces ($x \egalb 0\ \mathsf{m}$ and $x \egalb 0.1\ \mathsf{m}$) is shown in Figures~\ref{fig_AN2:Evolution_T} and \ref{fig_AN2:Evolution_Pv}, respectively. The vapour pressure varies according to the sinusoidal fluctuations of the boundary conditions until the rain hits the surface. In the first peak, it is not possible to observe high changes but in the second peak, as the material accumulates moisture, the vapour pressure is suddenly augmented and it diffuses through both layers. In fact, only when the vapour pressure is higher inside both material that is possible to observe the impact of the rain flux because of the material properties, more precisely due to the sorption isotherm. As can be observed in Figure~\ref{fig_AN2:Evolution_T}, the temperature at the boundaries of the composite wall varies according to the gradients of vapour pressure rather than the temperature gradients.

\begin{figure}
  \begin{center}
  \includegraphics[width=0.75\textwidth]{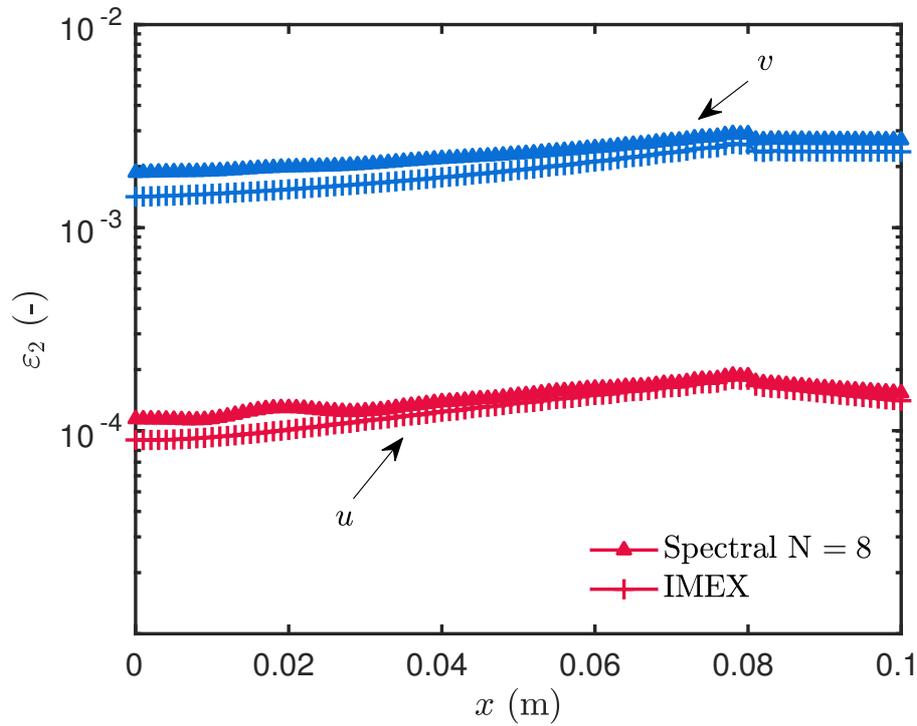}
  \caption{\small\em Error $\varepsilon_{\,2}$ given in function of $x\,$.}
  \label{fig_AN2:Error_fx}
  \end{center}
\end{figure}

By examining the profiles and evolution of temperature and vapour pressure profiles, it is possible to notice that the solutions of the Spectral and IMEX methods are both in a good agreement with the reference solution given by \texttt{Chebfun}. The distribution of the error $\varepsilon_{\,2}$ as a function of $x$ are shown in Figure~\ref{fig_AN2:Error_fx}. The $\L_{\infty}$ error of the spectral solution is $\varepsilon_{\infty}^{\,v} \egalb 2.9\cdot 10^{\,-3}$ for the vapour pressure and $\varepsilon_{\infty}^{\,u} \egalb 1.29\cdot 10^{\,-4}$ for the temperature. These values depend on the order of the time discretization, on the chosen tolerance of the solver and on the number of modes. For the IMEX solution, the $\L_{\infty}$ error is $\varepsilon_{\infty}^{\,v} \egalb 2.6\cdot 10^{\,-3}$ for the vapour pressure and $\varepsilon_{\infty}^{\,u} \egalb 1.1\cdot 10^{\,-4}$ for the temperature. To obtain the same accuracy of the solutions, the Spectral-ROM was $7$ times more efficient than the IMEX approach, as it can be seen in Table~\ref{table:CPU_time}. One should recall that the IMEX has the same accuracy of the \textsc{Euler} explicit and is more efficient, in terms of computer run time, than the largely used \textsc{Crank--Nicolson} scheme. Just for curiosity, the computational time of the \texttt{Chebfun} simulations have been added. As one can notice they are much higher than the values obtained by the other methods. In fact, \texttt{Chebfun} is made for numerical computing of a  wide range of problems. As it does not know about the specific problem, it does not take advantage of the problem structure. Anyway, it is a good tool for comparing purposes.

\begin{table}
\center
\small
\caption{\small\em Computer run time required for the numerical schemes to perform simulations.}
\bigskip
\setlength{\extrarowheight}{.3em}
\begin{tabular}[l]{@{}ccccc}
\hline
\textit{Case study} & \textit{Spectral} & \textit{IMEX} & \textit{Chebfun}   &\textit{Ratio (IMEX/Spectral)} \\
\hline
$1$ layer   &  $0.9\, \mathsf{s}$   & $6.4\, \mathsf{s}$ &  $46.6\, \mathsf{s}$ & $\mathbf{7.1}$\\
$2$ layers  &  $1.8\, \mathsf{s}$   & $12.5\, \mathsf{s}$ &  $227.9\, \mathsf{s}$ &$\mathbf{6.9}$\\
\hline
\end{tabular}
\label{table:CPU_time}
\end{table}

\begin{figure}
\centering
\subfigure[a][\label{fig_AN2:coef_spec_mat1}]{\includegraphics[width=0.48\textwidth]{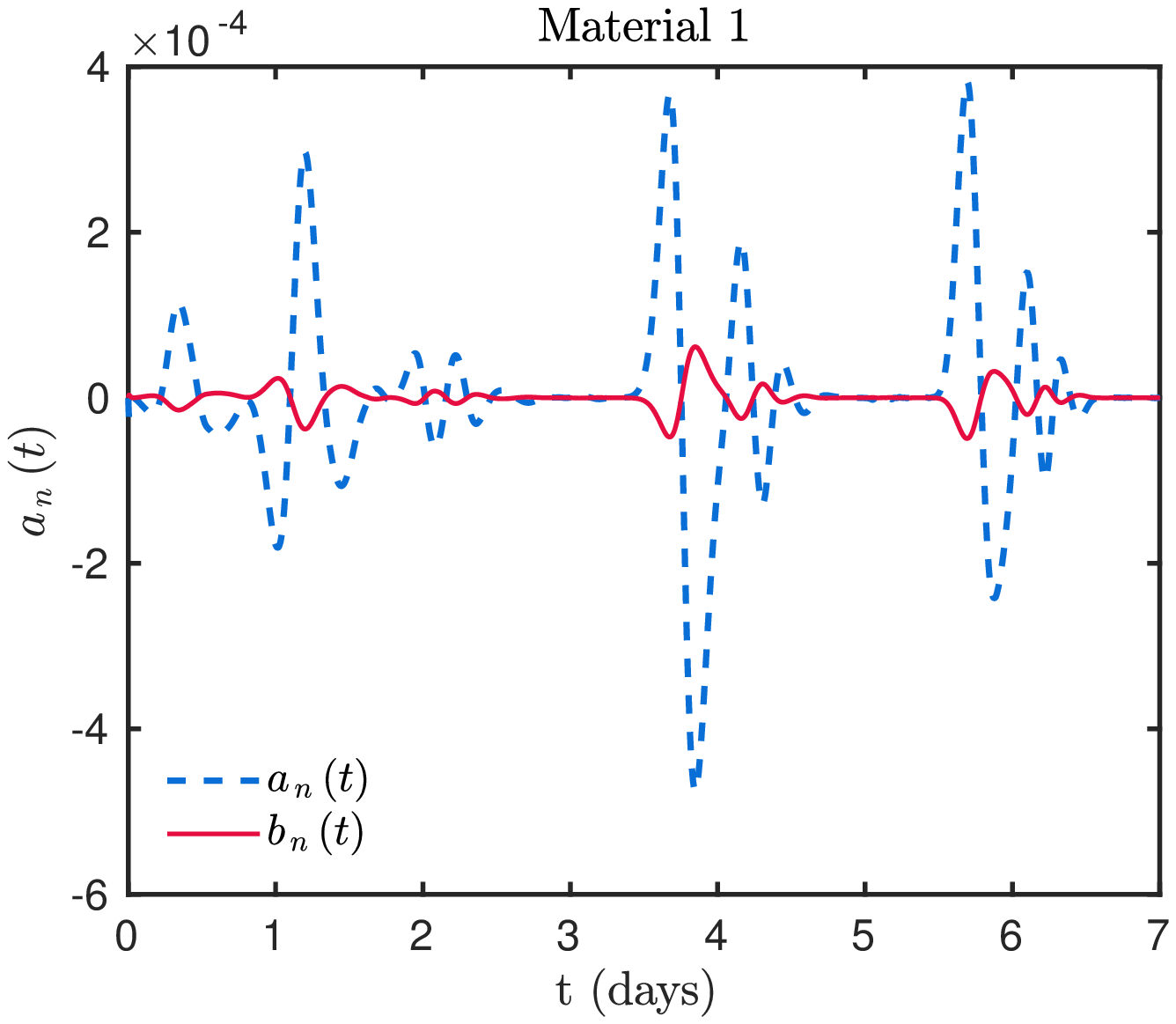}} \hspace{0.3cm}
\subfigure[b][\label{fig_AN2:coef_spec_mat2}]{\includegraphics[width=0.48\textwidth]{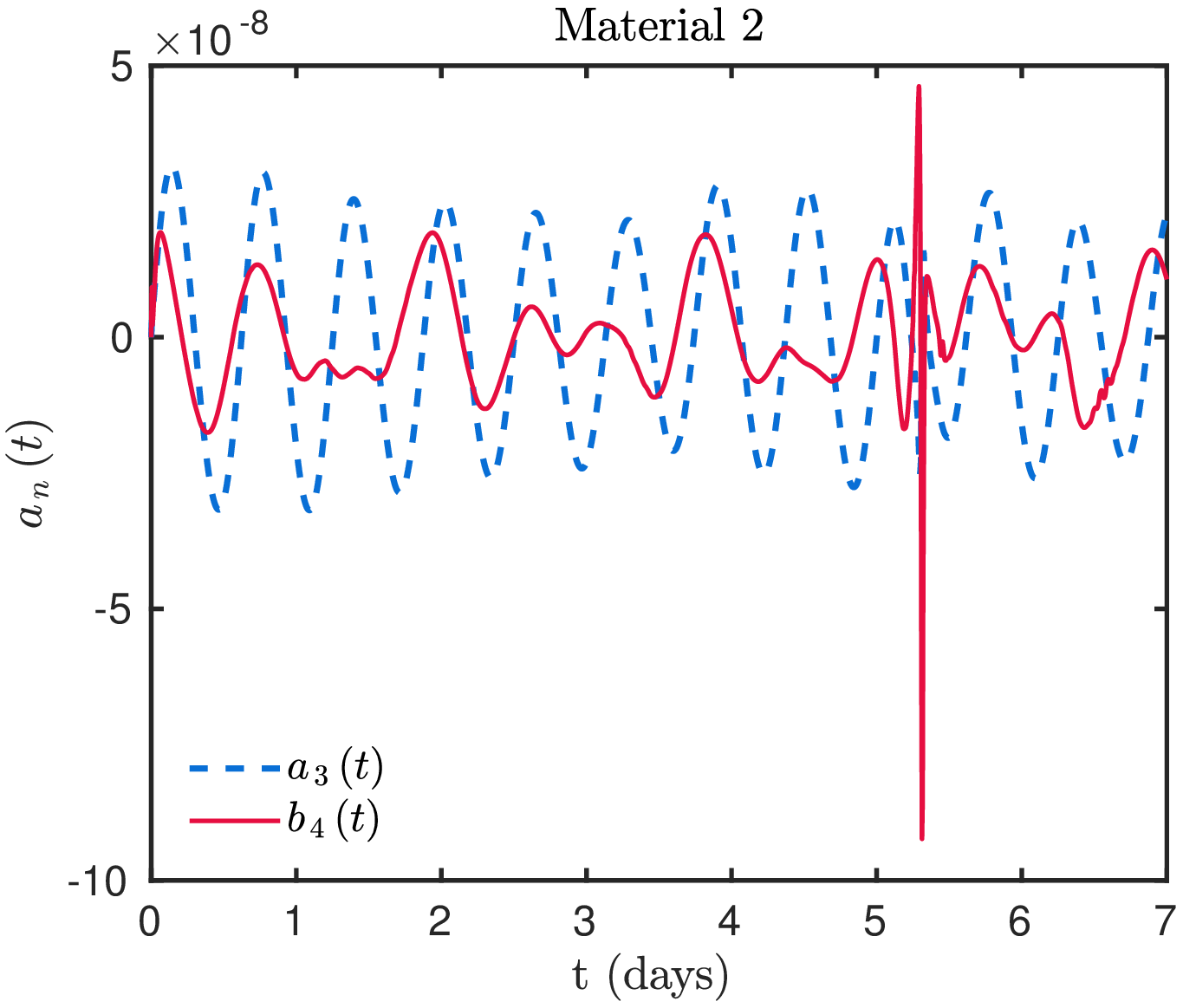}}
\caption{\small\em Spectral coefficients of material 1 (a) and of material 2 (b).}
\end{figure}

Figure~\ref{fig_AN2:coef_spec_mat1} presents the last spectral coefficients of $u$ and $v$ solutions for material $1$, while Figure~\ref{fig_AN2:coef_spec_mat2} presents the fourth and fifth spectral coefficients for material $2$. The magnitude of the last spectral coefficient acts as an error estimator, determining the error upper limit, as previous mentioned. In material $1$, spectral coefficients needed more modes to have an acceptable solution, while in the other material, an accurate solution can be built with a lower number of modes. This difference occurs due to the properties of material $1$ that are more nonlinear than those of material $2$. One may notice in Figure~\ref{fig_AN2:coef_spec_mat2} a peak after day $5$, which disappears shortly after. This may occur due to aliasing errors. However, the magnitude of this error is very small if compared with the magnitude of the solution and it does not occur on the other spectral coefficients. As the problem demands only a few modes (around $10$), this kind of error does not really affect the final solution.

\begin{figure}
\centering
\includegraphics[width=0.75\textwidth]{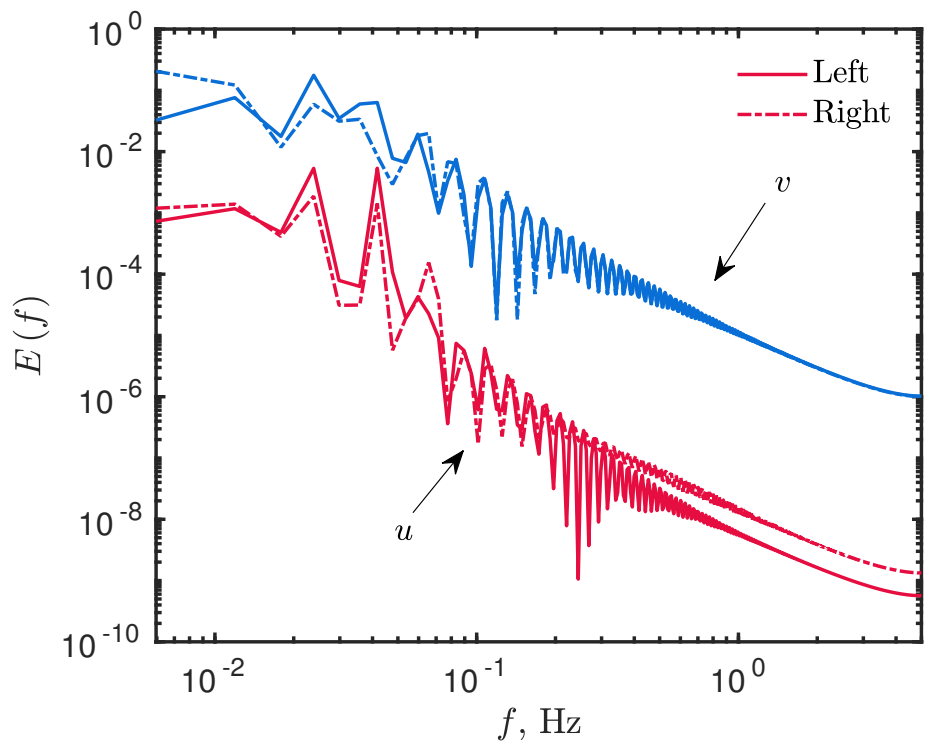} 
\caption{\small\em \textsc{Fourier} power spectrum of the Spectral solution computed in the right and left boundaries of the domain.}
\label{fig_AN2:FPS}
\end{figure}

In addition, Figure~\ref{fig_AN2:FPS} displays the \textsc{Fourier} power spectrum function of the signal frequency per unit of time, on the left and right boundaries, for $u$ and $v$ solutions. Oscillations occurring for $u\,$, from $2\cdot 10^{\,-\,1}\, \mathsf{Hz}$, are attributed to aliasing errors. However, the power of this frequency is very low if compared with the highest peak, corresponding to a difference of $6$ orders of magnitude.

\begin{figure}
\centering
\subfigure[a][\label{fig_AN2:flux}]{\includegraphics[width=0.48\textwidth]{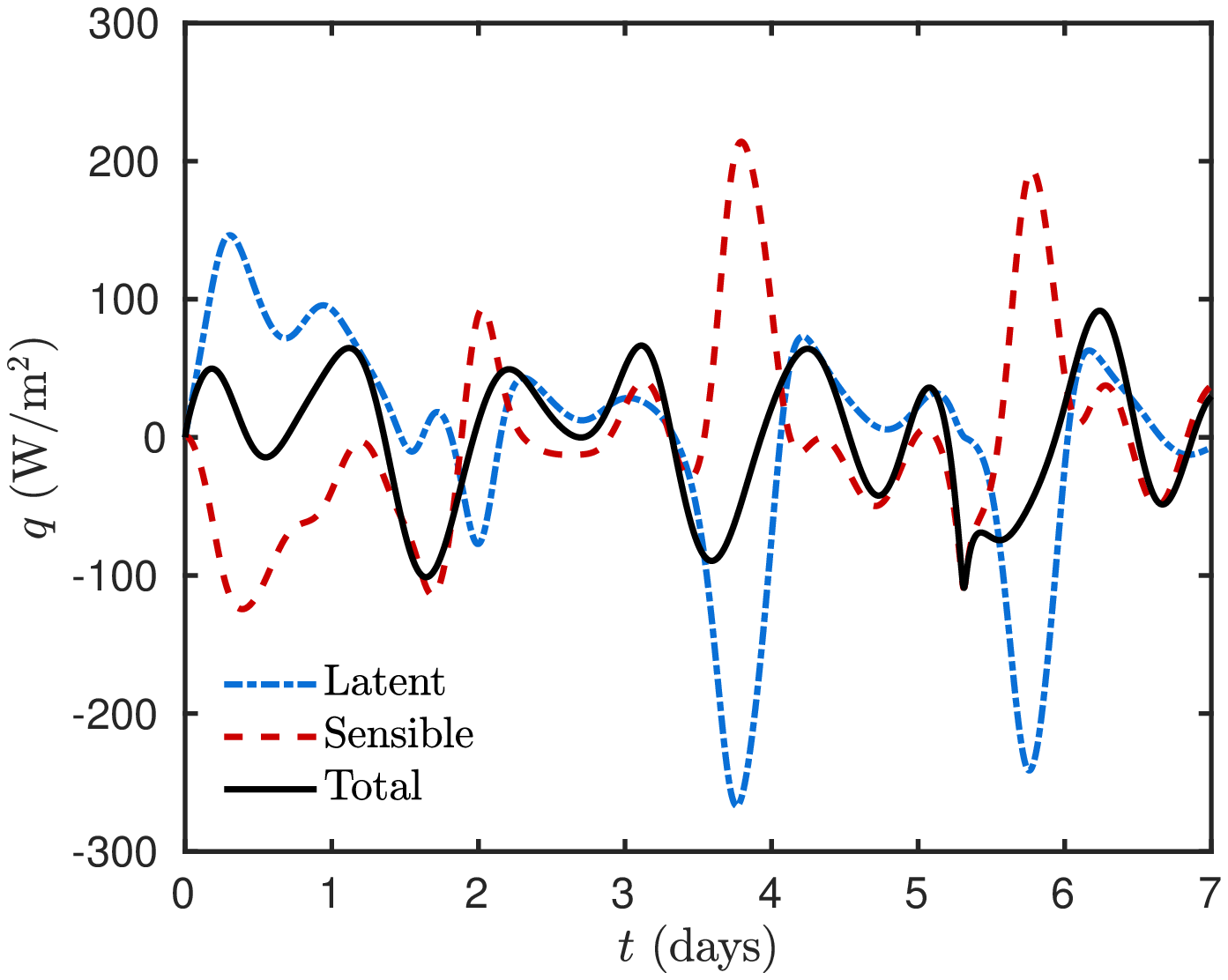}} \hspace{0.3cm}
\subfigure[b][\label{fig_AN2:flow}]{\includegraphics[width=0.48\textwidth]{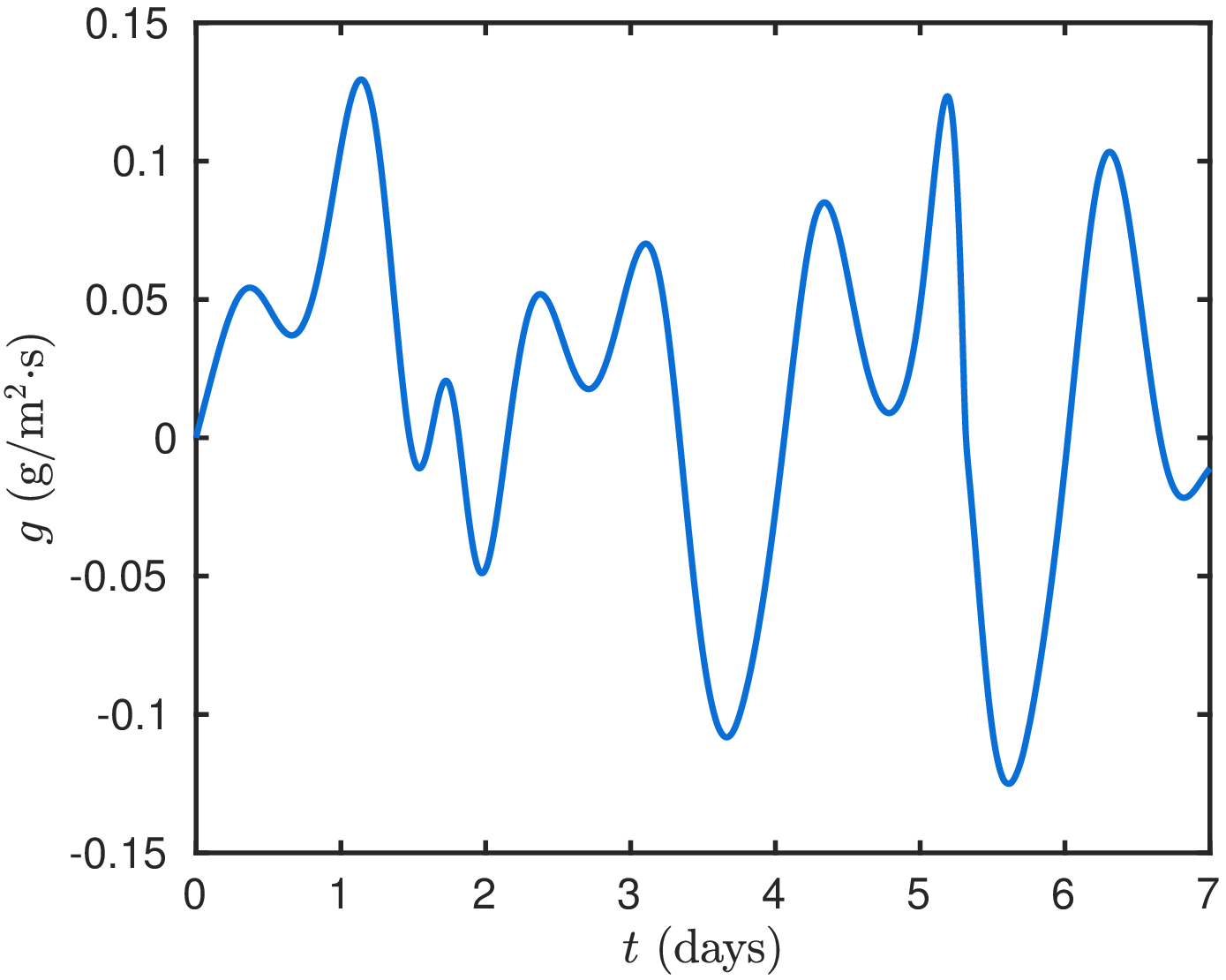}} 
\caption{\small\em Latent, sensible and total heat fluxes (a) and moisture flow (b) at the left boundary $(x\,=\,0)\,$.}
\end{figure}

The results for the sensible, latent and total heat fluxes at the left boundary $x \egalb 0\ \mathsf{m}$ are given graphically in Figure~\ref{fig_AN2:flux}, the boundary that receives the rain flow. The sensible and latent heat fluxes have high values but with opposite signs. Although, as they do not have the same value, they do not cancel each other, as shown for the total heat flux, which is the sum of the latent and sensible heat fluxes. Furthermore, Figure~\ref{fig_AN2:flow} presents the total moisture flow at the same boundary. The rain flow is observed in the two maximal peaks, one right after day $1$ and the other right after the day $5\,$. The other variations are caused by the boundary conditions.

\section{Validation of the model}
\label{sec:validation}

In this section, the physical model and the spectral solution are compared with experimental data gathered from the \textsc{French} project HYGRO-BAT \cite{HYGRO-BA2014}. The measurements were performed at the French laboratory LOCIE (Laboratory of Optimisation of the Conception and Engineering of the Environment) \cite{Rouchier2016}. One-dimensional coupled heat and moisture transfer through a single-layered wall is monitored by sensors placed inside of the material and on its surfaces. In this case, they have not considered the liquid transfer in the moisture balance equation -- Eq. \eqref{eq:M_equation}. Surface sensors provide boundary conditions for the coupled simulation, while the other sensors provide reference measurements for the model validation.

The relative error $\epsilon$ is computed to compare simulations with the experimental data and it is defined as: 
\begin{align*}
  \epsilon\, (\,t\,)\ &\eqdef\  \dfrac{\sqrt{\Bigl(Y_{\, k}^{\, \mathrm{num}}\, (\, x_{\,k}\,, t\,) \moins Y_{\, k}^{\mathrm{\, meas}}\, (\, x_{\,k}\,, t\,)\Bigr)^{2}}}{Y_{\, k}^{\mathrm{\, meas}}(\, x_{\,k}\,, t\,) }  \,, 
\end{align*}
where $Y_{\, k}^{\, \mathrm{num}}$ is the computed solution and $Y_{\, k}^{\mathrm{\, meas}}$ is the measured data.


\subsection{Experimental setup}

The wall considered in this study is composed of a $16-\mathsf{cm}$ layer of wood fibre material, which is subjected to variations in terms of temperature and relative humidity for a $14-$days period. The material properties are given in Table~\ref{table:properties_woodfibre}. The reference data for the evaluation is provided by temperature and humidity sensors \texttt{SHT75 Sensirion}, located at $x\egalb \{4,\,8,\,12\}\, \mathsf{cm}$ within the wall.

The sensors have a measurement uncertainty of $\pm\, 0.3 \gC$ for temperature and of $\pm\, 0.018$ for relative humidity. Furthermore, the uncertainty regarding the position of the sensors is of $\pm\, 1\, \mathsf{cm}$ for the sensors located at $ x \egalb \{ 4, 12 \}\,\mathsf{cm}$ and of $\pm\, 0.5\, \mathsf{cm}$ for the other sensor. The uncertainty on the positions are different because the sensor at $ x \egalb \{ 4, 12 \}\,\mathsf{cm}$ have been settled by perforating a whole in the material layer.

\begin{figure}
\centering
\subfigure[a][\label{fig_AN3:BC_T}]{\includegraphics[width=0.48\textwidth]{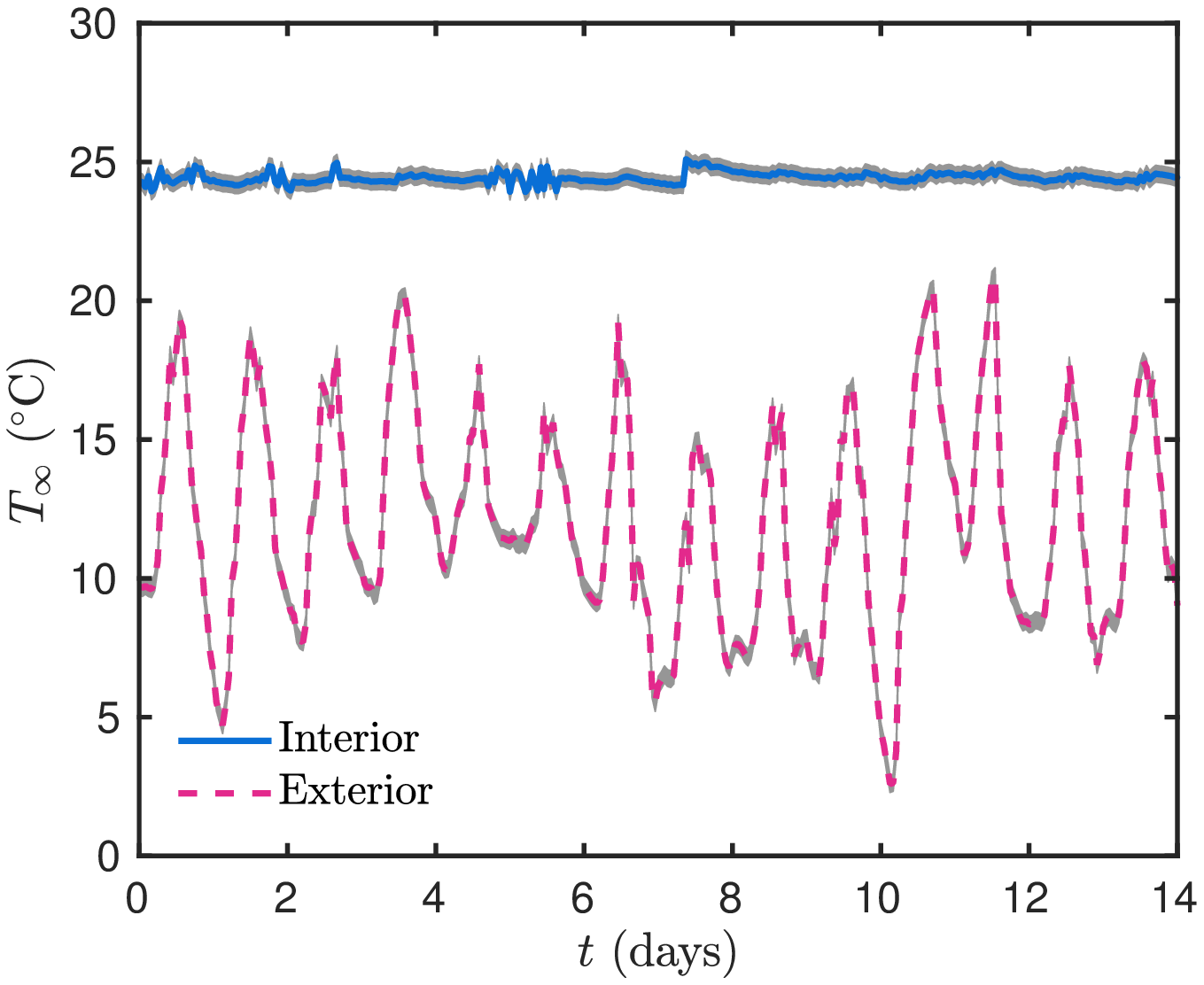}} \hspace{0.3cm}
\subfigure[b][\label{fig_AN3:BC_RH}]{\includegraphics[width=0.48\textwidth]{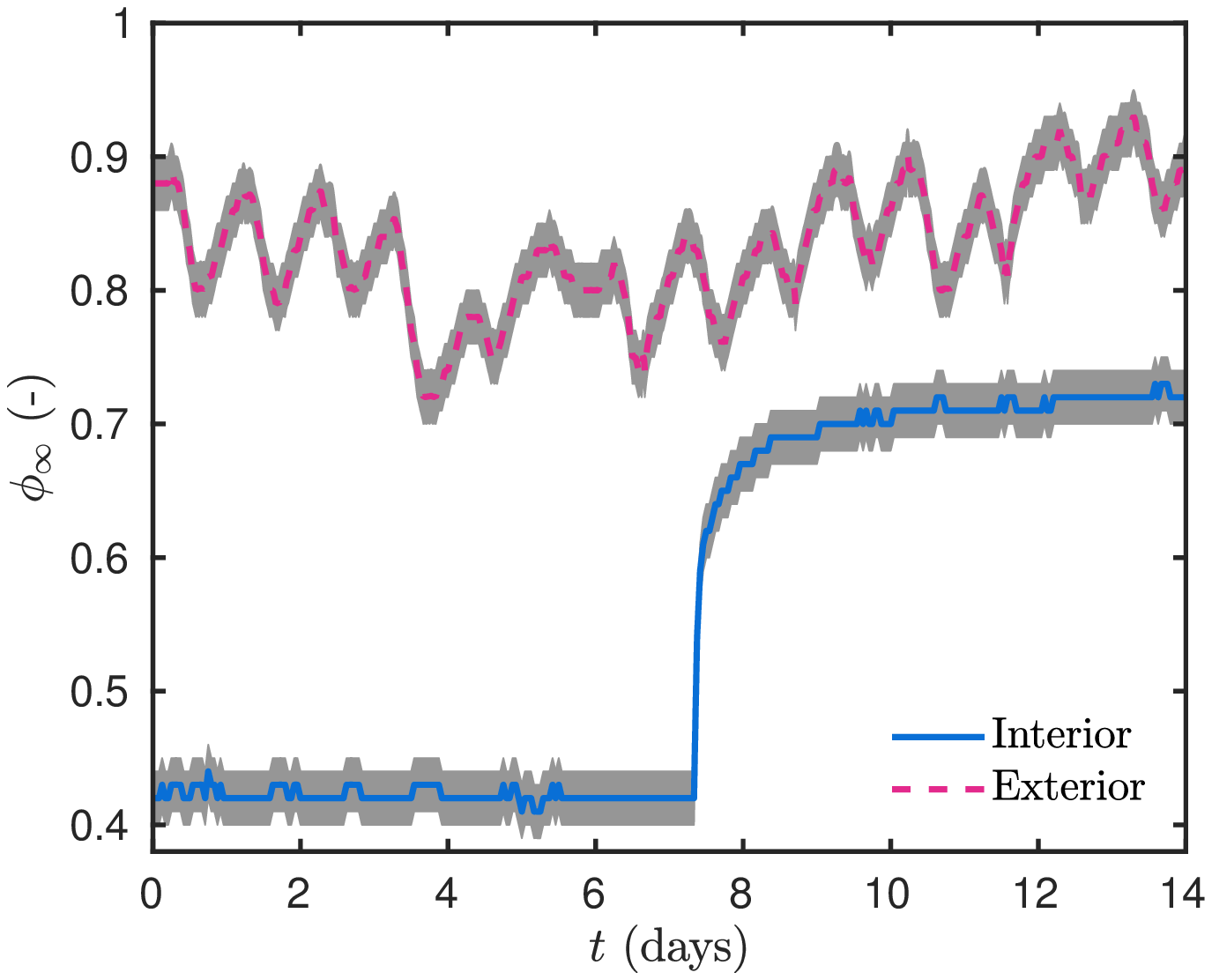}}
\caption{\small\em Ambient temperature (a) and relative humidity (b) of the experimental investigation.}
\end{figure}

The measured temperature and relative humidity at the interior and exterior boundaries are given in Figures~\ref{fig_AN3:BC_T} and \ref{fig_AN3:BC_RH}. The gray color around the curves represent the uncertainties related to the measurements. At the interior boundary, the temperature is set to approximately $24\gC$ and the relative humidity set to $40\%$ in the first week and to $70\%$ in the second week. The exterior temperature and relative humidity values are given by their measurement at the boundary, which is filtered by a $2\,\mathsf{cm}$ of a coating layer, excluding solar radiation and driven rain phenomenon. Thus, both boundaries are expressed as \textsc{Dirichlet}-type conditions for the model validation.

\subsection{Simulation}
\label{sec:validation_simulation}

Simulations are performed for a single-layer wall of a material that separates two well defined environments. The initial condition for temperature and vapour pressure fields are given by an interpolation of the measurements at $x \egalb \{0,\,4,\,8,\,12,\,16\}\, \mathsf{cm}$ at the first time instant.

The Spectral method is composed of $N \, = \, 8$ modes with $m \, = \, 13\,$. The \texttt{ODE15s} was used to solve numerically System~\eqref{eq:system_DAE}, with a tolerance set to $10^{\,-\,5}\,$. The solution has been computed with a time step of $\Delta \ts \,=\, 0.1\,$, the equivalent of $6\,\mathsf{min}$.

\begin{figure}
\centering
\subfigure[a][\label{fig_AN3:T4}]{\includegraphics[width=0.48\textwidth]{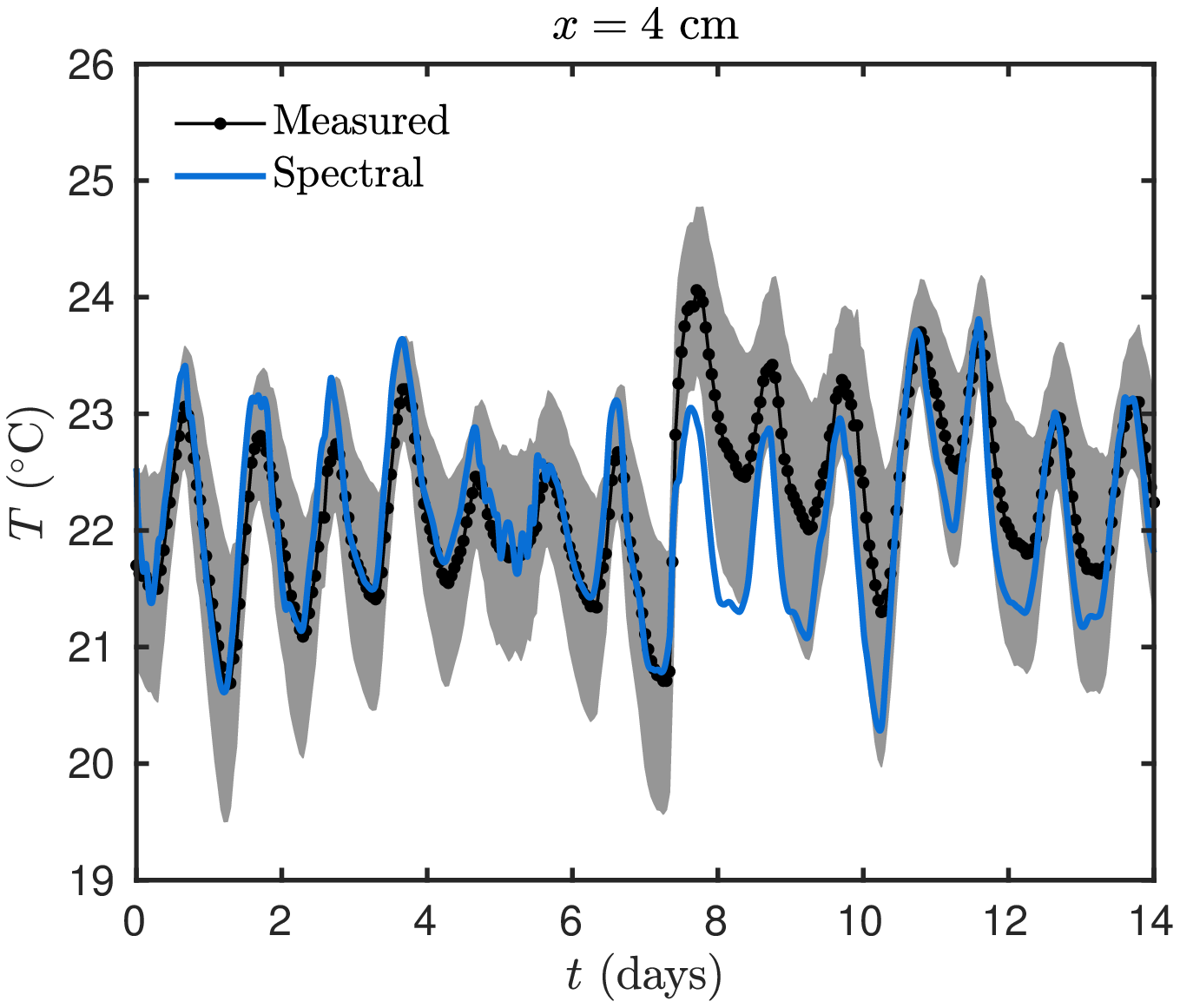}} \hspace{0.3cm}
\subfigure[b][\label{fig_AN3:RH4}]{\includegraphics[width=0.485\textwidth]{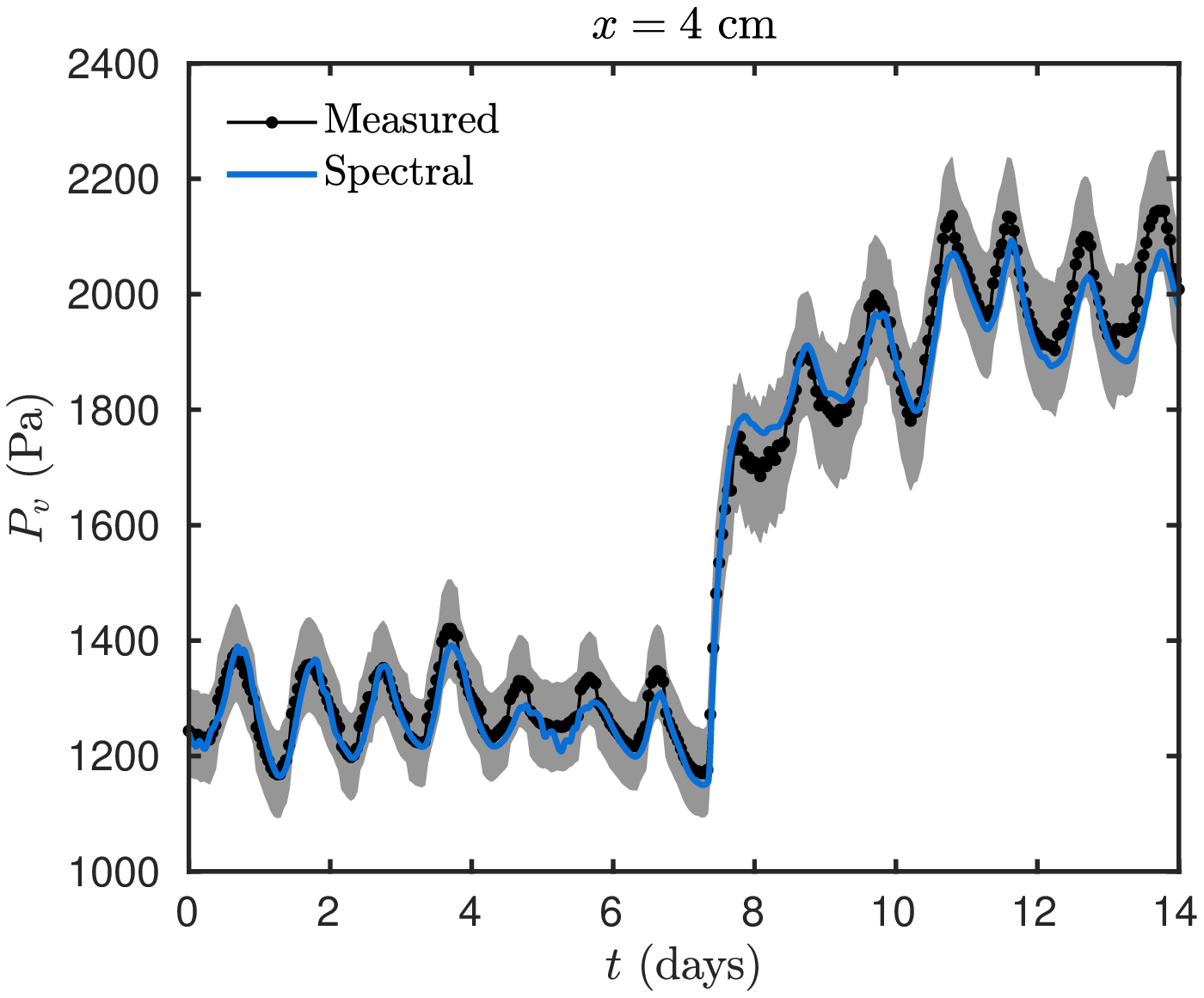}} \\
\subfigure[c][\label{fig_AN3:T8}]{\includegraphics[width=0.48\textwidth]{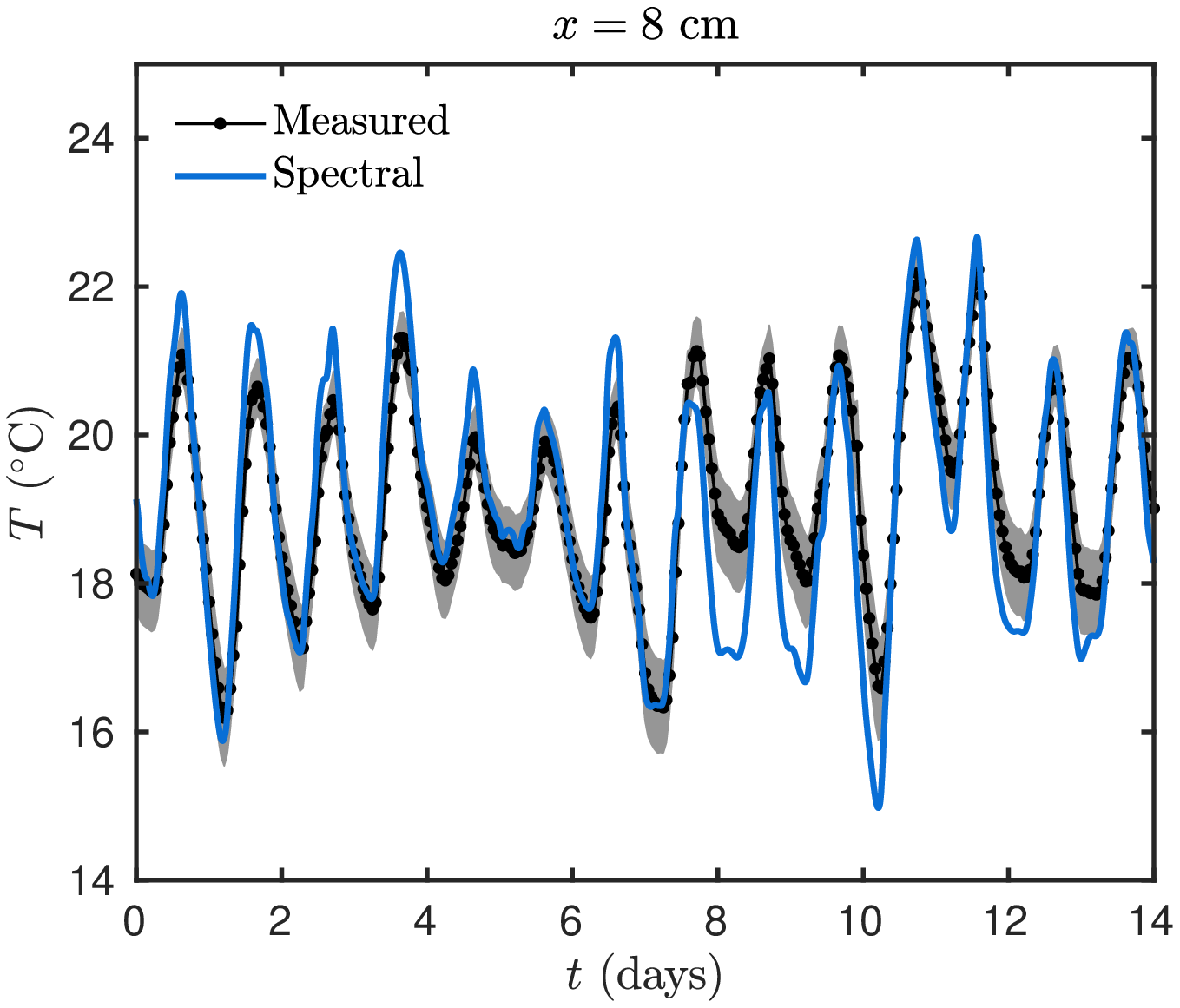}} \hspace{0.3cm}
\subfigure[d][\label{fig_AN3:RH8}]{\includegraphics[width=0.485\textwidth]{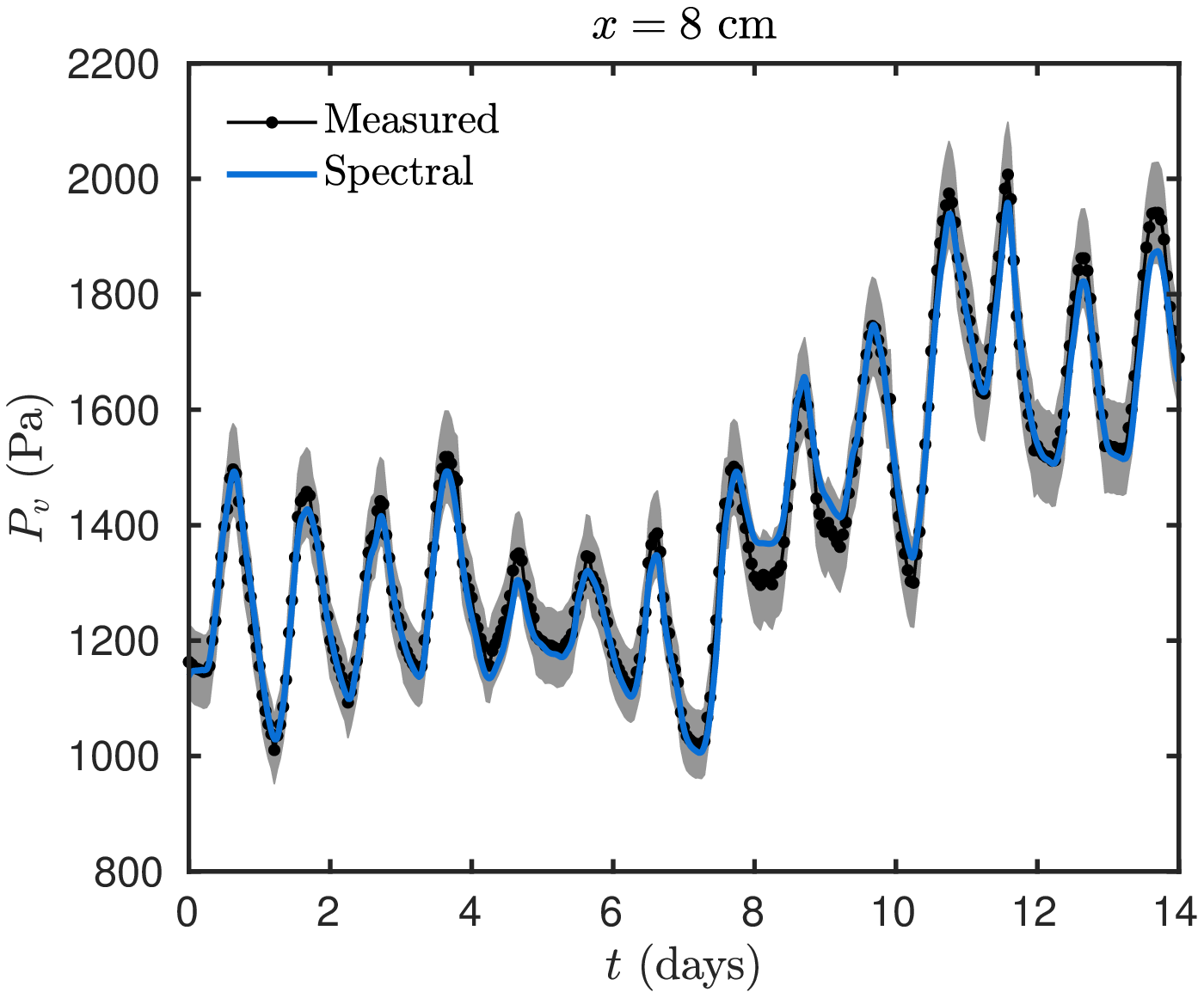}} \\
\subfigure[e][\label{fig_AN3:T12}]{\includegraphics[width=0.48\textwidth]{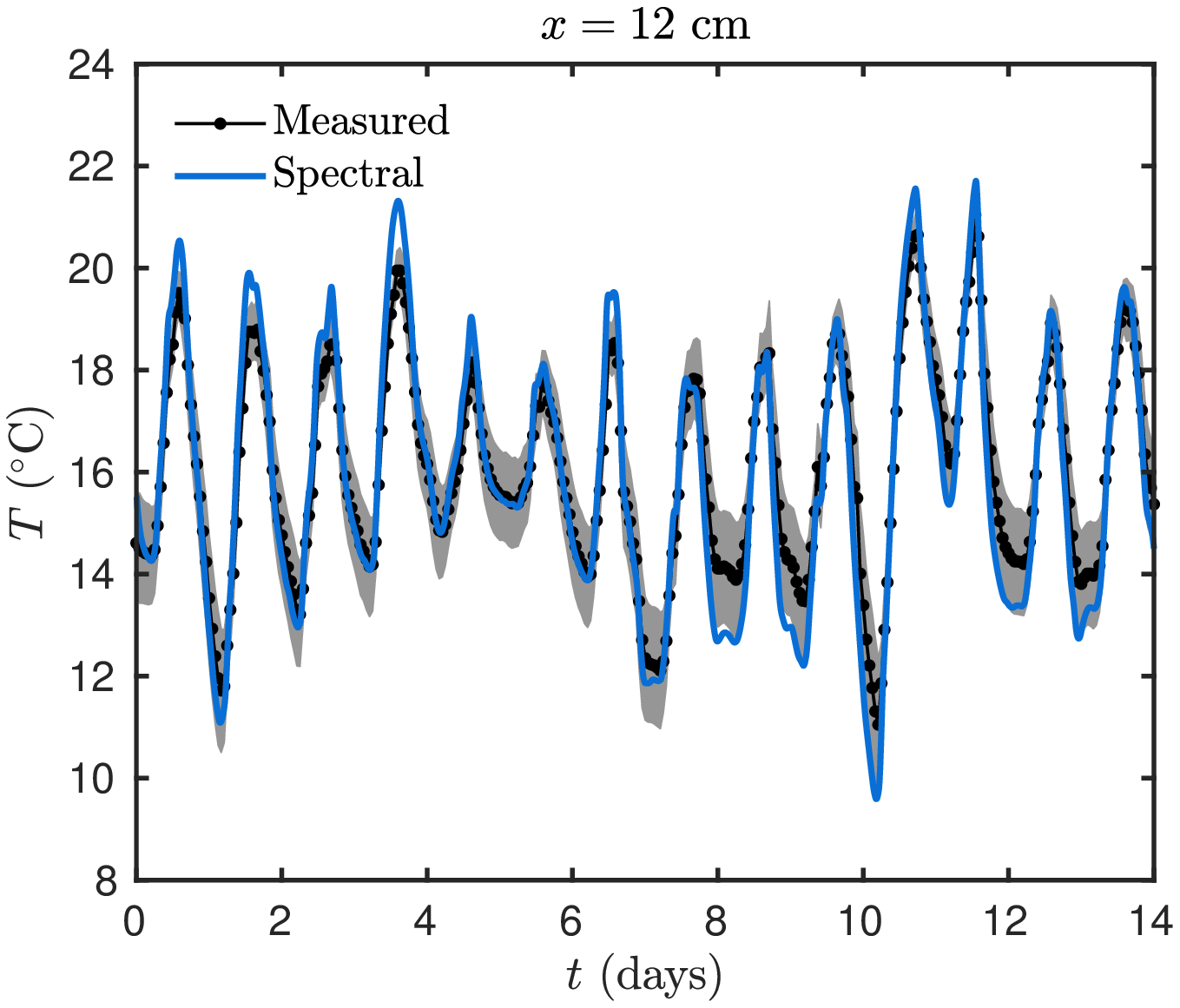}} \hspace{0.3cm}
\subfigure[f][\label{fig_AN3:RH12}]{\includegraphics[width=0.485\textwidth]{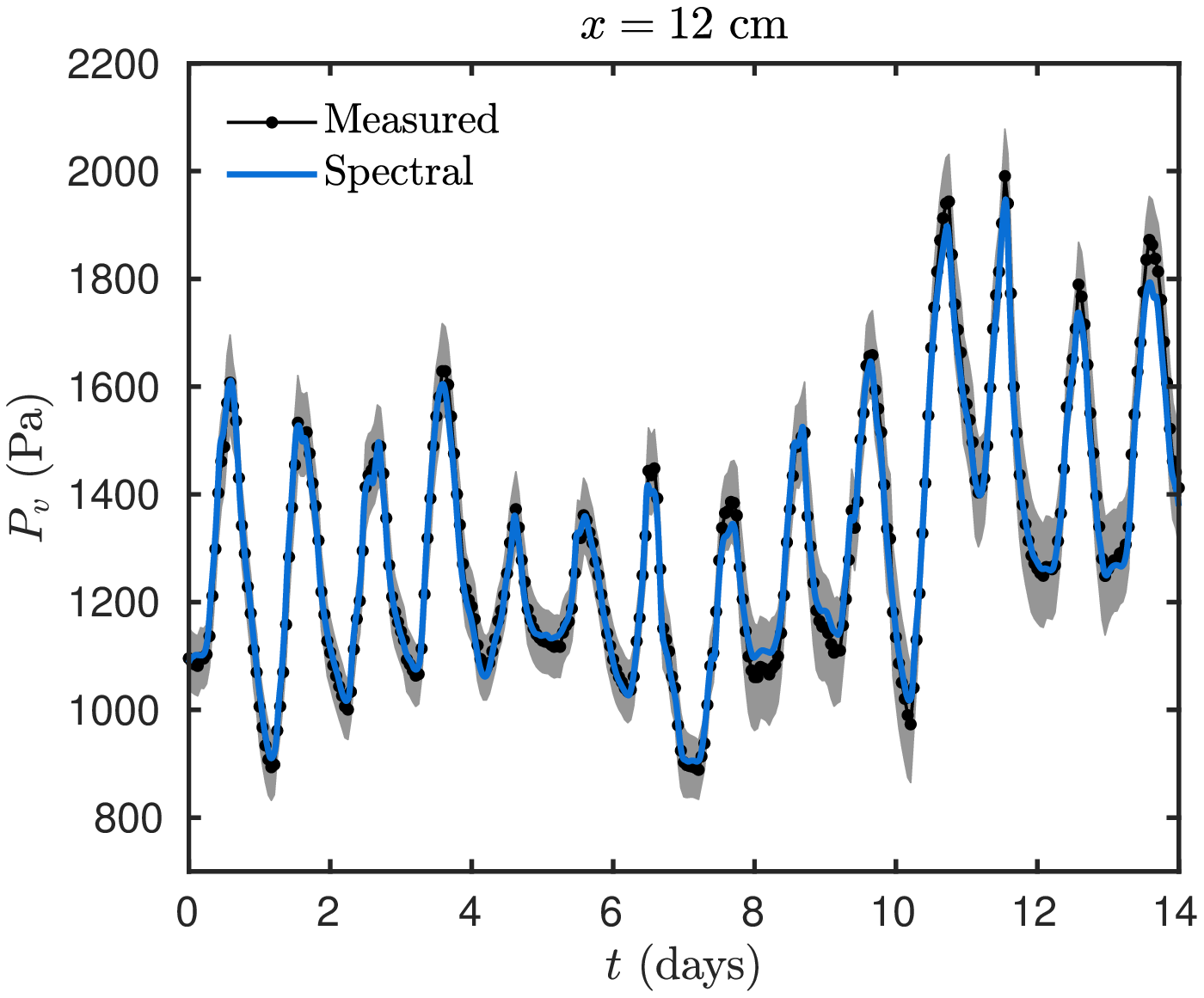}}
\caption{\small\em Measured and predicted temperature and vapour pressure values at $x \egalb \{4,\,8,\,12\}\, \mathsf{cm}$.}
\label{fig_AN3:evolution}
\end{figure}

Simulations are compared with the dynamic temperature and vapour pressure. Figure~\ref{fig_AN3:evolution} presents the measured data and the simulation results in each one location of the sensors, at $x \egalb \{4,\,8,\,12\}\, \mathsf{cm}$.

The error between the predicted solution and the experimental observations is between the uncertainties of the sensors during almost all simulation period. For the temperature evolution, the highest discrepancy is when the moisture at the left boundary changes from $40\%$ to $70\%$ of relative humidity. From the $7$\textsuperscript{th} day, the measured temperature rises more than the calculated one as shown in Figure~\ref{fig_AN3:T4}, which are higher than the uncertainties of the sensors. The difference between the simulated and measured temperature reaches a maximum of $2\gC$ in this period. The temperature simulated in at $x \egalb \{8,\,12\}\, \mathsf{cm}$ are presented in Figures~\ref{fig_AN3:T8} and \ref{fig_AN3:T12}. The predicted values are closer to the measurements than in $x \egalb 4\, \mathsf{cm}$, and they follow considerably well variations of the outside boundary. It seems that the total diffusion coefficient of the temperature used for the simulation is higher than the real one.

The influence of the step on the relative humidity can strongly be observed at $x\egalb 4\, \mathsf{cm}$ in Figure~\ref{fig_AN3:RH4} and less far from this boundary. Simulations are able to follow the variations of the measured vapour pressure better than the temperature. In Figures~\ref{fig_AN3:RH8} and \ref{fig_AN3:RH12}, the difference between simulations and measurements become more important, with the incoming moisture flow, reaching a maximum difference of $70\, \mathsf{Pa}$ and $80\, \mathsf{Pa}$, respectively. The absolute difference is higher for $x\egal 4\, \mathsf{cm}$ as it is closer to the left boundary and consequently to the incoming flow. The discrepancies, come from the physical model, which does not consider liquid transport neither hysteresis. Another explanation is the fact that the properties also were estimated with and admissible error, which influences the predicted solution. This difference on the vapour pressure simulations impacts on the predictions of temperature.

\begin{figure}
\centering
\subfigure[a][\label{fig_AN3:error_T}]{\includegraphics[width=0.485\textwidth]{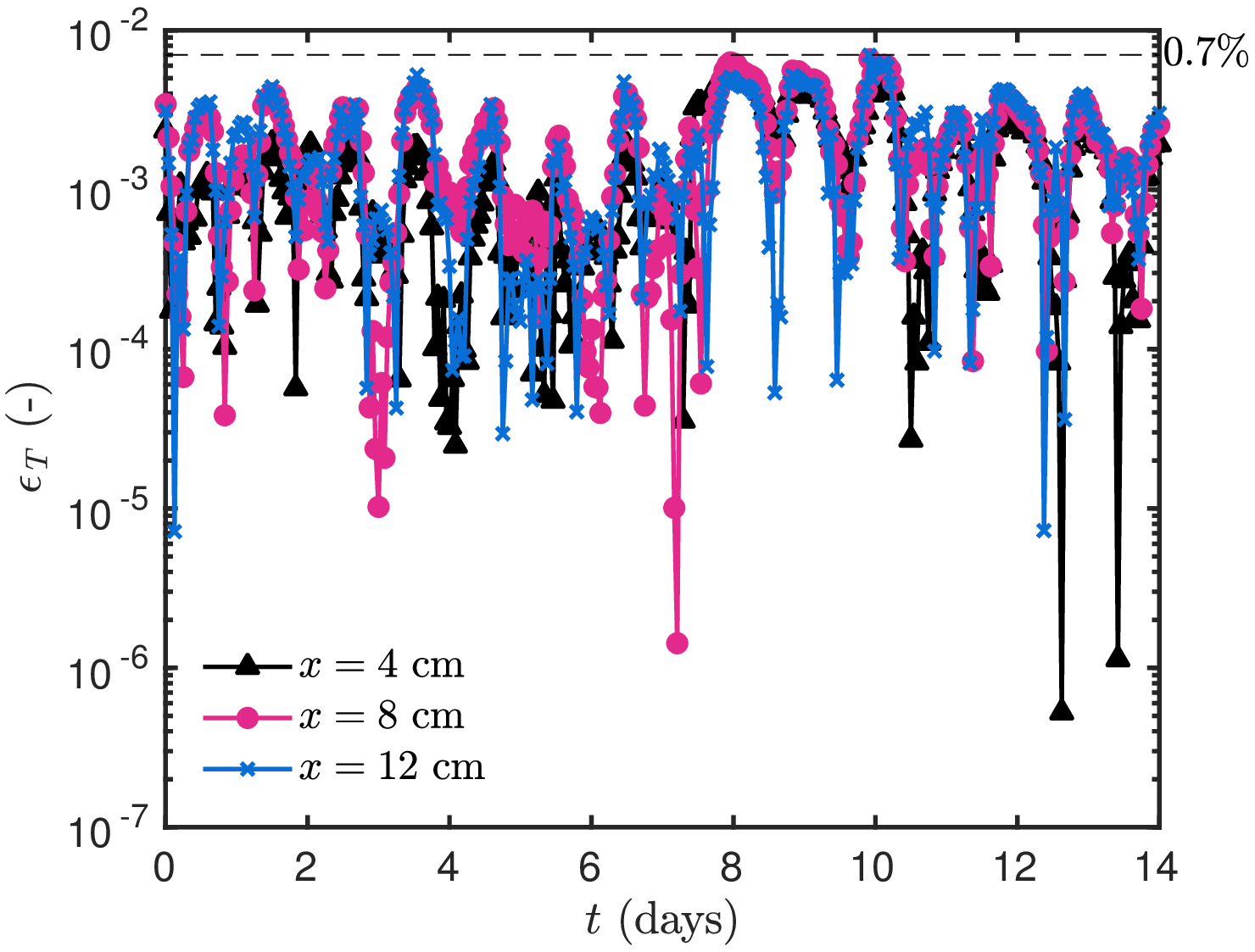}} \hspace{0.2cm}
\subfigure[b][\label{fig_AN3:error_Pv}]{\includegraphics[width=0.485\textwidth]{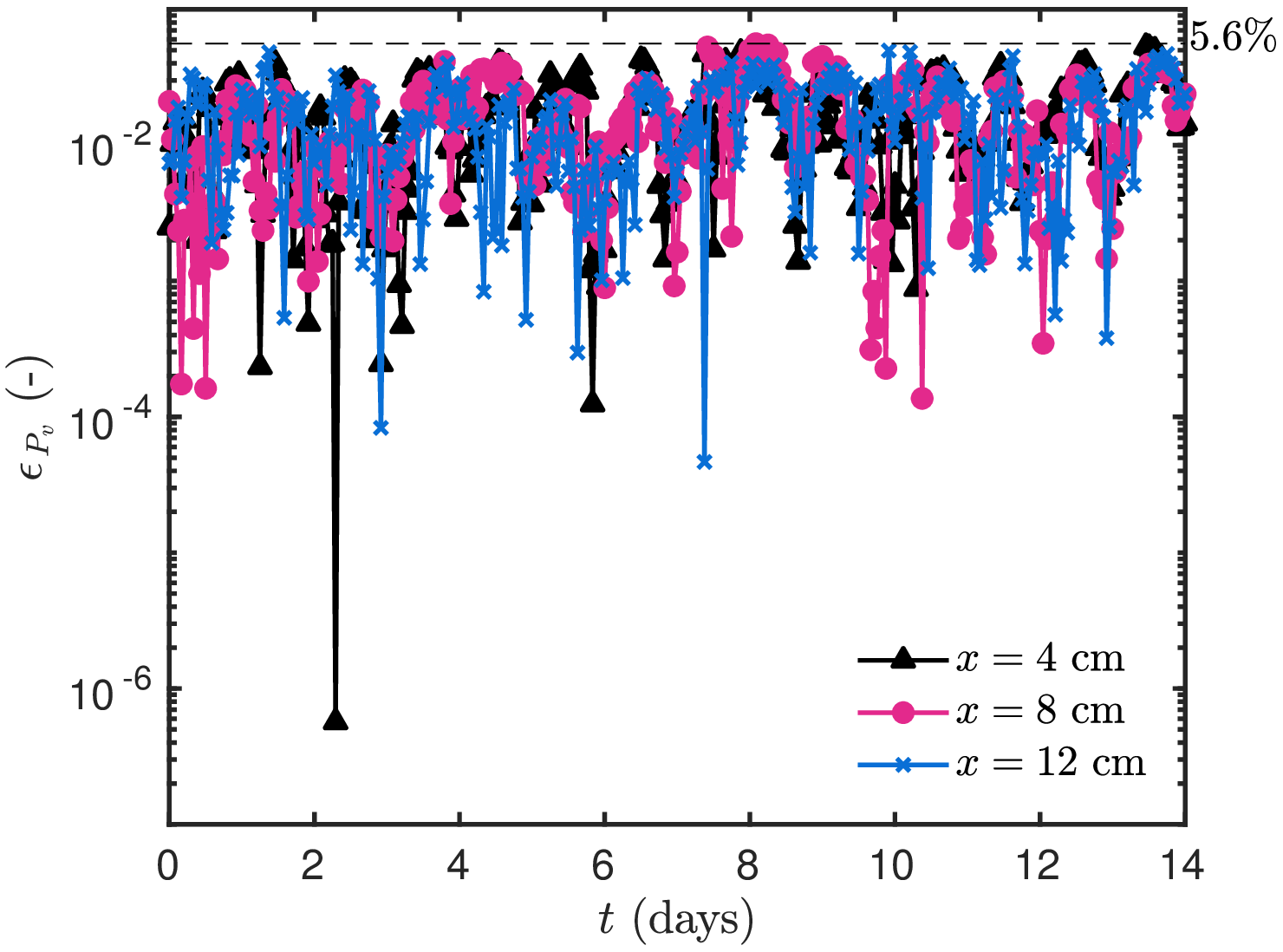}}
\caption{\small\em Relative errors regarding temperature (a) and vapour pressure (b) values at different positions within the porous material.}
\label{fig_AN3:error}
\end{figure}

The relative error $\epsilon$ of the computed solutions are present in Figure~\ref{fig_AN3:error}. Solutions simulated with the Spectral method showed a good agreement with the reference data. The maximum relative error for the temperature solution is of $0.7\%$ and for the vapour pressure solution is of $5.6\%$. The average error on the dynamic profiles is $0.2\%$ for temperature and $1.75\%$ for vapour pressure, which are close to the values obtained by \textsc{Rouchier} \etal \cite{Rouchier2016}.

The spectral method can provide a very accurate solution of the physical model. In this case, the last spectral coefficient for the temperature is of order of $b_{\,n}\, \simeq\, \O(10^{\,-\,5})$ and for the vapour pressure of order of $a_{\,n}\, \simeq\, \O(10^{\,-\,3})\,$, for $N \egalb 8\,$. The difference observed in Figure~\ref{fig_AN3:T4} can be caused by differences in the mathematical model and by the uncertainties of the estimated properties that are not taken in consideration \cite{Busser2018a}. Despite of this disparity, simulated results can be considered satisfactory to predict the heat and moisture transport.

As the boundary conditions are of the \textsc{Dirichlet}-type, the IMEX approach is more efficient in this case, taking only $4$ times more CPU time than the Spectral reduced-order model approach. For a single simulation, it may not appear important, however, when the direct model has to be simulated thousands of times as in \cite{Rouchier2016}, the gains become much more expressive.


\section{Conclusions}
\label{sec:conclusion}

In the present work, we showed that the unsteady heat and moisture transfers problem is solved efficiently with the spectral reduced-order model. This approach has been evaluated on two numerical unidimensional case studies of heat and moisture transfer in porous media. Each case aimed at exciting the nonlinear properties of the material to induce sharp profiles of temperature and vapour pressure. The first case considered a single material layer with sinusoidal boundary conditions. The second case took into account the rain effect at one boundary through a multi-layered material. For all cases, the spectral method has shown a high accuracy and perfect agreement with the respective reference solutions. The maximum global error was of the order of $\O(10^{\,-\,3})\,$. The advantage of the proposed method is the low computational burden $7$ times more efficient than the IMEX approach. In sensitivity analysis and inverse problems, when one has to simulate the system of differential equation many times, the spectral approach becomes very attractive.

Besides the numerical benchmark, one case with experimental data was performed to confirm the physical model and the spectral-based solution method. Results of the simulation showed a good agreement with the experimental data, which confirms all the process of simulation. Results can be improved by adding the hysteresis and the liquid transport effects. Although, this information was not available for this material and this experiment.

Further research should be dedicated to multiple space dimensions and to the application considering both the diffusive and advective transfer, which complexity increases considerably due to the nonlinearities of the problem.


\section*{Acknowledgements}

This study was financed in part by the Coordena\c{c}\~{a}o de Aperfei\c{c}oamento de Pessoal de N\'{i}vel Superior -- Brasil (CAPES) -- in the framework of the International Cooperation Program CAPES/COFECUB (Grant $\#774/13$). The Authors acknowledge also the support from CNRS/INSIS and Cellule \'Energie under the grant MN4BAT-2017. Finally, Professor~\textsc{Mendes} thanks the Laboratory LAMA UMR 5127 for the warm hospitality during his visits in 2018, which were supported by the project MN4BAT under the AAP Recherche 2018 programme of the University \textsc{Savoie Mont Blanc}.


\newpage
\addcontentsline{toc}{section}{Nomenclature}
\section*{Nomenclature}

\begin{table}[h!]
\centering
\begin{tabular*}{0.7\textwidth}{@{\extracolsep{\fill}} |@{} >{\scriptsize} c >{\scriptsize} l >{\scriptsize} l| }
\hline
\multicolumn{3}{|c|}{\emph{\textbf{Nomenclature}}} \\
\multicolumn{3}{|l|}{\emph{Latin letters}} \\
$\cz$ & material specific heat & $[\unitfrac{J}{(kg\cdot K)}]$ \\
$\cw$ & liquid water specific heat  & $[\mathsf{J/(kg\cdot K)}]$ \\
$\cM$ & moisture storage coefficient & $[\unitfrac{s^2}{m^2}]$ \\
$\cT$ & energy storage coefficient & $[\unitfrac{J}{(m^3\cdot K)}]$ \\
$\hM$ & convective vapour transfer coefficient & $[\unitfrac{s}{m}]$ \\
$\hT$ & convective heat transfer coefficient & $[\unitfrac{W}{(m^2\cdot K)}]$ \\
$H_{\,l}$ & liquid water enthalpy & $[\unitfrac{J}{kg}]$ \\
$g_{\,\infty}$ & liquid flow & $[\unitfrac{kg}{(m^2\cdot s)}]$ \\
$g$ & flow & $[\unitfrac{kg}{(m^2\cdot s)}]$ \\
$\kl$ & liquid permeability & $[\mathsf{s}]$ \\
$\kM$ & moisture transf. coeff. under vap. press. grad. & $[\mathsf{s}]$ \\
$\kTM$ & heat transf. coeff. under vap. press. grad.& $[\unitfrac{m^2}{s}]$ \\
$\kT$ & heat transf. coeff. under temp. grad. & $[\unitfrac{W}{(m\cdot K)}]$ \\
$L$ & length & $[\mathsf{m}]$ \\
$\Lv$ & latent heat of vaporization & $[\mathsf{J/kg}]$ \\
$\mathbf{n}$ & normal space, that assumes either $+1$ or $-1$ & $[-]$ \\
$\Ps$ & saturation pressure & $[\mathsf{Pa}]$ \\
$\Pv$ & vapour pressure & $[\mathsf{Pa}]$ \\
$q$ &  heat flux & $[\unitfrac{W}{m^2}]$ \\
$\Rv$ & water gas constant & $[\mathsf{J/(kg\cdot K)}]$\\
$T$ & temperature & $[\mathsf{K}]$ \\
$w$ & moisture content & $[\mathsf{kg/m^3}]$\\[5pt]
\multicolumn{3}{|l|}{\emph{Greek letters}} \\
$\delta_{\,v}$ & vapour permeability & $[\mathsf{s}]$ \\
$\phi$ & relative humidity & $[-]$ \\
$\rho$ & specific mass & $[\mathsf{kg/m^3}]$ \\
$\lambda$ & thermal conductivity & $[\mathsf{W/(m\cdot K)}]$ \\[5pt]
\multicolumn{3}{|l|}{\emph{Dimensionless parameters}} \\
$\mathrm{Bi}$ & \textsc{Biot} number & $[-]$ \\
$u$ & temperature & $[-]$ \\
$v$ & vapour pressure & $[-]$ \\
\hline
\end{tabular*}
\end{table}


\begin{appendices}

\newpage
\section{Material properties}
\label{annexe:material_properties}

\begin{table}[h!]
\centering
\caption{\small\em Hygrothermal properties of the load bearing material \cite{Hagentoft2004}.}
\bigskip
\setlength{\extrarowheight}{.3em}
\begin{tabular}{l p{9cm} l}
\hline
\textit{Property} & \textit{Value} & \textit{Unit} \\
\hline
Volumetric heat capacity & $\rhoz \,\cz \egal 2005\cdot 840 $  &  $ [\unitfrac{J}{m^3\cdot K}]$  \\
Sorption isotherm &  \parbox[t]{9cm}{$w\,(\phi) \egal 47.1 \, \biggl[1 + \Bigl(-1692.94 \cdot \ln(\phi)\Bigr)^{1.65}\biggr]^{\,-0.39} + \\ 109.9\, \biggl[1 + \Bigl(-2437.83\cdot \ln(\phi)\Bigr)^{6}\,\biggr]^{\,-0.83} $}  &  $  [\unitfrac{kg}{m^3}]$    \\
Vapour permeability & $\delta_{\,v}\,(\phi) \egal 6.413\cdot 10^{\,-9}\cdot \dfrac{\Bigl(1-\frac{w\,(\phi)}{157}\Bigr)}{0.503\, \Bigl(1-\frac{w\,(\phi)}{157}\Bigr)^{2}+0.497} $  &  $  [\mathsf{s}] $\\
Liquid permeability & $\kl\,(\phi) \egal 2.52\dix{-4}\cdot \exp(-1.55\dix{6}\cdot \phi) $  &  $  [\mathsf{s}] $\\
Thermal conductivity & $\lambda\,(\phi) \egal 0.5 \plus 0.0045\cdot w\,(\phi)$  &  $  [\unitfrac{W}{(m\cdot K)}]$\\
\hline                
\end{tabular}
\label{table:properties_loadbearing}
\end{table}

\begin{table}[h!]
\centering
\caption{\small\em Hygrothermal properties of the finishing material \cite{Hagentoft2004}.}
\bigskip
\setlength{\extrarowheight}{.3em}
\begin{tabular}{l p{9cm} l}
\hline
\textit{Property} & \textit{Value} & \textit{Unit} \\
\hline
Volumetric heat capacity & $\rhoz \,\cz \egal 790\cdot 870 $  &  $ [\unitfrac{J}{m^3\cdot K}]$  \\
Sorption isotherm &  $w\,(\phi) \egal 209 \, \biggl[1 + \Bigl(-2.7\dix{14} \cdot \ln(\phi)\Bigr)^{1.27}\biggr]^{\,-0.21} $  &  $  [\unitfrac{kg}{m^3}]$    \\
Vapour permeability & $\delta_{\,v}\,(\phi) \egal 6.413\cdot 10^{\,-9}\cdot \dfrac{\Bigl(1-\frac{w\,(\phi)}{209}\Bigr)}{0.503\, \Bigl(1-\frac{w\,(\phi)}{209}\Bigr)^{\,2}+0.497} $  &  $  [\mathsf{s}] $\\
Liquid permeability & \parbox[t]{9cm}{$\kl\,(\phi) \egal \exp[-33\plus 0.0704\cdot (w-120)\\-1.742\dix{-4}\cdot (w-120)^2 -2.795\dix{-6}\cdot (w-120)^3\\ -1.157\dix{-7}\cdot (w-120)^4 +2.597\dix{-9}\cdot (w-120)^5] $ } &  $  [\mathsf{s}] $\\
Thermal conductivity & $\lambda\,(\phi) \egal 0.2 \plus 0.0045\cdot w\,(\phi)$  &  $  [\unitfrac{W}{(m\cdot K)}]$\\
\hline                
\end{tabular}
\label{table:properties_finishing_mat}
\end{table}

\begin{table}[h!]
\centering
\caption{\small\em Hygrothermal properties of the wood fibre \cite{Rouchier2016}.}
\bigskip
\setlength{\extrarowheight}{.3em}
\begin{tabular}{l p{9.5cm} l}
\hline
\textit{Property} & \textit{Value} & \textit{Unit} \\
\hline
Volumetric heat capacity & $\cz\,\rhoz \egal 161.1 \dix{3} $  &  $   [\unitfrac{J}{(m^3\cdot K)}] $  \\
Sorption isotherm &  $f \, (\phi) \egal 7.063\dix{-5} \cdot \phi^{\,3} - 0.00736 \cdot \phi^{\,2} + 0.4105\cdot \phi + 0.2688 $  &  $  [\unitfrac{kg}{m^3}] $     \\
Vapour permeability  & $\delta_{\,v} \, (\phi)  \egal 6.36 \cdot \phi + 2.16\dix{-11} $  &  $  [\mathsf{s}] $\\
Thermal conductivity  & $\lambda\, (\phi,\,T)  \egal 0.038 + 0.192\cdot \dfrac{f\, (\phi)}{\rhol} + 1.08\dix{-4}\cdot T  $  &  $   [\unitfrac{W}{(m\cdot K)}] $ \\[5pt]
\hline 
\end{tabular} 
\label{table:properties_woodfibre}
\end{table}


\newpage
\section{Coefficients of the different materials}
\label{annexe:material_coefficients}

\begin{figure}[h!]
\centering
\subfigure[a][\label{fig:cm}]{\includegraphics[width=0.48\textwidth]{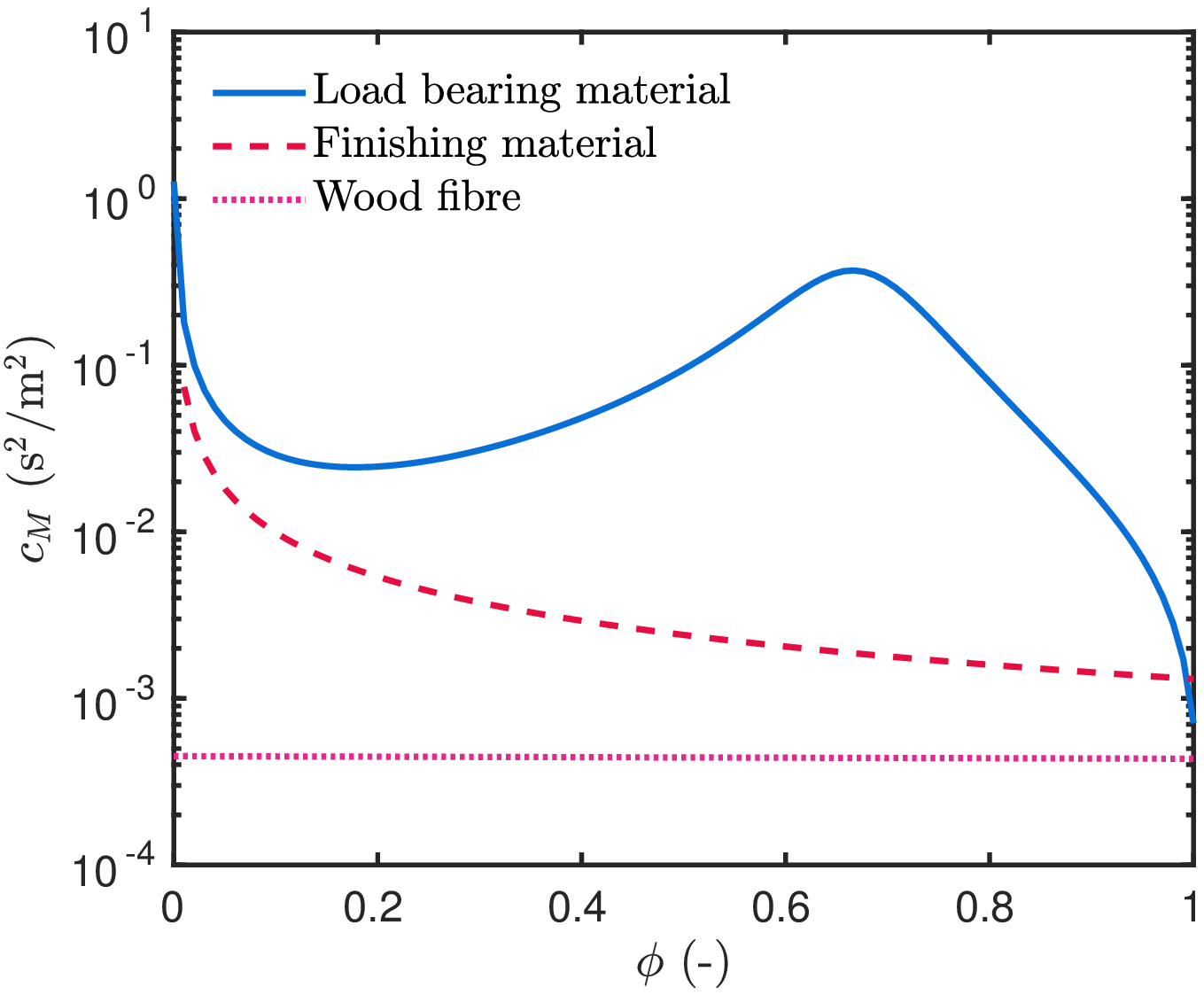}} \hspace{0.3cm}
\subfigure[b][\label{fig:cT}]{\includegraphics[width=0.48\textwidth]{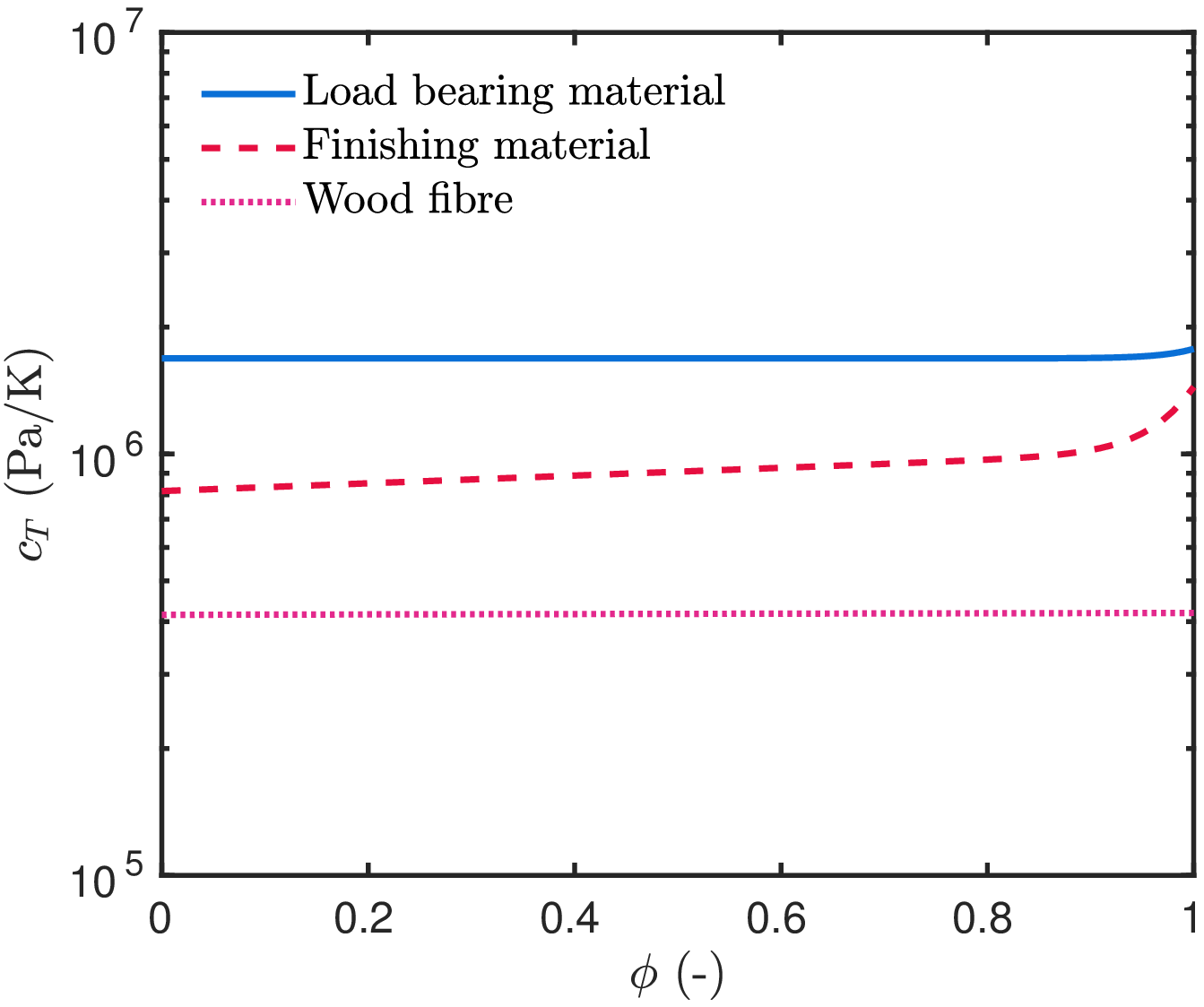}} \\
\subfigure[c][\label{fig:kM}]{\includegraphics[width=0.48\textwidth]{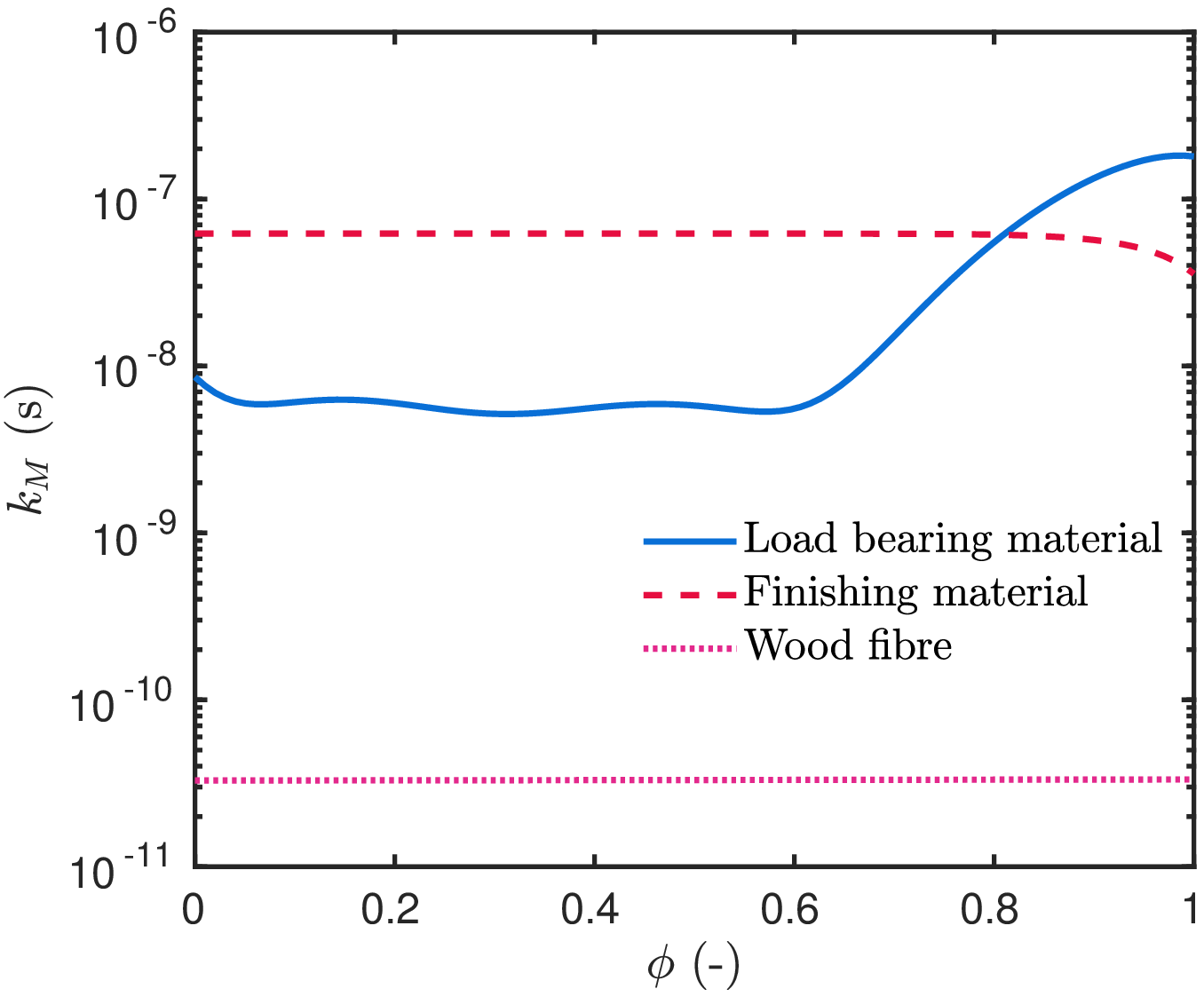}} \hspace{0.3cm}
\subfigure[d][\label{fig:kT}]{\includegraphics[width=0.48\textwidth]{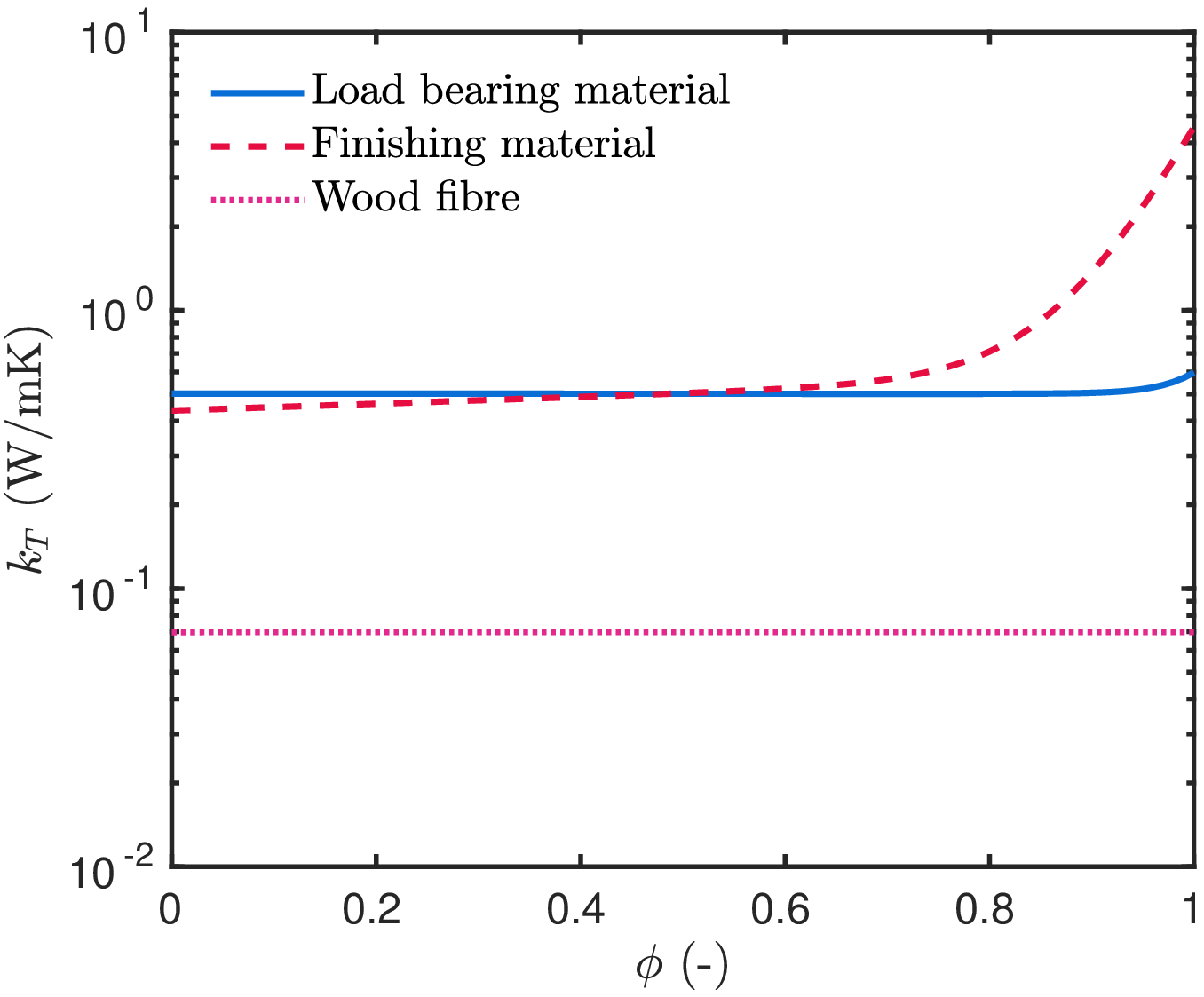}} \\
\subfigure[e][\label{fig:kTM}]{\includegraphics[width=0.48\textwidth]{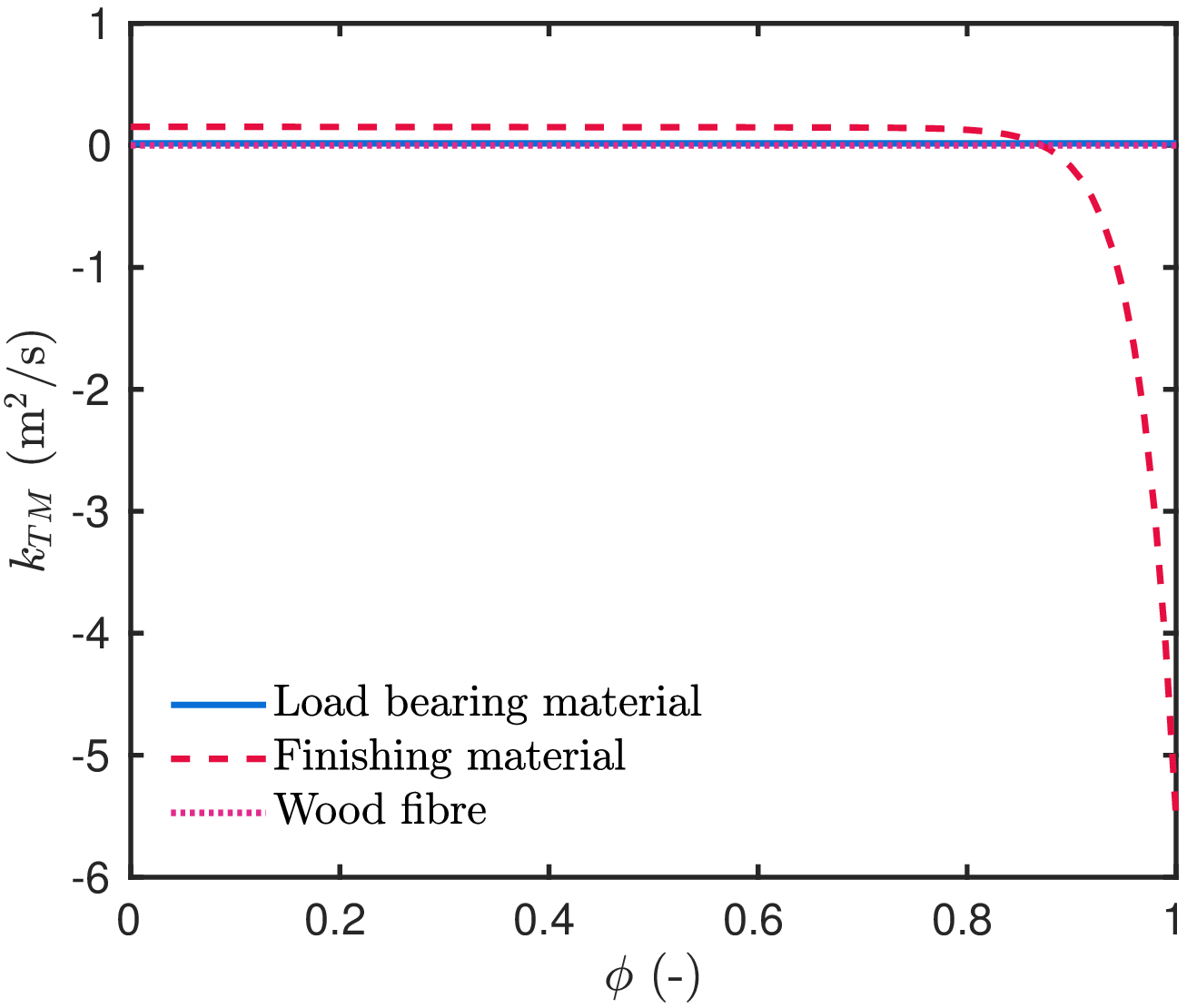}}
\caption{\small\em Coefficients $\cM$ (a), $\cT$ (b), $\kM$ (c), $\kT$ (d) and $\kTM$ (e) of the load bearing material, of the finishing material and of the wood fibre.}\label{fig_AN:properties_coeff}
\end{figure}

See Figure~\ref{fig_AN:properties_coeff}.

\section{Dimensionless values}
\label{annexe:dimensionless}

Some reference values were used on all the simulation cases. The reference time is $\tref \,=\, 1\, \mathsf{h}\,$, the equivalent to $3600\,\mathsf{s}$. The reference temperature was $\Tref \egalb 293.15\,\mathsf{K}$ and the reference of the vapour presure was $\Pvref \egalb 1166.9\,\mathsf{Pa}\,$. The reference length is the total length of the spatial domain $\Lref \egalb L \,\mathsf{m}\,$, so it is possible to have a dimensionless domain between $\xs \egalb [\,0,1\,]\,$.

\subsection{Case from Section~\ref{sec:case_1layer}}

The temperature boundary conditions are expressed as:
\begin{align*}
  \uinfL \, (\ts\,) &\egal 1 \moins 0.05\cdot \sin(\pi\, \ts/8760) \plus 0.01\cdot \sin(2\,\pi\,\ts/24) \,,\\
  \uinfR \, (\ts\,) &\egal 1 \plus 0.005\cdot \sin(\pi\, \ts/48) \,.
\end{align*}
and, of the vapour pressure boundary conditions as:
\begin{align*}
  \vinfL \, (\ts\,) &\egal \Bigl(0.5 \plus 0.45 \cdot \sin^{\,2} \left(\, 2\pi \, \ts/90\,\right) \Bigr)\cdot \Ps\Bigl(\uinfL \, (\,\ts\,)\cdot \Tref \Bigr)/\Pvref\,,\\
  \vinfR \, (\ts\,) &\egal \Bigl(0.5 \plus 0.4 \cdot \sin^{\,2} \left(\, 2\pi \, \ts/30\,\right) \Bigr)\cdot \Ps\Bigl(\uinfR \, (\,\ts\,)\cdot \Tref \Bigr)/\Pvref \,.
\end{align*}

For the dimensionless properties of the material, they can be written as:
\begin{align*}
  \cMs\,(v)\egal & \frac{ -4.15\dix{6} \cdot v^{\,3} \plus 7.76\dix{6}\cdot v^{\,2} \moins 5.44\dix{5}\cdot v \plus 3.53\dix{6}}{v^{\,4} \plus 9.18\dix{5}\cdot v^{\,3} \moins 2.47\dix{6}\cdot v^{\,2} \plus 1.68\dix{6}\cdot v \plus 5493} \,,\\
  \kMs\,(v)\egal & \moins 4.23\cdot v^8 \moins 4.901\cdot v^7 \plus 120.6\cdot v^6 \moins 340.7\cdot v^5 \plus 417.8\cdot v^4 \moins 255.9\cdot v^3 \\ &\plus 77.15\cdot v^2 \moins 10.33\cdot v \plus 1.57 \,,\\
  \cTs\,(v)\egal & 9.327\cdot \exp(-\ 2.4\dix{-4}\cdot v) \plus 1.457\dix{-14} \cdot \exp(15.58\cdot v) \,,\\
  \kTs\,(v)\egal & 0.9996\cdot \exp(-\ 8.813\dix{-4}\cdot v) \plus 5.65\dix{-15} \cdot \exp(15.58\cdot v) \,,\\
  \kTMs\,(v)\egal & 0.1276\cdot \exp(-\ 1.651\dix{-4}\cdot v) \,,
\end{align*}
with $\kMref \egalb 5.4712\dix{-9}\, \mathsf{s}$ and $\kTref \egalb 0.5021\, \mathsf{W/(m\cdot K)}\,$.

\subsection{Case from Section~\ref{sec:case_2layers}}

The dimensionless temperature and vapour pressure at the boundaries are written as in the previous case and also the materials properties of material 1. Thus, for the second material, properties are written as:
\begin{align*}
  \cMsd\,(v)\egal & 1.221\cdot v^{-0.878} \,,\\
  \kMsd\,(v)\egal & \moins 1.084\dix{-4}\cdot v^{15.44} \plus 11.34  \,,\\
  \cTsd\,(v)\egal & 4.52\cdot \exp(0.1058\cdot v) \plus 1.79\dix{-11} \cdot \exp(12.81\cdot v) \,,\\
  \kTsd\,(v)\egal & 0.8686\cdot \exp(0.1414\cdot v) \plus 9.498\dix{-7} \cdot \exp(7.968\cdot v) \,,\\
  \kTMsd\,(v)\egal & -1.884\dix{-11} \cdot \exp(14.25\cdot v) \plus 1.216 \cdot \exp(-\ 0.0284\cdot v)\,,
\end{align*}
with $\kMref$ and $\kTref$ equal to the previous case. In addition, the rain flow is expressed as:
\begin{align*}
  \gsinf\, (\ts\,) \egal 2.4\cdot \sin(\pi\,\ts/84)^{70}\,.
\end{align*}

\subsection{Case from Section~\ref{sec:validation_simulation}}

The boundary conditions are gathered from the experimental data and just admensionalized. The dimensionless form of the initial condition are:
\begin{align*}
  u_{\,0} \, (\xs) &\egal -0.08806 \cdot (\xs)^4 + 0.1688\cdot (\xs)^3 -0.1143\cdot (\xs)^2 - 0.01621\cdot \xs + 1.015 \,,\\
  v_{\,0} \, (\xs) &\egal -0.408\cdot (\xs)^4 + 1.188\cdot (\xs)^3 - 1.053\cdot (\xs)^2 + 0.08969\cdot \xs + 1.092 \,.
\end{align*}

The dimensionless properties of the material can be written as:
\begin{align*}
  \cMs\,(v)\egal & -0.663\cdot v\ +\ 37.52 \,,\\
  \kMs\,(v)\egal & 0.007289\cdot v\ +\ 0.9854  \,,\\
  \cTs\,(v)\egal & 0.08587\cdot v\ +\ 16.53 \,,\\
  \kTs\,(v)\egal & 0.0005546\cdot v\ +\ 0.9989 \,,\\
  \kTMs\,(v)\egal & 3.465\dix{-5}\cdot v\ +\ 0.004684 \,,
\end{align*}
with $\kMref \egalb 3.34\dix{\,-\,11}\, \mathsf{s}$ and $\kTref \egalb  6.98\dix{\,-\,2}\, \mathsf{W/(m\cdot K)}\,$.

\end{appendices}


\bigskip\bigskip
\addcontentsline{toc}{section}{References}
\bibliographystyle{abbrv}
\bibliography{biblio}
\bigskip

\end{document}